\documentclass[11pt]{article}
\usepackage{epsfig}
\usepackage{threeparttable}
\usepackage{amsmath}
\usepackage{amsfonts}
\usepackage{amssymb}
\begin{document}
\newcommand{\apj}{{ApJ}}
\newcommand{\pasp}{{PASP}}
\newcommand{\aj}{{AJ}}
\newcommand{\apjl}{{ApJL}}
\newcommand{\aaps}{{A\&A S}}
\newcommand{\mnras}{{MNRAS}}
\newcommand{\apjs}{{ApJS}}
\newcommand{\aap}{{A\&A}}
\newcommand{\prd}{{Phys. Rev. D}}
\newcommand{\nat}{{Nature}}
\newcommand{\physrep}{{Phys. Rep.}}

\centerline{{\em \bf Recent Research Developments in Astronomy \& Astrophysics - Vol. 2}}

\vspace*{5.00 cm}

\centerline{\bf\Large{Imaging the first light:}}
\centerline{\bf\Large{experimental challenges and future perspectives in the observation of}}
\centerline{\bf\Large{the Cosmic Microwave Background Anisotropy}}

\vspace*{2.00 cm}

\centerline{\em
  A. Mennella$^1$, C. Baccigalupi$^2$, A. Balbi$^3$, M. Bersanelli$^4$, C. Burigana$^5$, C. Butler$^5$, B. Cappellini$^4$,
}

\centerline{\em
    G. De Gasperis$^3$, F. Hansen$^3$, D. Maino$^4$, N. Mandolesi$^5$, M. Maris$^6$, G. Morgante$^5$, P. Natoli$^3$, 
}

\centerline{\em
   F. Pasian$^6$, F. Perrotta$^2$, P. Platania$^4$, 
   L. Valenziano$^5$, F. Villa$^5$ and A. Zacchei$^6$
}

\vspace*{1.00 cm}

\centerline{$^1$INAF -- IASF, Milano section, Via Bassini, 15, 20133 Milano (Italy)}
\centerline{$^2$SISSA, Via Beirut 2--4, 34014 Trieste (Italy)}
\centerline{$^3$Dipartimento di Fisica, Universit\`a di Roma, ``Tor Vergata'', 
  Via della Ricerca Scientifica, 1, 00133 Roma (Italy)}
\centerline{$^4$Dipartimento di Fisica, Universit\`a degli studi di Milano,
  Via Celoria 16, 20133 Milano (Italy)}
\centerline{$^5$INAF -- IASF, Bologna section, Via Gobetti 101, 40129 Bologna (Italy)}
\centerline{$^6$INAF -- OATs, Via Tiepolo, 11, 34131 Trieste (Italy)}

\vspace*{1.00 cm}

\begin{center}
{\em Short title}\\
{\bf Challenges and perspectives in the observation of CMB anisotropy}
\end{center}

\newpage

\centerline{\bf ABSTRACT} 

\vspace*{0.5cm}

\noindent Measurements of the cosmic microwave background (CMB) allow high precision 
observation of the Last Scattering Surface at redshift $z\sim$1100. After the success of the 
NASA satellite COBE, that in 1992 provided the first detection of the CMB anisotropy, 
results from many ground-based and balloon-borne experiments have showed a 
remarkable consistency between different results and provided quantitative estimates of 
fundamental cosmological properties. During 2003 the team of the NASA WMAP 
satellite has released the first improved full-sky maps of the CMB since COBE, leading 
to a deeper insight into the origin and evolution of the Universe. The ESA satellite Planck, 
scheduled for launch in 2007, is designed to provide the ultimate measurement of the 
CMB temperature anisotropy over the full sky, with an accuracy that will be limited only 
by astrophysical foregrounds, and robust detection of polarisation anisotropy. In this 
paper we review the experimental challenges in high precision CMB experiments and 
discuss the future perspectives opened by second and third generation space missions like 
WMAP and Planck.

\section{INTRODUCTION}
\label{sec:introduction}

    Quite often in the history of science major discoveries came from people
    who were not looking for them. This was certainly the case of the first
    observation of the Cosmic Microwave Background (CMB) radiation by Penzias
    and Wilson in 1964-65 \cite{penzias65}: they were lucky enough to find themselves at
    the right time and place, and also skilled enough to pursue to the very
    end the hint that Nature was offering them. But besides luck and skill two major
    factors were decisive in this discovery.
    
    First, the right technology was starting to become available. Advances in
    millimetre-wave technology after the end of World-War II gave a boost to
    the young field of radio-astronomy, begun with the pioneer works of Jansky
    and Reber \cite{reber40, reber44}. 
    The horn antenna and low noise receiver at Bell Labs was
    probably at the time the most sensitive microwave instrument on Earth. 
    
    Second, in the mid 60's the scientific stage was ready for this
    breakthrough. In 1929 Edwin Hubble's observation of the recession of galaxies
    revolutionised the vision of the Universe introducing the concept of
    ``evolution'' at cosmological scales \cite{hubble29}. About a decade before Hubble's
    discovery, Alexander Friedmann showed that Einstein's General
    Relativity equations were compatible with a variety of scenarios, both
    static (as developed by Einstein by introducing the ``Cosmological
    Constant'', $\Lambda$) and dynamic. 

    In the mid 1940's, George Gamow and
    his group developed a physical theory of the early Universe that was
    compatible with both General Relativity and Hubble's observations.
    They extrapolated physics back in time to the point where temperature and
    density were sufficiently high to support nuclear fusion, and
    studied the production of heavy elements from primordial protons. A
    remarkable side-prediction of the calculations of Gamow, Halper and
    Hermann (see, e.g., \cite{gamow46}) was 
    the presence of a photon field in the ``primordial
    fireball'', now red-shifted and cooled to very low temperatures by cosmic
    expansion.
    Penzias and Wilson, unaware of Gamow's results, observed for the
    first time this relic, isotropic, radiation at $\sim 3$~K in the range of
    microwaves. Their discovery was more than an experimental confirmation of
    a theoretical prediction: it showed that it was possible
    to measure directly (and precisely) properties of the Universe in a very
    young state, when all the processes were still in the linear regime,
    before structures had formed yet.
    
    Experimental and theoretical research in cosmology has been growing
    steadily since then, and CMB observations have been
    paying a major role in this growth. In the first decade or so after the
    CMB discovery, major efforts were aimed at understanding the nature of the
    CMB radiation itself and characterising its frequency spectrum. By the
    early 80's the Hot Big Bang prediction of a highly isotropic background
    with a nearly planckian spectrum was remarkably supported by observation.
    The scientific community gradually became more and more interested in the
    study of spectral distortions and spatial anisotropies of the CMB. In particular,
    anisotropies were believed to be present in order to
    explain the existence of local non-uniformities in the present-time matter
    distribution.
    
    A new phase was opened up by the COBE mission in the early 90's. The FIRAS
    instrument \cite{fixsen96} measured the CMB spectrum to be planckian at
    the level of 99.97\% in the frequency range 60-600~GHz with a temperature
    of $T_0 = 2.725\pm 0.002$~K. Coupled with sub-orbital measurements at low
    frequencies \cite{bersanelli94, deamici91, bensadoun93, smoot87, sironi91, salvaterra02} very
    stringent constraints to spectral distortion parameters were placed,
    leading to tight upper limits on energy injections
    in the early Universe. FIRAS demonstrated that ``precision cosmology'' is
    possible with accurate measurements of the CMB, and confirmed the maturity
    achieved by microwave and sub-mm technology.
    
    A second breakthrough came from COBE with the DMR instrument which
    provided the first unambiguous detection of anisotropies at the level
    of $\Delta T/T \approx 10^{-5}$ on large angular scales ($\sim
    7^\circ$) \cite{smoot92}. This result immediately stimulated
    many new experiments aiming at measuring the CMB angular
    distribution with increasing resolution and sensitivity. This explosion of
    experimental effort was motivated by the realisation that accurate
    measurements of the statistics of CMB anisotropies, reflected in its
    angular power spectrum from large scales to about $10'$, yield powerful
    constraints on fundamental parameters of cosmology such as the Hubble
    constant, $H_0$, the baryon density, $\Omega_b$, the dark energy density,
    $\Omega_\Lambda$, etc. To date, more than 20 independent projects have
    been carried out with different technologies and from a variety of ground-based
    and balloon-borne experiments, recently with remarkable precision.
    
    In early 2003 the NASA space mission WMAP has released the first full-sky map
    of the CMB with sub-degree angular resolution, setting tight constraints
    on many cosmological parameters. The ESA Planck mission, to be launched in
    2007, will exploit CMB temperature anisotropy measurements to its
    fundamental limits imposed by unavoidable cosmic variance and
    astrophysical foregrounds, and will likely open a new phase in CMB science
    aimed at a precise measurement of anisotropies in the CMB polarisation
    state.
    
    Although technological advances have allowed a steady increase in the
    sensitivity of CMB measurements, it has also increased
    the level of complexity of instruments and satellites. For example, the
    quest for high sensitivity leads to highly sophisticated cooling systems
    and to multi-feed arrays, which represent new challenges for the thermal
    and optical design of the instruments. The combination of instrument
    complexity and high sensitivity leads to very severe requirements in terms
    of rejection of systematic effects. In the second and third generation CMB
    space missions (namely WMAP and Planck) the systematic error control has
    become one of the most (if not {\em the} most) critical experimental
    challenge, often pushing the understanding of current technologies into
    poorly known grounds. 

    Many excellent CMB reviews have been published covering
    both theoretical and experimental aspects (see, e.g., \cite{bersanelli02} and 
    references therein); however,
    the rapid evolutions in CMB science and in
    its related technologies often renew the need of state-of-the-art analysis.
    In this paper we present an overview of the main
    experimental issues of current efforts devoted to
    precision CMB measurements and discuss the main implications for the
    future challenges represented by precision polarisation anisotropy
    measurements. In particular the paper presents a discussion
    about the control of systematic effects
    in second and third generation space CMB experiments.
    
    After a brief summary of the CMB theoretical background (Sect.~\ref{sec:theoretical_background}) 
    and of the main astrophysical
    limitations (Sect.~\ref{sec:astrophysical_limitations}) we will review
    the main issues concerning the control of systematic errors in
    high-precision CMB measurements from space, with examples taken from the
    (still growing) experience formed in the context of the Planck mission.
    Sect.~\ref{sec:cmb_experiments} gives a short account of the evolution of CMB
    experiments from COBE to Planck, highlighting the deep
    relationship between progress of technology  and increase 
    in the scientific achievements as well as experimental challenges.
    In the last section we provide a discussion of
    our foreseen scenarios in CMB science after Planck.


\section{THEORETICAL BACKGROUND}
\label{sec:theoretical_background}

    \subsection{The hot big bang model}
    \label{subsec:hot_big_bang_model}

        In the Hot Big Bang cosmological model, the expansion of the Universe started
        about 14 billion years ago from a state with high density and temperature
        ($\rho \simeq 10^{25}$gr/cm$^3$, 
        $T\simeq 10^2$GeV $\sim 10^{15}$~K at $t\simeq 10^{-8}$~s)
        and continues now with a progressive cooling.
        At primordial high temperatures matter and radiation were tightly coupled and can be
        regarded as a simple fluid. This holds up to $t\simeq 3.8\times 10^5$ years when temperature
        dropped to $T\simeq 3000$~K: nuclei could then capture electrons
        to form neutral hydrogen as well as other light elements ($^3$He, $^4$He, $^7$Li).
        This ``recombination'' process reduced the Thomson scattering cross 
        section and photons 
        were free to propagate and, since then, they have interacted only 
        gravitationally with matter.
        These photons are the Cosmic Microwave Background and the surface surrounding us at 
        recombination is where CMB photons have last directly interacted 
        with matter (the Last Scattering Surface, LSS).

        The Hot Big Bang model is the standard framework for cosmology since it is strongly
        supported by mainly three observational evidences: the abundances of the primordial
        light elements, the expansion of the Universe as observed by Hubble in 1926 \cite{hubble29}
        and the existence of the CMB.

        Although successful the Hot Big Bang model leaves many open issues: the so-called
        flatness problem (i.e. why the density of the Universe is so close to the critical one),
        the horizon problem (i.e. why regions separated by more the $\simeq 2^\circ$, the
        angular size of causal connected regions at LSS, have the same temperature) and the
        origin of structures in the Universe (i.e. why the Universe is so uniform on very large
        scales and indeed so clumpy on smaller ones).
        These issues are now solved by the inflationary paradigm, i.e. an exponential 
        expansion that happened at very early times ($t\simeq 10^{-34}$~s). 
        A causally connected region expanded to encompass the 
        observable Universe;
        furthermore quantum fluctuations present during
        inflation were stretched by the expansion to
        become, eventually, density perturbations which left their 
        signature in the spatial distribution 
        of the CMB. Such perturbations are predicted to have a power-law power spectrum, 
        $P(k) \propto k^n$, close to scale invariant $n=1$.
        
\begin{figure}
          \begin{center}
           \epsfig{file=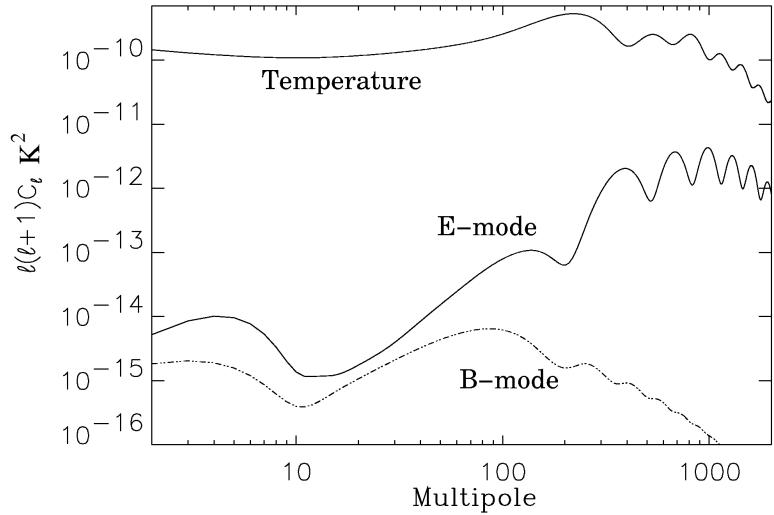,width=4.in}
          \end{center}
          \caption{A typical angular power spectrum from temperature (upper curve) and polarisation $E$ and
            $B$ modes (lower solid and dashed curves).}
          \label{cmbtpcl}
        \end{figure}

        A common representation of the CMB field is in terms of spherical-harmonic
        expansion:
        $\Delta T/T = \sum_{\ell m} a_{\ell m} Y_{\ell m}(\theta,\phi)$
        where the multipole $\ell \sim 180^\circ/\theta$ and the $a_{\ell m}$ coefficients are the 
        multipole moments which are predicted to have zero mean and variance given by
        $\langle |a_{\ell m}|^2\rangle=C_\ell$
        (the angle brackets means average over all the observers in the Universe).
        The set of $C_\ell$ are known as the angular power spectrum and comprise all the cosmological
        information if the fluctuations are Gaussian in origin (a typical example is shown
        in the upper curve of Fig.~\ref{cmbtpcl}). These coefficients are the key
        prediction of any theoretical cosmological model. Alternative theories start from non-Gaussian
        primordial fluctuations and therefore their predictions could be verified looking at higher
        order moments in the CMB field. 
        
        The plateau at low $\ell$ (angular scales larger that the horizon at
        recombination $\simeq 2^\circ$) is due to the Sachs-Wolfe effect \cite{sachs67} i.e. the gravitational 
        redshift and time dilation suffered by CMB photons when climbing potential wells induced by density 
        perturbations.
        If the initial power spectrum is scale-invariant then
        $C_\ell \propto \frac{1}{\ell(\ell+1)}$. It is therefore usual to plot the power spectrum
        in terms of $\ell(\ell+1)C_\ell$
        as a function of $\ell$ so that the Sachs-Wolfe plateau is easily recognised.
        
        On angular scales $0.1^\circ\lesssim \theta\lesssim 2^\circ$ causal processes in the photons-baryons
        fluid produce the oscillatory behaviour in the CMB power spectrum. 
        Since recombination is nearly instantaneous, different oscillation modes 
        were ``frozen'' at different oscillation phases. 
        The first peak is due to an oscillation
        with a density maximum at recombination; the other peaks are 
        harmonics of the main oscillation. 
        Note that between peaks velocity maxima (which are $90^\circ$ out of phase to density maxima) prevent the power
        spectrum to go to zero.

        Because in the short time of recombination CMB photons can diffuse a certain distance, anisotropies
        on scales smaller than this mean free path are erased by diffusion. This leads to the nearly exponential
        damping in the power spectrum (the so-called Silk damping \cite{silk68}). Little contribution is expected
        at such scales from intrinsic CMB anisotropies.
        
        CMB could be in principle also polarised due to Thomson scattering 
        (which should happen in optically thin regions)
        between CMB photons and
        electrons at recombination. Furthermore it is possible to show that 
        scattering produces net polarisation if CMB photons distribution has
        a quadrupole moment. This defines the main properties of CMB polarisation anisotropies
        which are expected (and recently detected) to be of the order of 5-10\% of the temperature anisotropies (less photons
        contribute) and to peak at scales
        smaller that the horizon at recombination since the process involved is causal (see Fig~\ref{cmbtpcl}). 
        Any polarisation field could 
        be decomposed into a curl-free and pure-curl components (the $E$ and $B$ modes). While density perturbations
        can produce only $E$ modes, primordial gravity waves instead give raise to $B$ modes, a very weak component expected to
        peak at large angular scales. 

    \subsection{CMB and cosmological parameters}
    \label{subsec:cosmological_parameters}

\begin{figure}[here!]
          \begin{center}
           \epsfig{file=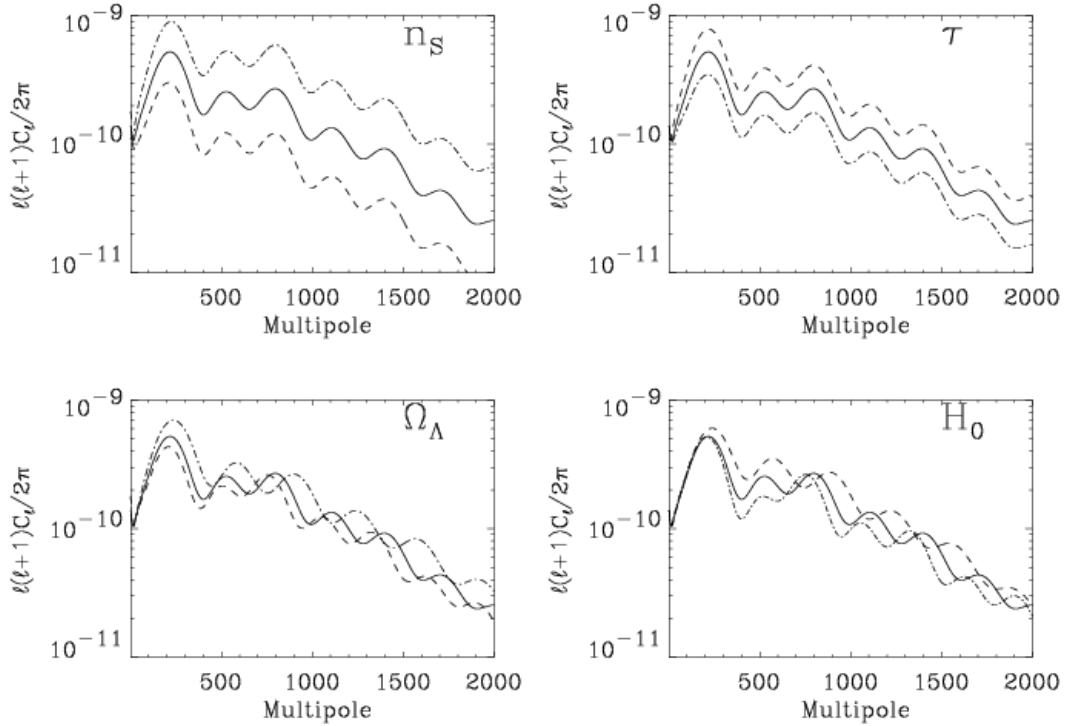,width=14.cm}
          \end{center}
          \caption{Examples of $C_\ell$ dependence on some cosmological parameters}
          \label{cosmoparam}
        \end{figure}

        The details of the acoustic features and the overall shape of the CMB angular power spectrum
        depend critically on the value of several cosmological parameters such as the baryon density $\Omega_{\rm b}$,
        the matter density $\Omega_{\rm m}$, the Hubble constant $H_0$, the vacuum energy $\Omega_\Lambda$ and
        many others, which can in principle be accurately determined by a precise measurement of the CMB power
        spectrum. Fig.~\ref{cosmoparam} shows the dependence of $C_\ell$ on some cosmological parameters.

        \begin{itemize}
          \item {\em Total Density}: decreasing $\Omega_0$ corresponds to a decrease in the curvature with a corresponding shift 
            of peaks toward high $\ell$s. In particular the position of the first peaks is $\ell_1 \approx 200\sqrt{\Omega_0}$.
          \item {\em Baryon Density}: increasing $\Omega_{\rm b}$ increases odd peaks (compression phase in the fluid) due
            to extra inertia from baryons with respect to even peaks.
          \item {\em Hubble constant}: a decrease in $h$ ($h\equiv H_0/100$~km s$^{-1}$ Mpc$^{-1}$) with constant 
            $\Omega_{\rm b}h^2$ means a delay in the epoch of matter-radiation equality and a different expansion rate. Peaks
            are boosted and slightly changed in the $\ell$ position.
          \item {\em Cosmological constant}: increasing $\Omega_\Lambda$ in a flat Universe, implies again a delay in
            matter-radiation equality with peak boosts and shifts (furthermore low $\ell$'s show the Integrated
            Sachs-Wolfe effect).
          \item {\em Spectral index}: increasing $n$ will raise the $C_\ell$ at large
            $\ell$'s with respect low ones.
          \item {\em Reionisation}: if Universe is re-ionised after recombination, the power spectrum at
            $\ell > 100$ would be suppressed by a factor $e^{-2\tau}$. Recent WMAP data suggest $\tau \simeq 0.17$
        \end{itemize}

        It is worth mentioning that some degeneracy between cosmological parameters do exist and different
        sets of cosmological parameters may produce nearly the same angular power spectrum.
        Some of these degeneracies
        (e.g. for the parameters $n$ and $\tau$) could be broken
        combining CMB temperature anisotropy data with polarisation data. 
        However information from other data sets
        (e.g. Supernovae Type Ia data, Large Scale Structure information, HST Project) are essential in order to be
        combined with CMB data and provide the complete picture.
        
        In general the best estimates of a set $\mathbf{p}=(p_1,p_2,\dots,p_n)$ of cosmological parameters 
        from the measured $C_\ell$ could be obtained in a Bayesian sense by maximizing the likelihood function 
        defined as $\mathcal{L}(\mathbf{p})\propto \mathcal{P}(C_\ell|\mathbf{p})\mathcal{P}(\mathbf{p}|{\rm prior})$.
        This is the probability of obtaining the observed $C_\ell$ given a set of $\mathbf{p}$ times
        the {\em prior} parameters knowledge.
        The actual situation is not so simple and some complications arise. One is due to the fact that the likelihood 
        is not Gaussian in $C_\ell$ and additional quantities have to be added for a proper likelihood characterisation \cite{bond00}. 
        Furthermore one is eventually interested in the probability distribution for a given parameter. In a
        Bayesian formalism the {\em posterior} probability of one or more parameters is obtained by
        marginalisation (integration) or maximisation over the other parameters. This
        is not a trivial task since it involves multi-dimensional integration and furthermore the
        overall shape of the likelihood influences the results even on a small subset of parameters.
       
        One way to proceed is to compute the CMB power spectrum on a grid of cosmological parameters.
        This is time consuming since with only 6 parameters and 10 steps for each parameter, a total
        number of 10$^6$ models must be evaluated. This approach is at limit for sub-orbital
        experiments but is probably unfeasible for space missions like WMAP and Planck. Another
        approach, applied to the analysis of WMAP data \cite{verde03}, 
        involves Monte Carlo Markov Chains in which
        the parameters space is sampled randomly and results are rejected/accepted according to 
        a precise acceptance rule. This has the advantage of scaling linearly with number of parameters allowing
        the inclusion of many parameters with low computational cost.

    \subsection{Secondary anisotropies}
    \label{subsec:secondary_anisotropies}

        The anisotropies generated during the travel of CMB photons from
        the last scattering surface to the observer are generally referred to as
        secondary anisotropies, which can be divided into two broad classes:
        (i) those due to scattering of CMB photons
        occurring after recombination, and (ii) those arising from gravitational
        effects taking place along the path travelled by the photons.
        
        One of the main causes of secondary anisotropies is 
        hydrogen reionisation by ultraviolet light from the first stars and quasars
        (see, e.g., \cite{madau03}).
        Observations of the Gunn-Peterson effect in distant quasars indicates that
        the Universe was completely reionised at a redshift $z\approx 6$
        \cite{Becker01}, although little is known about the detailed
        ionisation history at higher redshifts. The main effect on CMB
        anisotropies from early reionisation is the damping of power due to
        the scattering of CMB photons by the free electrons \cite{sugiyama93}.
        The parameter governing this effect is the reionisation optical
        depth, $\tau$: peaks in the power spectrum are
        suppressed by roughly a factor $e^{-2\tau}$ with respect to the low-l
        plateau. This effect, however, is similar to the effect of a tilted
        spectral index in the primordial power spectrum \cite{debernardis97}, 
        so that an accurate determination of $\tau$ by CMB temperature
        observations alone is not straightforward.  This degeneracy can be broken by
        polarisation information,
        since the scattering of CMB photons by the
        ionised medium causes the generation of large-angle polarisation anisotropy \cite{hu00},
        providing a unique signature of reionisation.  This effect has
        recently been used by the WMAP team to determine $\tau$, setting the
        epoch of reionisation at redshift $z=17\pm 5$ \cite{kogut03, spergel03}.
        
        Reionisation of the Universe is expected to be
        a highly inhomogeneous process.  Information on the detailed spatial
        distribution of the ionised medium may be obtained by studying the so-called 
        kinetic Sunyaev-Zel'dovich effect (see, e.g., \cite{valageas01}),
        arising from the bulk motion of ionised regions, which can
        provide information on the size of the ionised regions and
        correlations between them. The magnitude of the effect, however, is
        very small, requiring fine angular resolution (less than $10^\prime$)
        and very high sensitivity (1~$\mu$K) in order to be detected.
        
        Scattering of CMB photons by hot electrons in clusters of galaxies
        results in the thermal Sunyaev-Zel'dovich effect \cite{sunyaev72}.
        Energy is transferred from the hot electrons to the
        CMB photons by inverse Compton scattering, leaving a distinct
        signature in the CMB spectrum as photons are moved from the
        Rayleigh-Jeans region into the Wien tail. Recently, CBI and BIMA have
        detected  an excess of power at high multipoles ($\ell > 2000$) in the CMB angular
        power spectrum \cite{komatsu02} that could be explained by
        the cumulative effect of unresolved high-redshift clusters.
        
        CMB photons experience gravitational redshift as they pass through
        regions of varying gravitational potential. This effect is known as
        integrated Sachs-Wolfe effect \cite{sachs67} and is dominant at
        large angular scales.
        This effect is strongly dependent on the dark energy content of the
        Universe, but is unfortunately hard to detect due to the presence of
        cosmic variance uncertainties at low multipoles.
        
        Finally, when the CMB photons come across a mass concentration, their
        trajectories are deflected by gravitational lensing \cite{blanchard87},
        causing a distortion of hot and cold spots in the
        CMB. For the temperature, this results in a redistribution of power
        and a smoothing of features in the angular power spectrum, which is
        increasingly important at smaller angular scales \cite{seljak96}, while
        for polarisation, lensing is expected to be a spurious source of B-mode polarisation anisotropy.

    \subsection{Non Gaussianity}
    \label{subsec:non_gaussianity}

        Standard inflationary models predict the primordial fluctuations in the gravitational 
        potential $\Delta\Phi$ to be nearly Gaussian distributed, which reflects in a Gaussian
        distribution of the CMB temperature 
        fluctuations and of its spherical harmonic coefficients $a_{\ell m}$.
        As a Gaussian field is completely described by its mean and (co-)variances and the mean of the 
        $a_{\ell m}$ is zero, all the information in the CMB fluctuation field is completely contained in the variance 
        of the $a_{\ell m}$, given by the power spectrum $C_\ell$. However, if the field is to some degree 
        non-Gaussian, more information can be found in the higher order moments.

        There are several physical effects which could introduce some small non-Gaussianities in the CMB 
        (see, e.g., \cite{linde97, contaldi99, martin00, bartolo02, gangui02, gupta02, liguori03}). 
        We will mention here three 
        possible effects, (i) non-Gaussianities from inflation, (ii) presence of cosmic strings and (iii) a complex 
        global topology of the Universe. The first effect comes from the fact that the fluctuations in the 
        gravitational potential induced by inflation is a complicated function of a small Gaussian quantity 
        $\phi$ (related to the {\em inflaton} field). To second order in $\phi$, the gravitational perturbations can be written as
        \begin{equation}
          \Delta\Phi=\phi+f_{\rm NL}(\phi^2-<\phi^2>).
        \end{equation}

        The second term in this equation introduces a small non-Gaussian part into 
        the gravitational potential and thus into the 
        CMB temperature field. The coefficient $f_{\rm NL}$ is the non-linear coupling parameter used 
        to quantify the degree of 
        non-Gaussianity from inflation. With $f_{\rm NL}=1$, the non-Gaussian part of the temperature 
        field is of the order 
        $10^{-5}-10^{-6}$ times the Gaussian part. In standard inflationary models 
        $f_{\rm NL}\sim 10^{-1}$--$10^{-2}$ \cite{bartolo02, liguori03}, but when we take 
        into account evolutionary effects, 
        it might reach a value of about 3 \cite{bartolo03}. In more complicated inflationary models values of 
        $f_{\rm NL}$ up to 100 may be possible, but even in these extreme scenarios the non-Gaussian part will 
        be of the order $10^{-4}$ smaller than the Gaussian part, making this kind of non-Gaussianity hard to 
        detect.

        Another kind of non-Gaussianity could arise from topological defects, which could have 
        been created in 
        the early Universe during phase transitions of quantum fields predicted by high energy physics. 
        Particularly the so-called ``cosmic strings'' (see \cite{gangui01} for a review) are considered 
        a realistic possibility. The 
        cosmic string scenario was once competing with the inflationary theory for explaining the origin of 
        structure in the Universe. Although recent 
        observations of the CMB has excluded cosmic strings as the main contributor to structure formation,
        the possibility of a contribution from 
        cosmic strings has not been ruled out. In this case, they would show up as 
        small string-like discontinuities in the CMB temperature field.

        Finally, another scenario which could produce non-Gaussianity signatures in the CMB 
        is  a non-trivial global topology in the Universe.
        In fact Einstein's equations describe the geometry of the Universe, but leave the question 
        of the global topology open. Some models of complex topologies would give rise 
        not only to a non-Gaussian CMB field, 
        but also to a non-isotropic field. In particular they predict that structures seen in one direction could
        be repeated 
        in other directions on the sky (for a recent review see, e.g., \cite{levin02}).

        Several methods have been developed in order to detect non-Gaussianities in CMB maps. These methods can be
        divided into three classes:
        (i) pixel space methods, 
        (ii) spherical harmonic space methods and (iii) wavelet space methods. 
        
        The most basic tests for non-Gaussianity 
        in pixel space are tests which look at the higher order moments of the temperature field. In the simplest
        case, this means looking 
        at the skewness and kurtosis of the field. The more sophisticated methods look at the higher order
        correlation functions, 
        like the 3- or 4-point functions (see, e.g., \cite{eriksen02} for a detailed description of the method).
        A different method to search for non-Gaussianities in pixel space is by the use of the Minkowski
        functionals 
        (see, e.g., \cite{gott90, schmalzing98, novikov00}). There are three types of Minkowski functionals, 
        all depending on 
        a temperature threshold, $\nu$. The areas in the CMB sky maps being above above the level $\nu$ are called 
        ``hot spots'' and the areas below are called ``cold spots''. The first Minkowski functional 
        is just the total area of the hot spots for a given threshold $\nu$. The second is the contour
        length of all 
        the hot spots, and the third Minkowski functional (also called the ``genus'') is the number of 
        hot spots minus the number 
        of cold spots. For a Gaussian temperature field, the expectation value of these Minkowski
        functionals can be calculated 
        analytically and compared to the Minkowski functionals of an observed sky map. 
        Finally, one can also detect non-Gaussianity 
        by studying the peak-to-peak correlation functions \cite{heavens01}, the second 
        derivative of the temperature 
        field \cite{dore03} and other geometric estimators (as in \cite{barreiro01}).

        In spherical harmonic space, the most basic methods look at the higher order moments of the spherical harmonic coefficients 
        $a_{\ell m}$. In particular the bi- and tri-spectrum have been studied (e.g. 
        \cite{phillips01, winitzki00, hu01, kunz01, komatsu01, komatsu02a, detroia03}) and 
        used to limit the non-linear coupling parameter $f_{\rm NL}$. Another way to look for non-Gaussianity in 
        harmonic space is to analyse the complex phases $\phi_{\ell m}$ of the coefficients $a_{\ell m}$ \cite{chiang02}. 
        In a Gaussian model the phases should be uncorrelated between different $\ell$ and $m$ values, and non-Gaussianity 
        could be detected by searching for correlations. Using this method one can also localise the non-Gaussianity in multipole space. 
        Finally, one can look at the full distribution of the spherical harmonic coefficients, using the empirical process 
        method \cite{marinucci02, hansen02, hansen03}. By this method the distribution of the $a_{\ell m}$ is made uniform
        using the cumulative 
        distribution function, and then a Kolmogorov-Smirnov type test is used to check the degree of uniformity and 
        independence of the coefficients distribution.

        A third approach is the wavelet approach. As the wavelet transform is linear, the wavelet coefficients are 
        expected to be Gaussian distributed in a model with Gaussian temperature fluctuations. The skewness and kurtosis 
        of the wavelet coefficients are used to check for non-Gaussianity in CMB data 
        \cite{barreiro01a, barreiro00, martinezgonzalez02, mukherjee00}, and it 
        has been shown that this is a considerably stronger test than checking the skewness of the temperature field itself.

        From this summary one sees that the number of different tests to check for
        non-Gaussianity is large, but this is necessary as different types of non-Gaussianity tend to show up 
        in different tests. It is important to note that a detection of non-Gaussianity could very well be caused by 
        incorrectly subtracted foregrounds or systematic effects in the experiment. For the COBE experiment, 
        a non-Gaussian detection was 
        found using the bi-spectrum \cite{ferreira98, banday00}, that was later identified as a systematic effect. By applying the 
        non-Gaussianity tests to different frequency channels maps and by applying different galactic cuts, one can determine 
        whether a detection of non-Gaussianity is due to foreground contamination or if it could be of primordial origin.

        The maps produced by WMAP have been exposed to a set of different tests of non-Gaussianity. Unresolved 
        point sources show up as a non-Gaussian signal in CMB maps and the bi-spectrum was used to estimate the density 
        of point sources \cite{komatsu03}. Also, the bi-spectrum and the Minkowski functionals were used to constrain the 
        non-linear coupling parameter $f_{\rm NL}$ to $-58<f_{\rm NL}<134$ at the $2\sigma$ level \cite{komatsu03}. Tests for 
        phase correlations have been applied to the WMAP data and signatures of non-Gaussianity have been detected 
        \cite{chiang03, naselsky03a, coles03}.
        These non-Gaussianities have been found to be due to foreground contamination. An independent measurement of 
        the Minkowski functionals was performed by Park \cite{park03} who reports a non-Gaussian effect in the genus statistic. 
        Moreover the non-Gaussianity seemed to be confined to the southern galactic hemisphere. A similar asymmetry in 
        the higher order correlation functions and the power spectra were reported by Eriksen et al \cite{eriksen02}. 
        Also Vielva et al \cite{vielva03}
        found that the wavelet coefficients of the southern galactic hemisphere strongly reject the Gaussian hypothesis. This 
        asymmetry/non-Gaussianity appears to be independent of the frequency channel and galactic cut, excluding 
        foregrounds as the cause. The source of this non-Gaussianity has not yet been found and further studies of the current 
        WMAP maps and future CMB data sets are necessary to find its origin.

\section{ASTROPHYSICAL LIMITATIONS}
\label{sec:astrophysical_limitations}

    In any CMB experiment, the observed signal comprises several contributions other than
    the CMB itself and some are astrophysical in origin. In principle these foreground
    emissions can be separated from the CMB by multi-frequency observations exploiting the
    different spectral shapes of the signals to be separated. Due to the intrinsic smallness in the
    CMB signal and to the expected high sensitivity of present and future CMB experiments, the
    component separation process is quite delicate implying the need to disentangle the
    signals with few $\mu$K accuracy which is not trivial for both temperature and,
    even more, for polarisation measurements. Therefore microwave and sub-mm foreground emissions
    represent a challenge for CMB experiment although they provide interesting astrophysical
    information.
 
    \subsection{Galactic Synchrotron emission}
    \label{subsec:synchemission}
        Galactic diffuse synchrotron emission arises from cosmic rays electrons
        accelerated in the magnetic field showing large spatial variations due to the
        properties of the electrons energy distributions and to the structure of the
        Galactic magnetic field \cite{smoot99}. Synchrotron radiation dominates the
        microwave sky emission at $\nu \lesssim 10$~GHz. 
        A picture of the diffuse emission at low frequency is given by the 408~MHz full
        sky survey \cite{haslam82} and two wide coverage surveys at 1420~MHz \cite{reich86} 
        and at 2326~MHz \cite{jonas98}. However, large sky area surveys suffer significant 
        uncertainties related with calibration errors, zero level and scanning strategy 
        artifacts.
        Recent improvements on the analysis have minimised the
        systematic effects which depend on the specific experimental setup and observing
        strategy \cite{platania03}. The synchrotron brightness temperature is
        a power law function of frequency, $T_{\rm b}\propto \nu^{-\beta}$, with
        the spectral index $\beta$ that varies with frequency
        and position, with a mean value of 2.75 at $\nu < 10$~GHz \cite{platania03}.
        At higher frequencies the synchrotron map evaluated from the 
        WMAP data is now available with the synchrotron spectral index distribution between
        0.408 and 23~GHz \cite{bennett03}.
        In the radio, the angular dependence of the synchrotron emission has been 
        found to be $\ell^{-3}$ 
        \cite{tegmark96} with a flatter behaviour in some selected regions of the
        sky \cite{lasenby97}.
        Synchrotron radiation is expected to be, in principle, largely polarised
        (up to 70\%). This could be a potential issue for CMB polarisation measurements
        since the CMB signal is expected at few $\mu$K level. Furthermore, the
        knowledge of polarised synchrotron emission is poor and we lack information about
        both frequency and spatial scalings \cite{baccigalupi01, baccigalupi03}.
        
        \subsection{Galactic free-free emission}
        \label{subsec:ffemission}
        
            Thermal Bremsstrahlung (free-free) emission arises from hot electrons 
            ($T_e \gtrsim 10^4$~K) interacting with ions. The process is well known and the
            brightness temperature is a power law function of frequency with spectral
            index $\beta_{\rm ff} \simeq 2.15$. However due to the flatter spectrum
            with respect to synchrotron radiation, free-free becomes important with respect to synchrotron
            only at $\nu \gtrsim 30-60$~GHz where the
            Galactic signal is smaller than the CMB. This is why no direct templates
            of free-free emission are available.

            Galactic free-free emission consists of two independent components:
            diffuse and discrete. 
            The H{\sc ii} regions are the brightest discrete sources of free-free emission of our 
            Galaxy, produced when young, massive stars (O or B type) ionise the neutral 
            hydrogen in the interstellar medium with their ultraviolet radiation. Recently a
            catalog of 1442 Galactic H{\sc ii} regions has been produced \cite{paladini03a}
            together with a study on their 3D spatial distribution \cite{paladini03b}.
            The diffuse component can be well traced by H$\alpha$ (6563 $\AA$) 
            emission (see the recently published full-sky H$\alpha$ maps in 
            \cite{finkbeiner03, dickinson03}).
            At large angular scales, the free free angular distribution 
            from the available (pre-WMAP) data gives a dependence 
            $\propto\ell^{-3}$ \cite{kogut96}.
            By exploiting templates based on the H$\alpha$ data, 
            Bennett et al. \cite{bennett03} derived
            maps of the free free emission at the five frequencies 
            observed by WMAP.
            
        \subsection{Galactic dust emission}
        \label{subsec:dustemission}
            Dust grains heated by interstellar radiation are responsible for Galactic dust
            emission in the far infrared. The total intensity depends on the chemical
            composition, structure and dimension of the dust grains. The frequency dependence 
            is well modelled by a modified black-body
            emissivity law $I(\nu) \propto \nu^\alpha B_\nu(T_d)$ where
            the brightness $B_\nu(T_d)$ depends on the dust temperature $T_d$ 
            and $\alpha \sim 2$ is the emissivity. 
            Dust emission presents important spatial variations, as shown from the two
            full-sky surveys by COBE-DIRBE and IRAS. With a combination of the two surveys,
            a new map has been obtained \cite{schlegel98} with the IRAS angular resolution and 
            the DIRBE calibration accuracy.
            Exploiting these new data, a best fit model with two dust
            temperatures (16 and 9.5~K) and two emissivities (2.7 and 1.7) was proposed. 
            From the same map, a global dust angular power spectrum scaling $\ell^{-3}$ 
            has been derived. However in some high Galactic latitude regions this
            scaling is better represented by $\ell^{-2.5}$ but with amplitude differing
            from patch to patch.  
            
            Recent analysis of COBE-DMR data \cite{banday03, maino03} has shown
            a clear correlation between DMR data at 31.5~GHz with dust emission
            template after the expected dust thermal emission has been
            removed. This ``anomalous'' dust correlated component 
            is well fitted by a power law with a spectral index $\beta \simeq 2.5$,
            steeper than the expected free-free emission (although the latter
            is not ruled out). Further evidence of such correlation has
            been reported by several authors considering other data sets
            \cite{kogut96, deoliveira97, leitch97}. A possible explanation proposed by
            Draine \& Lazarian \cite{draine98} is the emission
            from spinning dust grains. However, no evidence of dust-grain emission 
            has been found in the WMAP data.
            
        \subsection{Extragalactic sources}
        \label{subsec:extrasources}

            Another unavoidable limitation for CMB measurements comes
            from extra-galactic sources. This issue has been long discussed in the 
            literature \cite{franceschini89} in the context of high resolution
            CMB experiments, but the effect on small-scale fluctuations has not been fully understood yet.
            Different populations of radio and dust
            sources show up below and above 200~GHz. At lower frequencies the main
            contribution comes from radio sources (``flat''-spectrum radio-galaxies,
            BL Lacs objects, blazars and quasars) while at higher frequencies the main role
            is played by dusty galaxies.
            Reasonable models of source population and number counts predict that 
            several hundreds of sources will be detected by future
            space missions like Planck in a wide frequency range \cite{toffolatti98}. 
            
            Although relatively bright point sources can be identified and removed (by $\sigma$-clipping or
            wavelet techniques), one is left with the background of unresolved 
            sources. Source Poisson distribution produces a simple white noise angular power
            spectrum with the same power at each multipole $\ell$. The expected level
            of source fluctuations is well below the CMB at frequencies between 50--200~GHz and
            for angular scales larger that few arc-minutes.
            
\begin{figure}
              \begin{center}
                \resizebox{14. cm}{!}{\includegraphics{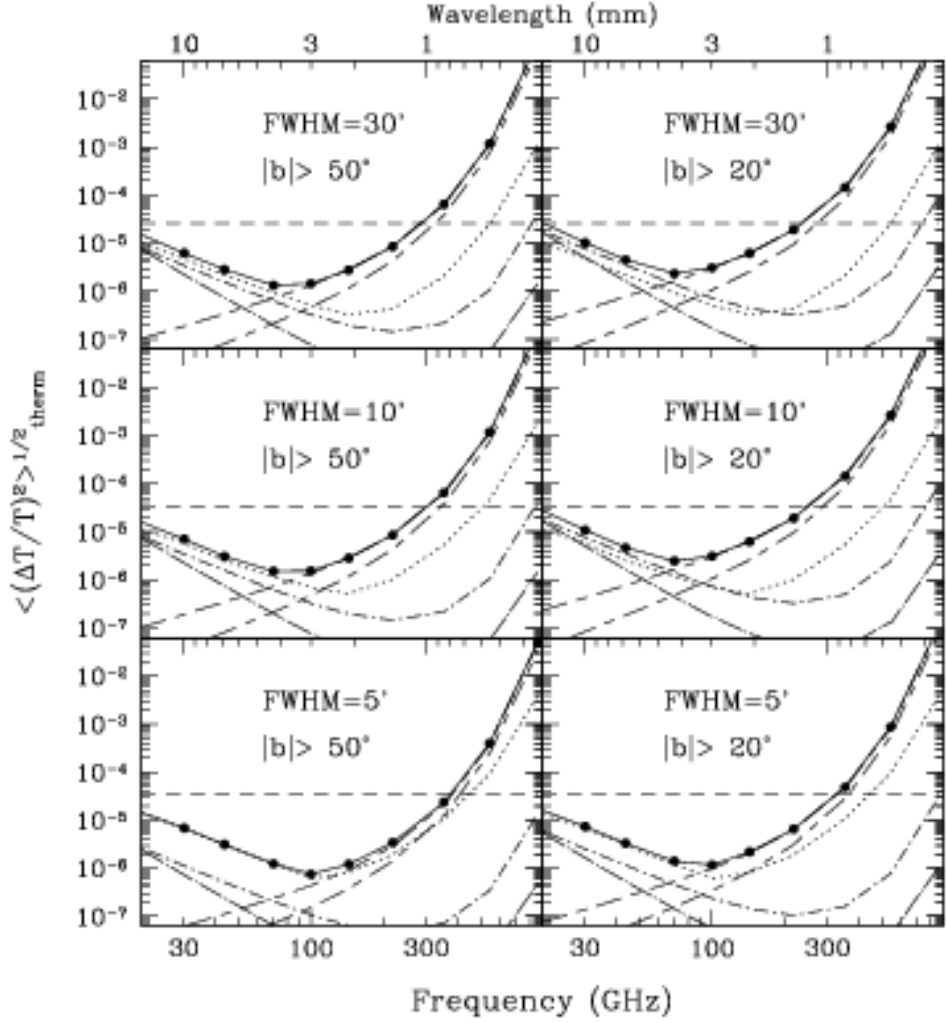}}
              \end{center}
              \caption{
                Temperature fluctuations for three different angular
                resolutions and two different cuts in galactic emission, as a
                function of frequency. The horizontal dashed line is the expected
                level of CMB fluctuations from a standard CDM model. Dot-short
                dashed, dot-long dashed and long-short dashed represent the mean
                contribution from Galactic free-free, synchrotron and dust
                emission respectively. Lower long/short dashed is a lower limit
                for Galactic dust fluctuations. The dotted curves represent the
                contribution from extra-galactic point sources fainter than 100
                mJy \cite{toffolatti98}. The solid curve is the quadratic sum of all
                the contributions and the filled circles are the selected 
                Planck frequencies (from \cite{dezotti99}).
              }
              \label{foresummary}
            \end{figure}

            In Fig.~\ref{foresummary} we summarise the relative importance of
            Galactic and extra-galactic foreground fluctuations compared to the
            level of CMB and for different angular scales. A clear window optimal
            for CMB observations arises between 70--100~GHz.
            
    \subsection{Cosmic and Instrument related variances}
    \label{subsection:cosmicvariance}

        ``Cosmic variance'' represents the ultimate accuracy limit of 
        any possible estimation of the CMB power spectrum. The CMB field is indeed a single
        realisation of a stochastic process and therefore we would not expect that
        our observable Universe follows the average over the ensemble of possible 
        realisation. This translates into the fact that the $a_{\ell m}$ coefficients
        are Gaussian distributed random variables, at a given $\ell$, and therefore
        their variance, $C_\ell$, is $\chi^2$ distributed with $2\ell +1$ degrees of
        freedom. The relative variance $\delta C_\ell / C_\ell$ is equal to 
        $\sqrt{2/2\ell+1}$ which is quite relevant at low $\ell$'s due to the 
        small number of available modes. 
        
        Beside cosmic variance, a basic limitation to power spectrum recovery concerns those
        experiments not probing the entire sky. Again the observed sky patch
        is not guaranteed to be representative of the whole sky and no direct information 
        can be gained on scales larger than the observed patch. This variance
        scales with the inverse of the covered sky fraction.
        
        If we also consider instrumental noise and angular resolution, assuming
        perfect foreground removal, 
        ideal calibration and no systematic effects, the relative
        uncertainty $\delta C_\ell / C_\ell$ can be written as:
        \begin{equation}
          \frac{\delta C_\ell}{C_\ell} = 
          f_{\rm sky}^{-1/2}\sqrt{\frac{2}{2\ell + 1}} \left[
            1 + \frac{A \sigma_{\rm pix}^2}{N_{\rm pix} C_\ell W_\ell^2}
            \right],
          \label{eq:delta_Cl_over_Cl}
        \end{equation}
        where $f_{\rm sky}$ is the fraction of ``useful'' CMB sky observed, $A$ is the surveyed area, 
        $N_{\rm pix}$ is the number of pixels and
        $W_\ell$ represents the beam window function which in the case of a Gaussian beam can be represented
        by:
        \begin{eqnarray}
          W_\ell^2 &=& \exp\left[ -\ell(\ell+1)\sigma_{\rm B}^2\right], \nonumber \\
          \sigma_{\rm B} &=& \frac{\theta_{\rm FWHM}}{\sqrt{8\ln(2)}}
          \frac{\pi}{60\times180} = (1.235\times 10^{-4})\theta_{\rm FWHM},
          \label{eq:W_l_sigma_b}
        \end{eqnarray}
        where $\theta_{\rm FWHM}$ represents the angular resolution in arc-minutes. 
        In Eq.~(\ref{eq:delta_Cl_over_Cl}) $\sigma_{\rm pix}$ is the final noise per pixel given by 
        $\sigma_{\rm pix} = \Delta T_{\rm rms}/\sqrt{n_{\rm det}\tau_{\rm pix}}$, where 
        $\Delta T_{\rm rms}$ represents the rms 1-second sensitivity, $n_{\rm det}$ the
        number of detectors in the focal plane and $\tau_{\rm pix}$ the average pixel integration time
        per detector.

        It is clear that the cosmic and sample variance terms dominate the uncertainty
        at large $\ell$'s while noise and beam related uncertainties dominate at higher
        $\ell$'s depending on the beam resolution and overall noise level.

    \subsection{Required accuracy for temperature and polarisation measurements}
    \label{subsec:required_accuracy}

        The main objective of an ideal CMB experiment is to map its temperature (and/or polarisation) anisotropy
        in the sky in order to accurately reconstruct the power spectrum, $C_\ell$, in the whole
        range of multipoles expected to contain information about physical processes acting at the Last Scattering
        Surface (i.e. $\ell \sim 1500 - 2000$).

\begin{figure}[here!]
          \begin{center}
            \resizebox{14.5 cm}{!}{\includegraphics{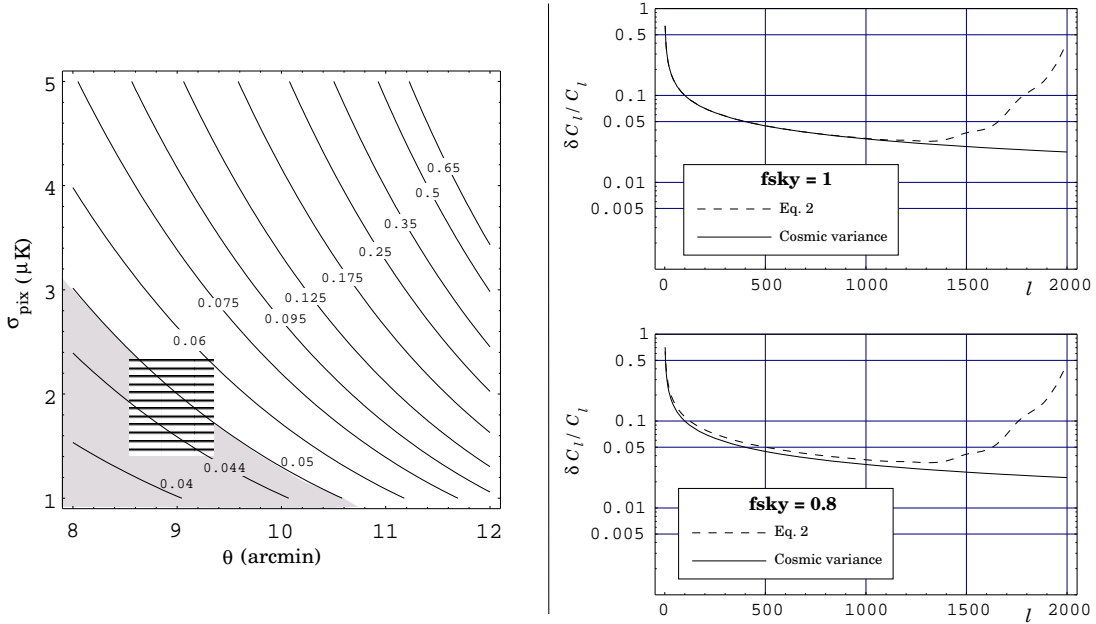}}
          \end{center}
          \caption{
            Left: values of $\delta C_\ell / C_\ell$ for $\ell = 1500$ as a function of the angular resolution,
            $\theta$, and the average noise per pixel, $\sigma_{\rm pix}$. The gray area represents the
            range of values for $\theta$
            and $\sigma_{\rm pix}$ for which  $\delta C_\ell / C_\ell \leq 0.05$; the striped
            box encloses the range of sensitivity and angular resolution 
            expected for Planck. Right:  $\delta C_\ell / C_\ell$ calculated considering the goal values
            of angular resolution and noise per pixel of the 100 GHz Planck-HFI channel (see 
            Tab.~\ref{tab:planck_performances}) and two values of $f_{\rm sky}$ (dashed curves). 
            The continuous curves show the contribution of the cosmic variance.
           }
          \label{fig:delta_cl_over_cl}
        \end{figure}

        In the left panel of Fig.~\ref{fig:delta_cl_over_cl} we show values of $\delta C_\ell / C_\ell$ 
        (calculated by Eq.~(\ref{eq:delta_Cl_over_Cl}) for $\ell = 1500$) 
        as a function of the angular resolution and of the average noise per pixel; this figure shows that
        to reach relative uncertainties $\lesssim 5\%$ at such high multipole values
        the instrument angular resolution and sensitivity must be within the gray area;
        for reference 
        the striped box encloses the
        range of expected performances for the 100~GHz Planck-HFI channel.
        In the right panel we show two $\delta C_\ell / C_\ell$ plots for $\ell \leq 2000$ calculated considering
        the goal performance values of the W-band Planck-HFI channel and two values of $f_{\rm sky}$ (0.8 and
        1); the figure shows that with
        these performances the uncertainty on the CMB power spectrum is dominated by the cosmic variance up to 
        $\ell\sim 1500$ even in the case in which only the 80\% of the sky is available.

        These requirements have profound implications on the design of an instrument aiming at reaching this kind of
        performances, in particular: (i) full sky coverage requires the experiment to be run from space, (ii) 
        a $\sim 10$ arc-minute
        angular resolution at 100~GHz requires the implementation of a reflector antenna with an aperture of the order
        of $\sim \lambda / \theta \sim$~1.5~m, (iii) $\mu$K sensitivity calls for multi-feed detector arrays operating
        at low temperatures with long integration times ($\gtrsim$1 year).

        If we consider also calibration, foreground removal and control of systematic effects we need to ensure that 
        the additional errors arising from their combination are of the order of $\lesssim$
        1--2\% of the noise per pixel in order not to dominate the final power spectrum and image 
        uncertainty. Calibration accuracies
        of the order of 1\% or better can be obtained using known astrophysical
        sources (as the dipole for photometric calibration and bright planets for beam calibration, see 
        Sect.~\ref{subsubsec:planck_mission_concept}), while accurate foreground removal is possible only
        with a wide frequency coverage in order to ``clean'' contributions
        from synchrotron, free-free (at frequencies $\leq$100~GHz) and dust (at frequencies $\geq$100~GHz)
        emissions. Similarly the level of spurious signals of instrumental and/or astrophysical signals
        must be at a level of $\lesssim$ 2 -- 3 $\mu$K, so that they do not become a dominant source of 
        uncertainty in the final results.
                
        For an experiment aimed at the measurement of E-- and B--modes 
        polarisation anisotropy with a comparable accuracy the
        instrumental requirements clearly become more stringent by 2--3 orders of magnitude (compare
        the levels of the temperature and the E-- and B--modes polarisation power spectra in 
        Fig.~\ref{cmbtpcl}), which implies the availability of ultra-high sensitivity detectors
        and a level of scientific understanding of polarised foregrounds that is currently at the beginning.
        The major technological and scientific issues involved in precise polarisation 
        anisotropy measurements can be summarised as follows:
        \begin{itemize}
          \item availability of detector arrays with sensitivity (and stability) of $\sim$2 order of magnitudes better 
            compared to those adopted in temperature anisotropy measurements;
          \item experimental setups able to guarantee a maximum level of systematic effects at the 
            sub-$\mu$K level;
          \item templates for removing polarised foreground components with an accuracy of $\lesssim$1~$\mu$K.
        \end{itemize}

        Furthermore in polarisation measurements the optical interface of the instrument
        needs to be more symmetrical in the response 
        than in the case of temperature anisotropy measurements, with
        minimal beam distortions and sidelobes/cross-polarisation requirements that 
        depend on the strategy assumed for polarisation measurements 
        (e.g. X-Y differencing receivers or correlation receivers).
        An higher level of accuracy 
        in optical simulations is also mandatory for systematic error control during the
        design phase of any polarisation sensitive instrument. 

        Further discussion of these issues in the context of future CMB polarisation research projects 
        can be found in Sect.~\ref{sec:future_challenges}.
        
\section{THE CONTROL OF SYSTEMATIC ERRORS IN CMB EXPERIMENTS}
\label{sec:systematic_errors}

    High sensitivity instruments
    provide accurate measurements of tiny signals independently from their source, so that
    great care must be taken to avoid experimental artifacts. The combination of increasingly sophisticated
    instruments and the quest for precise measurements of cosmological properties has made the
    control of systematic effects one of the main experimental challenges in
    present and future CMB experiments.
    
    The first non trivial issue is the definition of what has to be considered a systematic effect.
    A unique definition, in fact, is not easy to find, as the final result of a CMB experiment
    depends on the combination of many factors like the characteristics of the astrophysical signal and
    of the optical interface (i.e. the beam shape), the instrument response, the environmental
    thermal and electrical stability, the data reduction algorithms.
    
    In general we can identify three main classes of systematic effects: 
    \begin{enumerate}
    \item 
      signals in the instrument output that are not originated by the
      target astrophysical source (e.g. straylight effects, 
      periodic effects caused by thermal and/or electrical fluctuations in the 
      satellite and in the instruments, non-white components
      in the instrument noise spectrum);
    \item
      effects that cause a degradation of the instrument angular resolution (e.g.
      beam distortions and pointing effects);
    \item 
      spurious effects generated during data processing and analysis (e.g. quantisation effects).
    \end{enumerate}
    
    The second major difficulty is represented by the evaluation of the impact of a given systematic 
    effect on the scientific output of a CMB experiment (see 
    Sec.~\ref{subsec:impact_systematic_errors} for a more detailed discussion), which is generally characterised
    by scientific information at three different levels: (i) temperature (or polarisation) anisotropy maps,
    (ii) power spectra and, ultimately, (iii) derived cosmological parameters. If the calculation
    of the residual effect at map level is a relatively intuitive and simple task, it is increasingly
    difficult to perform such an assessment at the level of power spectra and of cosmological parameters,
    and no standard procedures currently exist to propagate the effect of systematic errors at these levels.
    
    The toughest challenge, however, is to set up high-precision CMB experiments that are 
    inherently free of systematic effects. One needs 
    to obtain {\em in hardware} the necessary signal stability using 
    highly stable detectors, optimising the thermal and electrical environment and 
    maximising the optical response in the main beam. However, when the accuracy requirements become highly
    demanding as in third generation space CMB experiments this ideal approach 
    can become unrealistic because of many technical and economical constraints
    imposed by the development of a complex space mission. Therefore a compromise between the hardware
    performances and the possibility to remove {\em in software} the residual effects from the data
    is often necessary.
    
    In the following sections we review the main categories of systematic effects with a particular
    attention dedicated to the experimental strategies that can be employed to reduce their impact to
    levels that are compatible with the scientific requirements of present and future high precision
    experiments.

    \subsection{Impact of systematic errors}
    \label{subsec:impact_systematic_errors}

        The impact of systematic effects on the mission scientific output, is generally dependent
        on the level of the spurious signal at the instrument output, the pixel size and the
        instrument scanning strategy. A relatively simple way to perform an 
        assessment of such impact is to simulate the measurement containing the systematic error and evaluate the residual
        peak-to-peak error on the generated maps.

\begin{figure}[here!]
          \begin{center}
            \resizebox{14.5 cm}{!}{\includegraphics{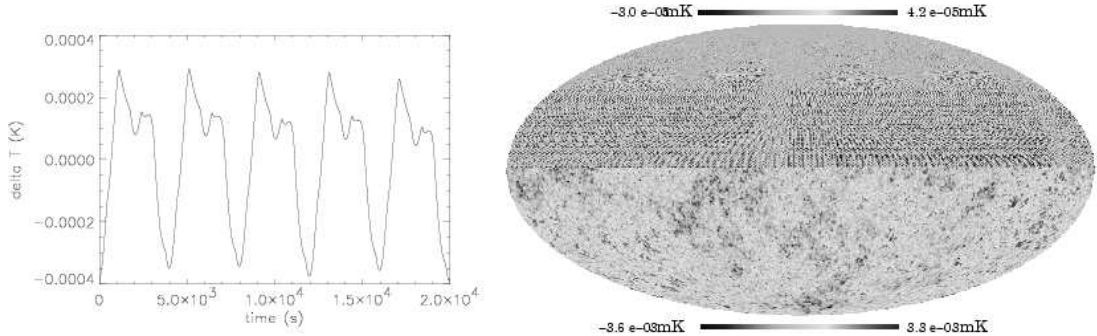}}
          \end{center}
          \caption{
            Simulated example of the effect of a periodic fluctuation on a CMB map. A periodic
            fluctuation (shown in the left panel) has been added to a simulated CMB signal; the resulting
            map is shown in the bottom half of the map in the right panel. The underlying systematic effect
            is revealed if we subtract the CMB signal (see top half of the map).
           }
          \label{fig:example_periodic_effect}
        \end{figure}

        Let us consider, for example, the periodic fluctuation in the left panel 
        in Fig.~\ref{fig:example_periodic_effect} representing a spurious sky signal caused by an 
        instrumental systematic variation (e.g. a thermal fluctuation of the telescope or a temperature
        variation of the radiometric receivers), which is characterised
        by a ``slow'' periodicity (i.e. with frequencies $\lesssim 10^{-2}$~Hz) and a peak-to-peak amplitude
        of the order of $\sim$1~mK. If we simulate a measurement where the CMB signal is 
        added to the periodic effect and then we project the time ordered data stream onto the  
        sky we obtain the lower half of the map in 
        Fig.~\ref{fig:example_periodic_effect}. This particular example has been calculated assuming
        a strategy similar  the one that will be used by the Planck satellite (see
        Sect.~\ref{subsubsec:planck_mission_concept}) i.e. a scanning in nearly great circles at the rate
        of 1 r.p.m. with each circle scanned $\sim$60 times before repointing.
        In the top half of the map we highlight the systematic effect signature lying about
        100 times below the astrophysical signal.
        
        Although the projection of a given periodic effect onto the sky is the most
        direct way to calculate the effect at map level it is also possible
        to estimate analytically the peak-to-peak value on the map by using a simple method that has been
        recently applied in the context of the Planck project to assess the impact of slow periodic 
        systematic effects in the measured signal \cite{mennella03}.

        If we consider a periodic fluctuation, $\delta T$, of general
        shape in the detected signal, we can expand it in Fourier series, 
        i.e.: $\delta T = \sum_{j=-\infty}^{+\infty}A_j \exp(i 2\pi f_j
        t)$, where $f_j$ represent the different frequency components
        in the fluctuation. For frequencies $f_j$ which are not synchronous with the
        instrument characteristic scanning frequency, $f_{\rm s}$,
        the measurement redundancy and the projection of the
        Time Ordered Data (TOD) onto a map with a pixel size $\theta_{\rm pix}$
        will damp the corresponding harmonic amplitude $A_j$ by a factor
        proportional to $\sin(\pi f_j / f_{\rm s})$. For
        frequencies synchronous with $f_{\rm s}$, instead, there will
        be no damping and these signals will be practically
        indistinguishable from the sky measurement. Therefore it is
        critical that any spurious signal
        which is synchronous with the characteristic scanning frequency
        is carefully controlled and kept at a negligible level {\em
          in hardware}. For a spinning experiment (for example like WMAP and Planck) we
        can estimate the final peak-to-peak effect of a generic signal
        fluctuation $\delta T$ on the map as follows:
        \begin{equation}
          \langle \delta T^{\rm p-p}\rangle_{\rm map} \sim 
          2\left[
            \frac{1}{N\times \theta_{\rm pix}/\theta_{\rm rep}}
            \left(
              \sum_{f_j < k\, f_{\rm s}}
              \left|
                \frac{A_j}{\sin(\pi f_j / f_{\rm s} )}
              \right|
              + \sum_{f_j < f_{\rm s},f_j \neq f_{\rm s}}
              \left|A_j\right|
            \right)
            + \sum_{f_j = k\, f_{\rm s}} A_j\right],
          \label{eq:p2p_map_general}
        \end{equation}
        where $\theta_{\rm pix}$ is the pixel size, $\theta_{\rm
          rep}$ is the {\em repointing angle} (i.e. the angle between two
        consecutive scan circles in the sky), $N$ is the number of times
        each sky pixel is sampled during in each scan circle and $f_{\rm s}$
        represents the scan frequency. Note that
        Eq.~(\ref{eq:p2p_map_general}) takes into account the damping
        provided only by the measurement redundancy and by the scanning
        strategy, without considering the possibility to detect and
        partially remove these spurious signals from the TOD. Several
        numerical strategies can be used to approach this issue and, as a
        general rule, the removal efficiency is greater with ``slow''
        fluctuations, i.e. with a frequency $f\ll f_{\rm s}$.
        If we denote with $F_j$ the damping factor obtained by applying a
        certain algorithm to a periodic signal with frequency $\nu_j$
        then we can write Eq.~(\ref{eq:p2p_map_general}) in the more
        general form:
        \begin{equation}
          \langle\delta T^{\rm p-p}\rangle_{\rm map}\sim
          2\left[
            \frac{1}{N\times \theta_{\rm pix}/\theta_{\rm rep}}
            \left(
              \sum_{f_j < k\, f_{\rm s}}
              \left|
                \frac{A_j/F_j}{\sin(\pi f_j / f_{\rm s} )}
              \right| +
              \sum_{f_j < f_{\rm s},f_j \neq f_{\rm s}}
              \left|A_j/F_j\right|
            \right)
            + \sum_{f_j = k\, f_{\rm s}} A_j\right].\nonumber
          \label{eq:p2p_map_with_destriping}
        \end{equation}

        The evaluation of the impact of a systematic effect at the map level is generally useful for
        quick assessments, but for a more thorough evaluation further analysis is necessary. The 
        next natural level is the comparison of the power spectrum of the systematic effect 
        with the expected level of the CMB power spectrum. This allows to identify the 
        angular regions most affected by the systematic error and provides useful guidelines
        to implement hardware and/or software strategies to reduce this impact if necessary.

\begin{figure}[here!]
          \begin{center}
            \resizebox{12.5 cm}{!}{\includegraphics{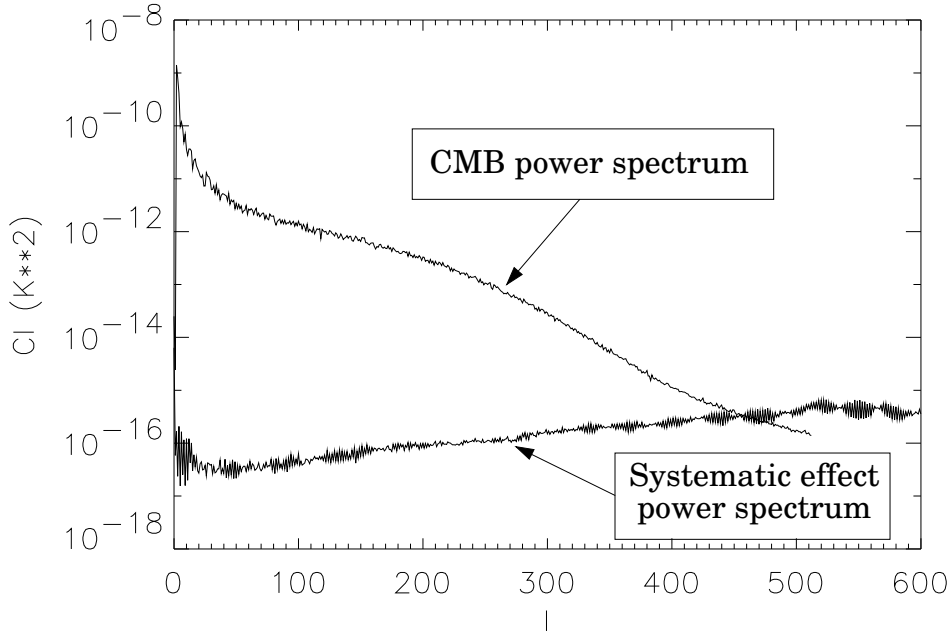}}
          \end{center}
          \caption{
            Simulated example of the effect of a periodic fluctuation on the CMB power spectrum.
           }
          \label{fig:example_periodic_effect_power_spectrum}
        \end{figure}

        In Fig.~\ref{fig:example_periodic_effect_power_spectrum} we show the comparison between the 
        angular power spectra of the simulated CMB map and of the periodic effect shown in
        Fig.~\ref{fig:example_periodic_effect}. This example clearly demonstrates that even if 
        a systematic effect appears to be very small in the visualisation of a map
        (about 100 times less than the CMB in this example) its impact on the angular power spectrum
        can be significant and needs to be carefully assessed.
        
        As the ultimate goal of any CMB experiment is to calculate the most likely set of cosmological
        parameters from the measured power spectrum, it is of primary importance that the evaluation
        process of the impact of experimental artifacts is fully carried out, in order
        to define confidence levels that are based both on the instrument sensitivity and on the
        expected level of residual systematic errors. This is not a trivial task since it requires 
        a full mission modeling for both the observation
        phase (instrument properties) and for the data reduction phase and therefore it is extremely
        time consuming and CPU intensive. Currently there are no complete studies and standard procedures 
        at this level; activities in this direction are in progress in the framework of the
        Planck Collaboration.

        In the simplest approach we start from the simulation of an ideal mission
        with perfectly symmetric Gaussian beams and pure white noise and finally extract the
        cosmological parameters in this ideal case. Then for each systematic effect the same procedure is repeated
        and the cosmological parameters are recalculated, now with the presence of the systematic error,
        before and after removal in the data reduction and analysis pipeline.
        Of course even after the removal some residual effect (that can also be dependent on the algorithm
        used to clean the data) will be present which needs to 
        be assessed. 

        As an example the $1/f$ noise after correction both with destriping or map-making techniques
        will made noise spectrum ``almost'' white with a small tail at low $\ell$. In principle
        this could be a problem for the detection of the Integrated Sachs-Wolfe effects and therefore
        for the determination of the cosmological constant $\Omega_\Lambda$. However at low $\ell$ the
        cosmic variance limits our knowledge of the true $C_\ell$. We have therefore to require
        residual of $1/f$ noise to be lower than cosmic variance in order to not have an impact on
        cosmological parameters. 

    \subsection{Optical and pointing effects}
    \label{subsec:optical_pointing}

        In any CMB experiment the optical system, which represents the interface between the detectors and the
        sky, often becomes an important source of systematic errors.

        High resolution experiments generally require detector arrays
        coupled with a reflector antenna system, generally an off-axis telescope.
        Typically, the coupling between the detectors and the telescope may be realised using 
        feeds such as corrugated horn antennas \cite{bersanelli98, barnes02, murphy02, villa03} or
        Winston cones \cite{welford78}.
        The angular response of the feed is modified by the telescope
        so that the $\left ( A_e \cdot\Omega_A \right )$ is preserved. $A_e$ is the feed (telescope) 
        effective aperture and $\Omega_A = \int_{4\pi}P_n(\theta,\phi)d\Omega$ is the feed (telescope) beam 
        solid angle, being $P_n(\theta,\phi)$ the normalised feed (telescope) beam pattern. 
        The greater the reflector diameter the smaller the beam solid angle and, clearly, the higher the 
        angular resolution.

        In the direction $\left ( \theta_0,\phi_0\right )$, the antenna temperature of the telescope/feed
        system seeing the sky is proportional to the convolution between the brightness temperature of the sky 
        and the normalised beam pattern:
        \begin{equation}
          T_A\left ( \theta_0,\phi_0\right ) = {\int_{4\pi}T_b \left (\theta,\phi\right ) 
            P_n \left (\theta-\theta_0,\phi-\phi_0\right ) d\Omega\over \int_{4\pi}P_n \left (\theta,\phi\right )d\Omega}.
        \end{equation}
        
        If $P_n(\theta,\phi)$ is not symmetric and not regular in shape, as in real cases, 
        the observed sky will be smeared and distorted by the optics. 
        
        The effects of the main beam distortions have been extensively studied in \cite{burigana98, fosalba02} 
        for elliptical Gaussian beams and in \cite{chiang02, mandolesi00,  page03a} 
        for complex distorted beams. Asymmetries degrade the angular resolution and increase the measurement uncertainty. 
        In terms of anisotropy power spectrum, asymmetries will decrease the maximum achievable multipole, $\ell$,
        and will increase the error on $C_\ell$.
        
        For elliptical beams, the increment in the r.m.s. temperature fluctuations
        due to ellipticity, $(\delta T_{\rm rms})_{\rm th}$,
        can be predicted easily by the following relationship \cite{burigana98}:
        \begin{equation}
          (\delta T_{\rm rms})_{\rm th}\simeq 1.11~10^{-3} \mu\mbox{K} \left [ -(r-1)^2+2.26\cdot(r-1)\right ]
          \cdot\left [-\theta_{\rm eff}^2+96.0\cdot\theta_{\rm eff}+1800\right ],
        \end{equation}
        where $\theta_{\rm eff} = \sqrt{\theta_M\theta_m}$ is the ``effective'' Full Width Half Maximum 
        (in arcmin) and $r = \theta_M/\theta_m$ is the ellipticity ratio, being $\theta_M$ and $\theta_m$ 
        the maximum and minimum beam angular resolutions (i.e. the major and minor axes of the elliptical 
        3\ dB contour of the beam).  
        For example for a beam with $r=1.2$ and $\theta_{\rm eff} = 10\,'$, we obtain 
        $(\delta T_{\rm rms})_{\rm th}=1.2~\mu\mbox{K}$.  
        In the case of complex beam shape, the amount of noise added by the distortions and the degradation of the 
        angular resolution can be predicted performing dedicated simulations. 
        To quantify the impact of the distortions in CMB anisotropy measurements, the effective angular 
        resolution $\theta_{\rm eff}$ and the additional noise $ (\delta T_{\rm rms})_{\rm th}$ 
        need to be calculated, also considering the
        instrument scanning strategy.

        The impact of main beam distortion on the angular power spectrum 
        is particularly relevant at the highest range of multipoles accessible
        to the resolution and sensitivity of a given experiment.
        The effect of typical scanning strategies is a  
        beam rotation in the sky, so that the 
        convolution (and the deconvolution) between the beam pattern and the sky 
        signal cannot be in general performed through a simple Fourier transform.
        A method based on the iterative solution with the Jacobi 
        algorithm of a linear system derived by the joint exploitation of 
        (possibly multi-beam) time ordered data and coadded maps to recover 
        the deconvolved sky map have been proposed by Arnau and S\'aez \cite{arnau00}. In this framework
        and under quite general scanning strategy assumptions and pixelisation 
        schemes, encouraging results have been recently found by Burigana and S\'aez \cite{burigana03a}
        by combining this fully blind (with respect to the sky 
        properties) approach with Monte Carlo simulations to 
        correct for the noise, for main beam distortion and noise 
        levels typical of Planck. 

        \begin{figure}[h!]
          \begin{center}
            \resizebox{15.5 cm}{!}{\includegraphics{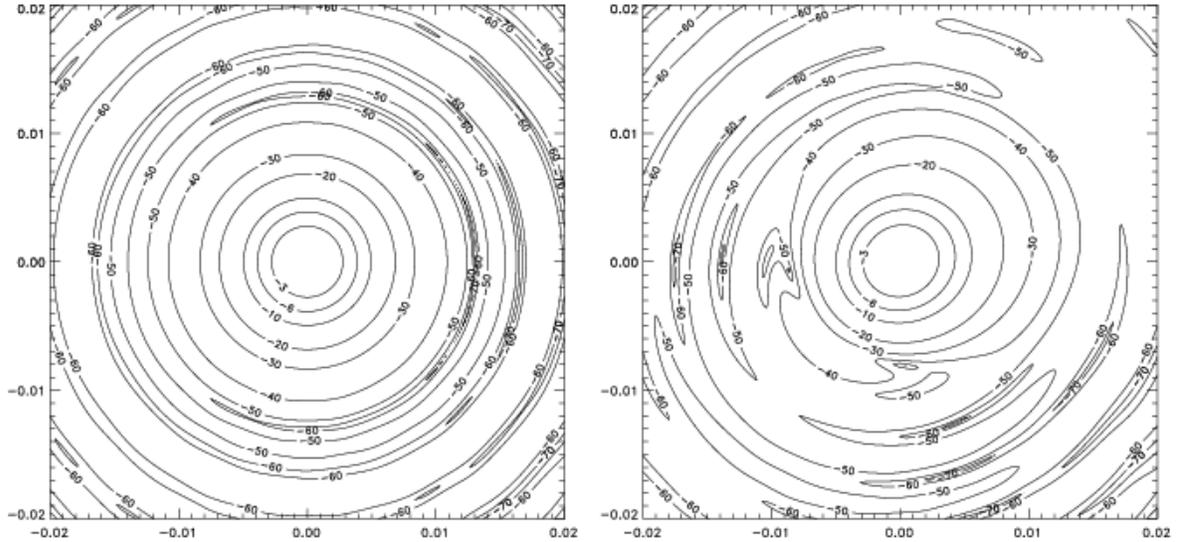}}
          \end{center}
          \caption{
            Left: Contour plot of a symmetric beam; typically this is the response of a detector located 
            at the centre of the focal plane, i.e. at the nominal focus of the telescope. Right: distorted beam 
            with mainly coma and spherical aberrations. This kind of aberration increases with
            the distance of the detectors from the optical axis.
          }
          \label{fig:beams}
        \end{figure}
        
        Even if the main beam response is symmetric, the detector may collect unwanted power (straylight) 
        that is not originated by sources in the main beam. 
        Variations in the straylight signal cause systematic errors in measurements. Specifically, the antenna 
        temperature can be divided into two terms:
        \begin{eqnarray}
          T_A\left ( \theta_0,\phi_0\right ) &=& {\int_{\Omega_{\rm MB}}T_b \left (\theta,\phi\right ) 
            P_n \left (\theta-\theta_0,\phi-\phi_0\right ) d\Omega\over \int_{4\pi}P_n \left (\theta,\phi\right )d\Omega}
          + {\int_{\Omega_{\rm SL}}T_b \left (\theta,\phi\right ) P_n \left (\theta-\theta_0,\phi-\phi_0\right ) 
            d\Omega\over \int_{4\pi}P_n \left (\theta,\phi\right )d\Omega}\nonumber\\
          &=& T_A^{(\rm MB)}\left ( \theta_0,\phi_0\right ) + T_A^{(\rm SL)}\left ( \theta_0,\phi_0\right ),
        \end{eqnarray}
        where $\Omega_{\rm MB}$ is the solid angle of the main beam region and $\Omega_{\rm SL} = 4\pi - \Omega_{\rm MB}$ 
        represents the solid angle of the side lobe region (i.e. the antenna response in the region outside the main beam).
        The signal is then picked-up by the antenna sidelobes and detected as features in the observed signal, $T_A^{(\rm SL)}$,
        that may be indistinguishable from signals induced by CMB fluctuations recorded by the main beam. 
        
        Although the real beam shape and the straylight contamination signal can be taken into account during 
        data analysis, an optical system dedicated to high precision CMB measurements must be designed 
        in such a way that aberrations are 
        minimised and the straylight contamination, $T_A^{(\rm SL)}$, is well below the instrument 
        final noise per pixel. Aberrations can be controlled 
        by implementing appropriate optical schemes in the telescope 
        (see for instance \cite{dubruel00, villa02, page03b}), while 
        straylight can be reduced by minimizing the antenna response in the angular regions outside the main beam, 
        or, equivalently, by minimizing the diffraction effects of the antenna system by ``under-illuminating'' 
        the reflectors. This can be achieved by optimizing the design of the feeds in the focal plane (see, for example,
        \cite{sandri03a, sandri03b, burigana03} for the Planck case).
        Clearly the under-illumination has a negative impact on the angular resolution, 
        since the telescope aperture will not be optimally used; this implies that 
        a trade--off between angular resolution and 
        straylight rejection is generally mandatory and this is often one of the major issues 
        in instrument optimisation studies.   
        
        For high resolution experiments, the uncertainty in the pointing 
        accuracy may be another serious source of degradation. 
        A typical requirement is that the pointing error, $\sigma_p$, should be 
        a small fraction of beam angular resolution. Typically $\sigma_p \leq 
        0.1\cdot \theta_{\rm eff}$. 
        The uncertainty in pointing direction and its knowledge has a 
        direct impact on the effective angular resolution and this 
        translates into a restriction of the multipole range effectively 
        probed.
        In fact, the ``real'' beam shape is determined by the convolution of the beam with the
        statistical distribution of the pointing uncertainty and this
        affects the sky angular power spectrum at high multipoles.
        Usually, the shape of the main beams will be reconstructed during the 
        observation phase by means of bright point sources such as planets or 
        supernova remnants. The accuracy of beam reconstruction is strongly 
        dependent on the pointing accuracy, being the pointing accuracy a 
        limiting factor for the proper knowledge of beam resolution and shape. 
        This leads to an additional uncertainty in the evaluation and subtraction 
        of the systematic effects introduced by main beam distortions in the 
        angular power spectrum at different multipoles.
        
        It is worth mentioning that pointing reconstruction and, 
        therefore, beam reconstruction, can be obtained only with the combination 
        of scientific data (e.g. data from bright point sources like
        Jupiter) together with attitude data. 
        These include CCD-camera, star-mapper or star-tracker 
        systems to map the sky for bright known stars which 
        would eventually yield the attitude reconstruction.
        From attitude it is possible to derive pointing information
        for each feed in the focal plane.
        The process is usually iterative where
        improved solutions for both pointing and beam shape are progressively
        up-dated until the final accuracy is achieved. 

    \subsection{1/f noise and thermal effects}
    \label{one_over_f_noise_and_thermal_effects}

\begin{figure}[h!]
          \begin{center}
            \resizebox{10.5 cm}{!}{\includegraphics{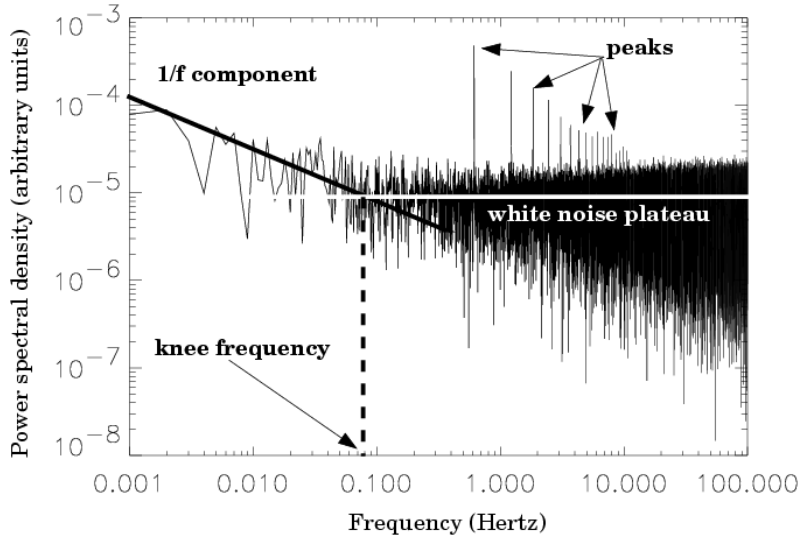}}
          \end{center}
          \caption{Typical noise power spectrum showing the white noise plateau, a 1/$f$ component at
            low frequencies and periodic peaks.}
          \label{fig:noise_example}
        \end{figure}

        Detector noise is generally represented by a spectral density \cite{ziel76} characterised by the following features:
        a wide, flat plateau at high frequencies (white noise plateau), a steep rise at low frequency 
        ($1/f^{\alpha}$ region), some lines at particular frequencies, which are generally caused by
        periodic fluctuations in the thermal and/or electrical environment
        (see Fig.~\ref{fig:noise_example}).

        Let us now examine the main causes of instrumental noise:
        Johnson (thermal) noise, temperature noise, shot noise, 1/$f$ noise, photon noise.

        Voltage fluctuations across a resistor, a linear function of its physical temperature, are the cause of 
        the so-called Johnson noise \cite{einstein06, johnson28}, caused by 
        random motion of carriers in a conductor, which results in a fluctuating 
        voltage across the terminals, $\delta v_f$. 
        Its power spectrum is white \cite{nyquist28} and given by:

        \begin{equation}
          \left\langle \delta v_f^2 \right\rangle df = S_V (f) df = 4 \, k \, T\, R \, df,
        \end{equation}
        where $k$ is the Boltzmann constant, $T$ is the physical temperature, $R$ is the element resistance.

        Shot noise (sometimes called ``Schottky noise'' \cite{schottky18}) 
        occurs when a phenomenon may be described as a series of 
        independent events occurring randomly. For instance, crossing of a p-n junction by electrical carriers 
        in a transistor produces a current, $\delta i_f$, 
        which is affected by shot noise. This kind of noise, which does not
        depend on the physical
        temperature, is observed in semiconductor devices. Also in this case
        the power spectrum is white. The following formula 
        illustrates the case of a transistor \cite{ziel76}.
        
        \begin{equation}
          \left\langle \delta i_f^2 \right\rangle df =S_I(f) df = 2eI df,
        \end{equation}
        where $e$ is the electric charge, $I$ is the current.
        Generation-recombination noise, due to the random fluctuating rates
        of generation and recombination of free carriers in semiconductor devices, may be considered 
        a form of shot noise. However, since it modulates the current passing through the device by 
        changing its impedance, it is usually treated in a separate way.

        Flicker noise is characterised by a 
        $1/f^{\alpha}$ (where  $\alpha \approx 1$) behaviour in the power spectrum
        below the so-called ``knee frequency'', which is the frequency at which the power of the 1/$f$ 
        component equals the white noise plateau (see Fig.~\ref{fig:noise_example}).
        It can be produced by many different sources \cite{wong03},
        and in some cases it is due to generation-recombinations noise modulated 
        by other effects. It also arises in semiconductor/metal junctions, but its origin
        remains partly unknown.
        This type of noise represents one of the major issues in the development of HEMT-based
        (HEMT = High Electron Mobility Transistor)
        radiometric receivers for CMB observations. Its effect can be minimised by means
        of differential measurements strategies in which the sky signal is compared to
        a signal received from another point in the sky (as in COBE and WMAP 
        \cite{smoot90, jarosik03}) or from
        a stable internal cryogenic reference load (as in Planck-LFI \cite{mennella03, seiffert02}).

        A fundamental noise is the photon noise, due to the statistics in the arrival time of photons
        and resulting in a fluctuating detected power. This kind of noise is strongly dependent on the temperature:
        thermal emission from telescope mirrors (which are gray bodies with few percent emissivity $\epsilon $)
        in a room temperature environment, for example, may be a concern for sensitive observations. It can be shown that the power 
        fluctuation spectrum is:
        \begin{equation}
          \sqrt{\left\langle (\Delta W)^2\right\rangle}df={\rm const.} \times  T^{\frac{5}{2}}f(\epsilon),
        \end{equation}
        where the constant depends on the particular experimental setup,
        while $f(\epsilon)$ is zero when $\epsilon = 0$. In space telescopes, which may be passively cooled to 
        50~K or below, this effect is strongly reduced.

        As a general rule, as it is evident from the noise spectral densities, a general benefit is obtained by cooling 
        detectors and optics at the lowest possible temperature. This is obtained by using cryogenic detectors 
        (radiometers between 20~K and 4~K, bolometers between 300~mK and 100~mK) and cold optics 
        (for instance, a passively cooled telescope in space may reach 35~K).
        However, this is not sufficient to reach the stability required in CMB observations. 
        Temperature in all the detection sub-systems must be not only low but also stable.

        The response of receiver components generally depends on their temperature,
        so that a variation in physical temperature necessarily induce a variation in the 
        receiver output. If temperature fluctuations are random then the effect is an
        increase in the receiver white noise level; if, on the other hand
        peculiar frequency components are present then the effect is a systematic variation of the
        final measurement.
        
        For example, in differential radio receivers, any fluctuation which impacts asymmetrically
        on the two receiver legs upstream of the differential element 
        (e.g. a variation in the reference signal or a fluctuation in the
        telescope temperature) will mimic a ``true'' sky signal; similarly
        bolometric detectors must be operated with constant power background. 
        If any element between the sky and the detector 
        (e.g. a bandpass filter) is characterised by a non-ideal transmission then any temperature
        fluctuation will mimic a ``true'' signal on the detectors.

        It is therefore mandatory to reduce temperature fluctuations of the receiving system to the lowest possible values,
        taking into account several aspects in the instrument design, like
        shielding receivers from external sources, minimising internal temperature fluctuations, 
        providing a stable reference signal.

        Thermal fluctuations from the external environment are generally minimised by placing shields
        around the focal plane and by designing orbits and scanning strategies that provide a thermal
        environment as stable as possible (e.g. see the examples of WMAP in Sect.~\ref{subsec:wmap} and
        Planck in Sect.~\ref{subsec:planck}).

        Stabilisation of the intrinsic payload and instrument temperature is generally obtained by a combination
        of passive (with low thermal conductivity and high thermal capacity materials like carbon fiber) and
        active (with PID\footnote{PID = Proportional Integral Derivative} control loops) strategies.
        
        Trade-offs must often be reached between instrument feasibility (in terms of technology and costs) and 
        control of systematic effects. A careful study of the receiving system can 
        also lead to split requirements between different perturbing components, 
        in order to control systematic effects at the 
        maximum possible level.

    \subsection{Effects from data handling}
    \label{subsec:data_handling}

        \subsubsection{Signal quantisation and telemetry losses}
        \label{subsec:signal_quantisation}

            In space CMB experiments the analog instrument output is digitised
            and stored during the acquisition
            in an on-board buffer memory which is periodically downloaded
            to ground. Data are usually sent as a sequence of packets 
            containing the relevant data and ancillary information 
            (time, voltages, temperature sensors) that allows the
            on-ground reconstruction of the data stream.
            Depending on their content, packets are usually defined 
            ``scientific'' packets, carrying measured data, and 
            ``house-keeping'' packets, containing sensor data,
            alarms and telecommands\footnote{
              See ECSS-E-70-41A standard (www.ecss.nl)
              for packets definitions and use in space missions}.
 
            The effective amount of scientific data which may be recovered by
            the receiving station during an observation is
            \begin{equation}\label{eq:max:data:flow}
              I_{\mathrm{eff}} =
              C_{\mathrm{r}}
              (1-f_{\mathrm{hk}} - f_{\mathrm{cnt}})
              \left[\frac{B_{\mathrm{ch}}}{1 \mathrm{bit/sec}}\right]
              \left[ \frac{T_{\mathrm{c}}}{1 \mathrm{sec}} \right],
            \end{equation}
            where $B_{\mathrm{ch}}$ represent the communication bandwidth,
            $T_{\mathrm{c}}$ the connection time,
            $f_{\mathrm{cnt}}$ and $f_{\mathrm{hk}}$ the fraction of telemetry
            allocated for contingencies.
                                
            The term $C_{\mathrm{r}}$ represents a compression factor applied to 
            scientific data before telemetry in order to optimise the available bandwidth and which 
            is usually obtained through a combination of lossy (data quantisation) and 
            lossless (arithmetic compression algorithms) strategies. 

            Lossy compression
            schemes clearly can introduce additional noises and systematic effects into the
            scientific data stream.
            For CMB data acquired by a stable instrument, and for short
            times, data are roughly normal distributed, and the distribution is
            dominated by the instrumental white noise, plus minor
            contributions from the CMB dipole, the Galaxy and  point sources.
            In these conditions the loss-less compression factor 
            $C_{\mathrm{r}}^{\rm lossless} < {N_{\mathrm{bits}}}/{\log_2
              (\sqrt{2\pi e} \sigma/q})$ being $\sigma$ the data RMS, $q$\ the
            digitisation step and $N_{\mathrm{bits}}$ the number of bits used
            by the on-board electronics
            (see \cite{maris00} and references therein). This relationship
            shows that this factor increases by decreasing the value of 
            $\sigma / q$ which can be optimised by adding a re-digitisation step on the
            on-board processing before compression.
 
            The impact of digitisation in CMB experiments has been studied in detail
            (see \cite{maris03} and references therein).
            The main result is that digitisation adds a constant
            baseline to the derived CMB power spectrum, 
            which can be estimated and removed once $q$
            and the scanning strategy are known.  
            Digitisation also introduces spurious noise into time ordered data
            and maps which may possibly affect CMB non-Gaussianity studies.

            Packet loss or corruption (generally of the order of $\sim$ 5\%) may cause errors in the
            data reduction with the consequent introduction of additive
            noises and systematic effects (see \cite{maino03a} for a review of 
            the various causes that may induce telemetry loss). This effects are generally reduced
            by requiring the maximum degree of independence (at the cost
            of some redundancy loss) among packets, so that
            the interpretation of the content of each packet 
            can be done independently from the other packets\footnote{
              Packets independence is often required also because 
              in some circumstances they may be sent randomly in time to the receiving station}.

            Random loss of scientific packets acts as a source of noise with
            a limited impact, if re-observation of the same region
            of sky in different periods is possible. In particular it has been
            demonstrated \cite{maino03a} that under general
            conditions a higher observation redundancy with higher telemetry losses is favourable with respect to 
            a lower redundancy with lower telemetry losses.
            The strongest impact from this effect is in the 
            study of foregrounds, which is dependent on the local details of the 
            signal distribution in the sky rather than on the full sky statistics.

            Telemetry losses of large portions of scientific or housekeeping data 
            may cause effects which are not easy to predict. An example is the loss
            of a packet which carries the confirmation of execution of a
            telecommand which changed an instrumental parameter; in this case it may be non trivial to understand on ground
            from which sample in the data stream
            the parameter change has to be applied by the data reduction pipeline.
            Depending on the level of telemetry redundancy, this
            uncertainty may affect one or tens of packets. The effect is
            particularly important if more parameters are changed in turn.
            Packets independence may allow to mitigate it.

        \subsubsection{Map making}
        \label{subsec:map_making}

            Optimal algorithms for producing maps from time ordered data are a primary task in CMB
            data analysis.
            The large data sets involved, the low signal-to-noise per sample and the complex
            correlation structure
            require computationally challenging and advanced statistical methods 
            to solve the problem. Furthermore, map making can also be used to
            flag and cure systematic effects in the time ordered data which may have negative impacts
            on the scientific analysis.
            
            A special kind of systematic errors are those produced during data analysis itself. 
            Although not immune by data
            handling errors, as explained below, state of the art map making codes produce 
            very modest artifacts.
            Making maps out CMB experiment timelines
            allows for a lossless compression of
            the raw data with minimal assumptions.

            Map-making {\em a la COBE}
            (see, e.g., \cite{lineweaver94}) has been extended 
            to the  differential, high resolution WMAP experiment 
            \cite{bennett03, wright96}.
            In experiments 
            like COBE and WMAP (in which the difference is taken between two essentially equal sky signals)
            the noise is nearly white  \cite{hinshaw03}, so that the map making problem
            reduces in this case to reconstructing the field from a set of differences. A more complicated
            problem arises in the case of differential measurements where the difference is taken between
            two slightly different signals (such as in Planck-LFI) which are characterised by a somewhat simpler 
            experimental structure but contamination
            by 1/$f$ noise (see, e.g., \cite{wright96, tegmark97} and 
            Sect.~\ref{subsubsec:planck_mission_concept}) needs to be taken into account.

            The instrument measurement can be modelled as the re-projection over time
            of sky observations (the scan), plus a contribution from noise. In doing so,
            we assume that the data, {\bf d},  depend linearly on the map, {\bf m}\footnote{
              For the sake of simplicity,
              we restrict to the case of temperature-only measurements. For an instrument with linear
              polarisation capabilities, a very similar formalism can be written to account
              for the Stokes linear polarisation parameters.
            }:
            \begin{equation}
              \label{measures} {\bf d} = {\bf P}{\bf m} + {\bf n},
            \end{equation}
            where ${\bf n}$ is a vector of random noise with (in general) non diagonal covariance. 
            The rectangular,
            $\mathcal{ N}_d \times
            \mathcal{ N}_p$ matrix ${\bf P}$ is called {\em pointing matrix} and
            models the scan by ``unrolling'' the map over the timeline. Because of
            efficiency considerations, it is normally assumed that beam smearing effects
            are taken into account at the map level, i.e. hidden in ${\bf m}$. This
            is meaningful only if the beam profile is symmetrical. In this case, the
            structure of $\mathbf{P}$ for a one-horned experiment would
            then be very simple. Only one element per row would be different from
            zero, the one connecting the
            observation of $j$-th pixel to the $i$-th element of the time stream. 
            
            Many methods have been
            proposed to estimate ${\bf m}$ in Eq.~(\ref{measures}) (for a review
            see, e.g., \cite{tegmark97}).
            Since the problem is linear in ${\bf m}$, the use of a Generalised
            Least Squares (GLS) method appears well suited. This leads to a solution
            of the form:
            \begin{equation}\label{map} {\tilde {\bf m}} = {\bf \Sigma}^{-1} \, {\bf
                P}^T {\bf N}^{-1} {\bf d},
            \end{equation}
            where
            \begin{equation}
              \label{cov} {\bf \Sigma} = {\bf P}^T {\bf N}^{-1} {\bf P}.
            \end{equation}
            In the above equations $N$ is the noise covariance matrix, that must be estimated from the
            data themselves.
            
            Although optimal in a statistical sense (if the noise is Gaussian
            Eq.~(\ref{map}) is a Maximum Likelihood solution), the method
            outlined above is computationally unfeasible for large datasets, because
            of the size of the matrices involved (of order $10^5$ to $10^7$).
            The solution is to keep the same equations, recasting them into a different
            manner to allow for the use of an iterative solver. The latter is
            much less computationally demanding than ordinary linear algebra techniques,
            because it only requires matrix to vector products. The map making
            problem is naturally suited for such a scheme. In fact, application
            of the matrix $\Sigma$ to a vector can be thought as the ``unrolling'' of
            a tentative solution map over the timeline (application of $P$), the
            subsequent application of $N^{-1}$ (a convolution if the noise is
            stationary) and, finally, the application of $P^{T}$ which sums
            back the convolved data into a map. This approach, usually named ``unroll,
            convolve and bin'' \cite{natoli01, dore01} can be
            successfully embedded into an iterative scheme (usually relying on a conjugate
            gradient solver) to converge efficiently to the required solution.
            
            The ``unroll, convolve and bin'' approach has solved the problem of producing maps
            for large datasets, a problem deemed unfeasible until a few years ago.
            Moreover, its computational efficiency has allowed this method to be inserted into a
            Monte Carlo chain, thus providing a powerful tool to check the robustness
            of a data analysis pipeline and flag sources of errors. In particular, it
            has been shown that in most applications the map making stage itself introduces
            little, and usually negligible, artifacts in the data.
 
            In this respect, the potentially more dangerous source of error is to be
            located in the way the beam is handled. 
            The scheme outline above assumes
            symmetric beams and is prone to inject systematic errors should
            this assumption fail. Removing this limitation, which is more a limitation
            of the data model rather than of the map making scheme itself, is at the
            present time the frontier in this area of CMB data analysis.

        \subsubsection{Destriping}

            In the previous section we have addressed the issue of correlated 1/$f$ noise
            in the production of optimal maps.
            When coupled to the observing
            strategy, 1/$f$ noise will generally produce ``stripes'' in the final maps.
            Besides map-making there are other approaches that can deal with $1/f$ noise striping
            as well as with a large variety of systematic effects,  
            usually called 
            ``destriping'' \cite{maino02a}.

            These algorithms do not
            solve for the map but eventually obtain time ordered data cleaned from the 
            low frequency systematic effect. 
            They can be 
            then used for many applications: in-flight main beam reconstruction \cite{burigana01a, terenzi02}
            and calibration \cite{cappellini03}, studies of (and cleaning from) radiosource variability 
            \cite{terenzi02} and Solar System objects, map production from
            cleaned time ordered data. 
            Although destriping does not guarantee that the final map will be 
            the minimum variance map, in terms of final
            noise power spectrum, the two approaches are virtually identical.
            In addition, accurate numerical simulations \cite{maris03a, maris03b} have shown that 
            the quality of destriping results, both in terms of de-drifting of 
            time ordered data and of map and power spectrum 
            recovery, is largely unaffected by the presence of
            optical distortions (both in the main beam and in the
            sidelobes), as the differences
            between the measurements at the same crossing points are largely
            dominated by the noise.
            
            It is however worth mentioning the approximations for which the destriping
            approach is valid. Destriping assumes that the overall effect of $1/f$ noise
            can be approximated by an additive constant. This condition is verified as long as the
            natural spinning/observing frequency is large compared to the 
            $1/f$ noise knee-frequency (see, Sect.~\ref{one_over_f_noise_and_thermal_effects}), which
            in the case of Planck-LFI this implies to work with
            averaged (over 1 hour period) scan circles,
            while for the HFI instrument (which shows lower knee-frequencies) it is possible
            to work directly with elementary (1 minute) rings.
            
            If we assume that the $1/f$ noise effect is well approximated by a single
            baseline for each scan circle $A_i$, it is possible to recover the
            baselines which 
            are obtained using all the possible crossing points between
            different scan trajectories. Formally this means to solve the linear system:
            \begin{equation}
              \label{destri}
              \sum_{\pi=1}^{n_c}\left[
                \frac{[(A_i - A_j) - (T_{il} - T_{jm})]\dot[\delta_{ik} - \delta_{jk}]}
                {E_{ij}^2 - E_{jm}^2}\right]_\pi = 0,
            \end{equation}
            for all the circles $k=1,\dots,n_s$. Here $\delta$ is the usual Kronecker symbol,
            $A_i$ are the baselines to be recovered, $T_{jl}$ is the observed signal in the 
            ring $j$ at the position $l$ along the ring. The terms in the denominator are the
            level of white noise for the same pixel. This system simply translated into:
            \begin{equation}
              \label{finaldestri}
              \sum_{h=1}^{n_s} C_{kh}A_h = B_k, \, k=1,\dots,n_s\, ,
            \end{equation}
            which can be easily solved. The matrix of the coefficients $C_{hk}$ is
            in general positive defined, symmetric and non singular provided that
            there are enough intersections between different scan circles.

\section{CMB EXPERIMENTS}
\label{sec:cmb_experiments}

    In this section we review briefly the evolution of CMB anisotropy experiments from COBE to
    WMAP and describe the next planned mission of the European Space Agency, Planck. A more detailed
    and complete review of experiments before WMAP can be found in \cite{bersanelli02}.

    \subsection{From COBE to WMAP}
    \label{subsec:from_cobe_to_wmap}

        The first detection of anisotropies was achieved by COBE-DMR with three
        pairs of Dicke-switched radiometers at 31.5, 53 and 90 GHz.  As in any CMB
        experiment, the success of DMR depended on the accurate control of
        systematic effects and calibration \cite{kogut96a}. The DMR 4-year frequency-averaged
        map has a signal-to-noise ratio of $\sim 2$ and gives a visual impression
        of the actual large scale CMB structure in the sky. The observed power
        spectrum, limited to $\ell < 20$, was found to be consistent with a
        scale-invariant spectral index ($n_S \sim 1$) of the primordial density
        fluctuations \cite{gorski96}, and a power-law fit yielded an extrapolated quadrupole
        term of 15.3~$\mu$K \cite{hinshaw96}. The DMR results proved
        the presence of temperature fluctuations at a
        measurable level $\Delta T/T_0 \sim 10^{-5}$, and started the
        race to unveil new features in the angular power spectrum.
        Increasingly ambitious sub-orbital programs were carried out mostly aiming
        at sub-degree scale detections in limited sky regions.
        
        \subsubsection {Ground-Based Experiments}
        \label{subsubsec:ground_based}

            High altitude, dry sites have provided excellent results, in spite of the
            limitations of atmosphere and ground emissions. With only a few exceptions
            \cite{romeo01}, most ground based experiments take advantage of the atmospheric
            windows below 15 GHz, around 35 GHz and 90 GHz.  The Tenerife experiment,
            carried out from the Teide site (2400~m) used similar technology as DMR in
            the 10-33 GHz range \cite{davies96}. Soon after COBE, it was possible to detect 
            CMB anisotropy at a
            level $\sim$ 40~$\mu$K in the multipole range $\ell \sim 10$--$30$,
            overlapping with DMR. More recently, a 33~GHz interferometer was installed
            \cite{dicker99} at Teide scanning a sky portion largely overlapping the original
            program. The detection of CMB anisotropy was unambiguous, with $\Delta
            T_\ell \sim 43$~$\mu$K at $\ell \sim 110$.
            
            Even if harsh and isolated, such as the Antarctic Plateau, ground based
            sites offer the enormous advantage of reachable instruments and long
            integration times. The UCSB degree-scale measurements from the
            Amundsen-Scott South Pole Station carried out in the period 1988--1994 with
            HEMT cryo receivers at 30--40 GHz \cite{gundersen95} showed correlated structure with
            amplitude $\Delta T_\ell \sim 33$~$\mu$K at $\ell \sim 60$. The polar site
            also hosted the Python/Viper program, leading to detection at $\sim 1\deg$
            scales using an off-axis parabolic telescope that was coupled either
            to bolometers at 90 GHz \cite{platt97} or to a HEMT radiometer at 40 GHz \cite{coble99}. The
            results were consistent with DMR at low $\ell$'s, and suggested an increase
            at $\ell > 40$. Viper was a follow up of Python at higher angular
            resolution \cite{peterson00} with observations at 40 GHz with a large 2.15 m telescope,
            yielding sensitivity in the range $100 < \ell < 600$.
            
            The Saskatoon instrument \cite{wollack97} used an off-axis parabolic reflector and 
            total power receivers based on low-noise cryogenic HEMT amplifiers, and
            produced one of the first convincing evidences of a raising spectrum (from
            $\Delta T_\ell \sim 49$~$\mu$K at $\ell \sim 87$ to $\Delta T_\ell \sim 85$~$\mu$K 
            at $\ell \sim 237$).  As a follow-up, the MAT program probed smaller
            scales, $40 < \ell < 600$.  The Saskatoon instrument was installed at a
            site near Cerro Toco (5240 m) with the addition of a 144 GHz channel (0.2
            beam), taking advantage of both HEMT and SIS technologies. The MAT/TOCO 31
            and 42~GHz instruments were flown twice in 1996 (QMAP) to perform
            degree-scale observations providing detection in both flights \cite{devlin98}.
            
            At small angular scales, a series of filled-aperture observations at OVRO
            lead in the late 90's to an unambiguous detection \cite{leitch00} in the range $7'$
            to $22'$. Interferometers have provided high-quality observations at high
            $\ell$'s. The CAT three-element array operating at 13--17 GHz showed
            detection at $\ell \sim 420$ \cite{baker99}. The VLA has been used to set limits to
            CMB anisotropy at sub-arcminute angular resolution since the early 80's
            and established new upper limits in the range $0.2'$ to $1.3'$ \cite{partridge97}. ATCA
            \cite{subrahmanyan98} yielded upper limits in polarised intensity and 
            $\Delta T_\ell < 25$~$\mu$K for multipoles $3300 < \ell < 6000$. 
            Finally, the SuZIE bolometer
            array \cite{holzapfel97} operated at Mauna Kea gave upper limits to CMB primary
            anisotropy. The combination of all data at high resolution gives evidence
            for a downturn in the power spectrum at sub-degree scales, as expected by
            standard models.
            
            Three high quality ground-based data sets have been recently obtained. The
            DASI 13-element interferometer \cite{halverson02} used cryogenic HEMTs in the 26--36 GHz
            spectral window and observed from the South Pole in the 2000 austral
            winter. The results mapped the power spectrum in the range $100 < \ell <
            900$. The first peak at $\ell \sim 200$ was evident and in good agreement
            with results from other experiments. In addition, the DASI data suggested
            the presence of further peaks at $\ell \sim 550$ and $\ell \sim 800$. 
            DASI was also sensitive to polarisation anisotropy and produced 
            maximum likelihood EE, BB, and TE power spectra,
            probing interesting upper limits on EE and BB modes for some multipole ranges
            and showing also interesting detections
            in a quite good agreement with theoretical estimations \cite{kovac02}.
            The CBI interferometer was sensitive to scales from $5'$ to $1\deg$ ($300 <
            \ell < 3000$). Excellent observations were obtained \cite{pearson03} with power
            spectrum results in good agreement with other experiments at $\ell < 2000$
            but extending up to $\ell \sim 3000$. An excess power was observed at high
            $\ell$'s suggesting a Sunyaev-Zel'dovich effect. Even more
            recently, the Viper 2.1 m telescope was used at the South Pole in the
            ACBAR experiment \cite{kuo02} to obtain two deep fields of the CMB, 3 degrees in
            size, with an rms of 8~$\mu$K per $5'$ beam.

        \subsection {Balloon-Borne Experiments}
        \label{subsubsec:balloon_borne}

            Although the reduction of atmospheric emission from balloon altitudes
            (35--40~km) is great (factor $\sim 10^3$), in conventional flights the
            available observing time is only about 10--12 hours. In recent years,
            long-duration balloon (LDB) flights (10--15 days) have been successfully
            flown. Early balloon experiments confirmed the DMR detection. The FIRS
            bolometers covered about 1/4 of the sky at $\sim 3.8\deg$ resolution \cite{meyer91}
            in the range 170--680 GHz. The ARGO degree-scale bolometric experiment 
            \cite{debernardis94}
            lead to a detection of CMB anisotropy $\Delta T_\ell \sim 39$~$\mu$K. BAM
            obtained statistically significant detection at degree scales ($\Delta
            T_\ell \sim 56$~$\mu$K at $\ell \sim 75$) \cite{tucker97} using a cryogenic
            differential Fourier transform spectrometer coupled to a 1.65 meter
            reflector. The MAX and MSAM balloon-borne multi-band bolometric receivers
            contributed to the progress at sub-degree scale. MAX was flown 5 times
            between 1989 and 1994 and observed in nine different sky regions yielding
            seven positive anisotropy detections \cite{tanaka96}. The results also suggested a
            band power at degree-scales higher than DMR. MSAM \cite{fixsen96} observed a 10
            deg$^2$ region with $30'$ resolution. MSAM was launched 3 times in the
            period 1992--95, each flight detected a clear CMB anisotropy signature. A
            combination of the three flights yielded data points in three power bands
            centred at $\ell \sim 34, 101, 407$.
            
            Boomerang and Maxima represented a major breakthrough. Boomerang was
            launched for its first LDB flight around Antarctica in December 1998 and
            landed roughly 10.5 days later. An array of bolometric detectors at 90,
            150, 240 and 410 GHz were cooled by a $^3$He refrigerator to 0.28~K. The first
            results \cite{debernardis00} were based on the data of a single detector at 150 GHz, while
            a more detailed analysis covering four 150 GHz detectors and a larger data
            set was later presented \cite{netterfield02}. Maxima, conceived as a follow-up of MAX, was
            a bolometric receiver, cooled at 100~mK by an adiabatic demagnetisation
            refrigerator, sensitive to multipoles $80 < \ell < 800$ with frequency
            bands centred at 150, 240, and 410 GHz. The first flight in its full
            configuration \cite{hanany00} was launched in August 1998, and covered 0.3\% of the
            sky (3200 independent pixels) during a 7-hour conventional flight. The
            Boomerang and Maxima power spectrum measurements were in agreement for
            what concerns the first-order key features: both showed clear evidence of
            a peak $\ell \sim 200$ (corresponding to a density parameter
            $\Omega_0\sim 1$), and indications of further acoustic features at
            higher multipoles \cite{balbi00}. More recently, the Archeops experiment \cite{benoit03}
            obtained accurate measurements over the multipole range 15--350 with an
            array of 21 bolometers in the 140--550 GHz range cooled at 100 mK with an
            artic balloon flight covering 30\% of the sky, and yielded accurate power
            spectrum data bridging between COBE and the first acoustic peak.
            
    \subsection{WMAP}
    \label{subsec:wmap}

        \subsubsection{Mission summary}
        \label{subsubsec:wmap_summary}
        
            The original idea of the Microwave Anisotropy Probe was conceived in
            the early 90's.
            After the loss of Dr. David Wilkinson, a member 
            of the science team and pioneer in the study of the CMB
            occurred during the first year data acquisition, the
            satellite was renamed Wilkinson Microwave Anisotropy Probe (WMAP). 
            
            The primary science goal of WMAP is to map the relative CMB
            temperature over the full sky with an angular resolution corresponding 
            to a pixel side of about 0.3 degrees, a sensitivity of 30 $\mu$K 
            per square pixel, with a contribution from systematics limited to 5 $\mu$K per pixel. 
            As we discuss below, the entire design of the instrument was conceived in order 
            to achieve control of foreground emission and systematics. 

\begin{figure}[here]
              \begin{center}
                \includegraphics{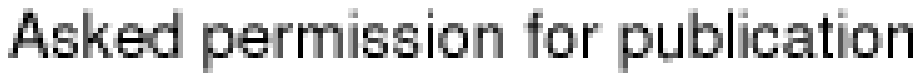}
                \caption{An overview of the WMAP satellite, showing the
                  double-telescope system, the large radiators, and the satellite
                  service module with deployable solar panels on the bottom. One of
                  the two focal arrays of feed horns is also partially visible on
                  the right. The overall height is 3.6~m, the mass is 830~Kg and
                  the diameter of the deployed solar panels reaches 5.1~m. The
                  solar arrays supply a 400~Watts power to the spacecraft and
                  instruments 
                  (with permission, courtesy of the NASA/WMAP Science Team).}
                \label{fig:map_satellite}
              \end{center}
            \end{figure}

            WMAP (see Fig.~\ref{fig:map_satellite}) consists of two back-to-back Gregorian telescopes 
            (1.4~m~$\times$~1.6~m primary and 0.9~m~$\times$~1.0~m sub-reflector)
            coupled to two symmetric focal planes with 10 radiometers each. 
            Two sky signals coming from directions separated by a total angle of
            $141^{\circ}$ are compared by pseudo-correlation receivers at five different frequency bands
            centred at 22.8, 33.0, 40.7, 60.8, 93.5~GHz with FWHM beamwidths
            respectively of 0.82, 0.62, 0.49, 0.33, 0.21~degrees \cite{jarosik03}.

            The signal collected by a
            corrugated feed horn is split into two orthogonal polarisation 
            modes by an Ortho-Mode Transducer (OMT). Each polarised component is
            combined with the orthogonal component coming from the symmetric channel 
            in the opposite focal unit and then amplified by passively cooled (at 
            $\sim$ 100~K) High Mobility Electrons 
            Transistor (HEMT) amplifiers. After a second warm amplification,
            signals are phase-shifted and then split back into the two original
            input signals, detected and then differenced. This differential setup is common to many
            high sensitivity CMB microwave detectors and provides reduced
            1/$f$ noise and contamination by amplifiers 
            instabilities with respect to other
            radiometer chain configurations.
            For example the $1/f$ knee frequency, $f_{\rm k}$, of WMAP radiometers
            is typically $<$ 50~mHz and for half of them it is below 1~mHz. WMAP main instrumental
            performances are summarised in Tab.~\ref{WMAP_summary}; 
            for more details, see \cite{bennett03a} and 
            visit http://map.gsfc.nasa.gov. 
            
            \begin{table}[here!]
              \begin{center}
                \caption{A summary of the WMAP nominal features.}
                \label{WMAP_summary}
                \begin{tabular}{l l l l c c c c c c}
                  \hline
                  \hline
                  Band & K & Ka & Q & V & W \\
                  \hline
                  Centre frequency (GHz) & 23 & 33 & 41 & 61 & 94 \\
                  \hline
                  Bandwidth (GHz) & 5.5 & 7 & 8.3 & 14 & 20.5 \\
                  \hline
                  Equivalent FWHM (deg.) & 0.93 & 0.68 & 0.53 & 0.35 & 0.30 \\
                  \hline
                  Sensitivity ($\mu$K 0.3 deg. $\times$ 0.3 deg. pixel) & 35 & 35 & 35 & 35 & 35 \\
                  \hline
                  Number of detectors & 4 & 4 & 8 & 8 & 16 \\ 
                  \hline
                  \hline
                \end{tabular}
              \end{center}
            \end{table}
            
\begin{figure}[here!]
              \begin{center}
                \resizebox{10.5 cm}{!}{\includegraphics{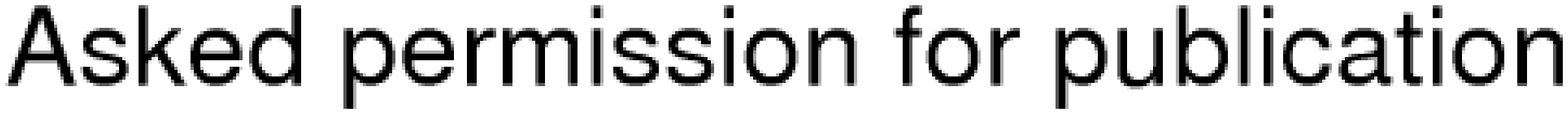}}
              \end{center}
              \caption{The WMAP scan strategy (with permission, courtesy of the NASA/WMAP science team).
              }
              \label{scanning}
            \end{figure}

            The observation orbit is located at the L2 Sun-Earth Lagrange point, 1.5
            million km from Earth. The WMAP observation direction is constantly away
            from the Sun, Earth and Moon, with a scanning strategy that allows to observe
            the full sky 
            every six months.
            This ensures a high thermal stability with no active cooling stages and negligible
            radio frequency interference from the Sun, Earth and Moon.
            The observational strategy (see Fig.~\ref{scanning}) 
            is based on fast precessions ($\sim 0.34$~mHz)
            of the spacecraft, spinning at 0.464 rpm (7.57~mHz); as a consequence 
            approximately 30~\% of the sky is observed each hour and the 
            sensitivity  pattern is effectively smoothed.  
            
            The data were calibrated by comparing the raw voltage output from the 
            differential radiometer with the known cosmic dipole signal; the WMAP 
            map making procedure is based on a technique already 
            used for COBE \cite{wright96}. The final map, $X$, is obtained 
            iteratively:
            \begin{equation}
              \label{WMAP_mapmaking}
              X^{n+1}=X^{n}+D^{-1}[B-AX]\, ,
            \end{equation}
            where $A=\sum_{t}P^{T}(t)P(t)$ is sum over all the observation times
            of the time dependent pointing vector $P(t)$ times its transposal,
            $B=\sum_{t}V(t)P(t)^{T}$ with $V(t)$ representing the time ordered
            data and $D$ a diagonal matrix with the 
            number of observations at each sky pixel in the diagonal elements. WMAP adopted the HEALPix 
            sky pixelisation scheme \cite{gorski98}. 
            
        \subsubsection{Discussion of first-year results}
        \label{subsec:wmap_discussion_results}
            
            The recent data release of the first year of WMAP observation is, no doubt, 
            the most important
            event in CMB anisotropy study since COBE,  a few months after the
            first evidence of a polarised signal by
            DASI \cite{kovac02}.
                  
            The WMAP frequency coverage 
            allows a quite good removal of 
            the astrophysical foregrounds, in particular of the 
            Galactic components that are more crucial at the multipoles
            ($\ell \lesssim 800$) where the WMAP sensitivity is high enough
            to accurately measure the CMB power spectrum and main beam
            distortion effects are not crucial.
            
            The first step \cite{bennett03b} in this subtraction has been the construction
            of several masks, based on asymmetric tail at positive 
            signals of the histogram of the pixel values at 
            22.8~GHz, to be applied 
            to the full sky maps according to the desired maximum level 
            of signal to be excluded from the analysis. This step is then
            complemented by excluding sky regions selected for high
            contamination by bright discrete sources, identified 
            by exploiting several source catalogues.

            Then, three different and complementary approaches
            were pursued to separate CMB anisotropy from Galactic foregrounds,
            providing quite consistent results.
            The first one, which has the advantage to be based only on WMAP data themselves,
            searches a linear combination of the frequency maps using weights
            aimed at reproducing the planckian shape of the CMB spectrum.
            The second one, recognised by the WMAP team as the best to reconstruct 
            the CMB anisotropies, is based on on fitting the
            WMAP frequency maps with the templates of Galactic emission obtained at radio 
            (for the synchrotron emission)
            or at Far-IR (for the thermal dust emission) bands and on the $H\alpha$ emission
            (for the free-free emission).
            Finally, these templates can be used as inputs for a MEM component separation
            of WMAP frequency maps, a third method claimed by the authors as the best
            to derive Galactic diffuse emission maps.

            A good correlation has been found between the free-free WMAP template
            and the map of $H\alpha$ emission. The relatively flat spectral index of the synchrotron
            emission ($\sim -2.5$) close to the Galactic plane supports the idea of 
            a significant diffusion and convection of cosmic rays associated to
            recent star formation, while the steeper spectral index ($\sim -3$)      
            at high Galactic latitudes suggests the relevance of diffusion 
            and that cosmic-ray electrons are trapped in the halo for a timescale
            long enough to lose a relevant fraction of their energy.
            The steep thermal dust spectral index ($\sim -2.2$) found at WMAP frequencies
            possibly implies a significant contribution from silicate grains.
            The overall level of the Galactic emission at WMAP scales has been confirmed
            to be minimum at frequencies about 70--80~GHz while upper limits
            of about 5~\% (at 22.8~GHz) have been estimated for the contribution of spinning 
            and magnetic dust emission. 
            Finally, the angular power spectrum of Galactic emission at all
            WMAP frequencies is approximately $\propto \ell^{-2}$ 
            and the same holds for each of the three most relevant components,
            indicating a strong correlation among them. 
            This slope, in agreement with that derived from Far-IR maps,
            is flatter than that derived
            from the Haslam map (about $\propto \ell^{-3}$).
            
            In spite of its quite poor sensitivity at small angular scales
            ($\ell \gtrsim 700$), WMAP provided also interesting results on
            discrete extragalactic sources and on their confusion noise.
            By cross-checking on several radio catalogues or directly by 
            identifying sources in the WMAP maps by dedicated filters,
            density fluxes of about two hundreds radio sources 
            (the large majority of them showing a quite flat spectrum) 
            have been measured.
            The corresponding level of confusion noise derived by  
            the authors is in good agreement with the $C_\ell$ value 
            at $\sim 40$~GHz of the model by 
            Toffolatti et al. \cite{toffolatti98},
            as well as the number counts at fluxes above 1--2~Jy
            that seems to favour the lowest predictions of this model,
            while smaller scale experiments (CBI, VSA), sensitive 
            to fainter sources, are in a more strict agreement 
            with its mean predictions. 
            Finally, although some thermal SZ effect toward 
            some bright cluster has been clearly detected by WMAP, 
            the global SZ contamination is found to be negligible in
            the WMAP first-year maps.

            The main scientific product of the WMAP data consists in the angular
            power spectrum of CMB temperature anisotropies and of the cross
            correlation between temperature and $E$ polarisation modes (see
            Fig.~\ref{WMAP_cl} \cite{spergel03}). 
            The sky maps used for the computation of the power spectra have been
            obtained by cross correlating the 60 GHz and the 90 GHz channels, in 
            order to reduce  systematics effects, and excluding the
            galactic plane from the analysis to minimise the foreground signal 
            contamination.
            
            The basic approach of the WMAP team in getting the best 
            cosmological model fit is a Bayesian analysis assuming flat 
            priors, obtaining the parameters characteristic of the best fit model 
            from the maximum of the likelihood surface \cite{spergel03}. 
            For each parameter, the one 
            dimensional likelihood is also constructed, by marginalising over all the 
            other parameters.  A first parameter estimation is done with respect to 
            a basic cosmological model ($\Lambda$CDM), with zero spatial curvature, 
            radiation, baryonic matter, cold dark matter and a cosmological
            constant.
            
            It turns out that the temperature-temperature (TT) and 
            temperature-polarisation (TE) cross correlations, as measured from 
            WMAP only, are well fitted by a set of values for the basic cosmological 
            parameters which is summarised in Tab.~\ref{cosmopar} (each value is given 
            at 2 $\sigma$ confidence level). It is important to note that the 
            combination of polarisation and temperature data has a 
            fundamental role in reducing the uncertainty in the parameter estimation.
            One of the most relevant results is that the 
            TE cross correlation appears to favour high optical depth values,
            indicating an early reionisation process; however, if 
            the spectral index $n_s$ is assumed to be scale-independent, the 
            likelihood is quite flat with respect to variations of these two 
            parameters, because the observed low amplitude of fluctuations at low 
            multipoles disfavours the additional large scale anisotropies which would 
            be produced by reionisation. The degeneracy is alleviated by assuming a 
            running spectral index.
            
            \begin{table}[here!]
              \begin{center}
                \caption{Main cosmological parameters estimated from WMAP
                  first year data}
                \label{cosmopar}
                \begin{tabular} {llll}
                  \\
                  Parameter   &  & Mean ( $68\%$  confidence range) & Maximum
                  Likelihood   \\
                  \hline
                  \hline
                  Baryon Density & $\Omega_b h^2$ &${0.024 \pm 0.001}$ &
                  0.023\\
                  Matter Density & $\Omega_m h^2$ &${0.14 \pm 0.02}$ &
                  0.13 \\
                  Hubble Constant & $h$ & ${0.72 \pm 0.05}$
                  &0.68 \\
                  Amplitude &$A$ &${0.9 \pm 0.1}$ & 0.78 \\
                  Optical Depth &$\tau$ &${0.166^{+ 0.076}_{- 0.071}}$ & 0.10 \\
                  Spectral Index &$n_s$ &${0.99 \pm 0.04}$  & 0.97 \\
                  &$\chi^2_{eff}/\nu$ &   &1431/1342\\
                  \hline
                  \hline
                \end{tabular}
              \end{center}
            \end{table}

            The WMAP temperature and polarisation data have been combined with recent 
            CMB data on smaller angular scales, and with complementary and 
            independent data from large scale structure (weak lensing, velocity 
            fields and clusters abundance, Lyman-$\alpha$ forest) and Type Ia 
            Supernovae, as well with measurements of the Hubble constant
            (see \cite{spergel03} and references therein). 
            In addition to the simpler $\Lambda$CDM, more complex 
            cosmological models (non-flat universes, Dark Energy models, 
            models with tensor perturbations, and models with massive neutrinos) 
            have also been considered; each of them required Monte Carlo 
            simulations to explore the likelihood surface.  Obviously, 
            increasing the complexity of the cosmological  model introduces additional 
            parameters, and possibly new degeneracies in the parameter space at
            the WMAP resolution. 

            When WMAP data are combined with independent datasets, it is possible to constrain 
            significantly the additional parameters in the model under 
            investigation, as in the case of a Dark Energy model with constant 
            equation of state $w$: in this case, the combination of CMB data and 
            Supernovae or large scale structures, place a $2 \sigma $ limit on $w <  
            -0.78$.
            In general, it has been found a noticeable agreement between WMAP 
            data and most of the independent astrophysical datasets. 

            The remarkable observational convergence made possible to build a 
            ``concordance model'' of the Universe, i.e. the currently accepted 
            cosmological scenario: 
            incontrovertible evidence has accumulated in favor of an almost flat 
            Universe, with 70\% of its energy density made by the Dark Energy, a
            cosmological component responsible for the acceleration in cosmic expansion,
            opening a new era for theoretical and fundamental physics; 
            the quite high value of the Hubble constant, as determined 
            by independent observations, forces the present Dark Energy equation of 
            state to be close to $-1$ (the value typical of the cosmological 
            constant); the standard Big Bang scenario received further powerful confirmation by 
            the remarkable agreement between the baryon density inferred from the 
            Deuterium/Hydrogen abundances ratio and the relative height of the CMB 
            temperature anisotropy. 
            
            Despite the agreement between WMAP data  and the predictions of the 
            ``concordance'' model, a note of caution should be done about the 
            ``anomalies'' found in the power spectrum: in particular, WMAP
            confirmed the COBE evidence for a lower-than-expected amplitude for the
            temperature quadrupole and octupole 
            anisotropies. It is still a 
            controversial question whether this  apparent discrepancy between data 
            and predictions requires a new physics, or it can be simply interpreted 
            as a systematics artifact or just a 
            low-probability realisation of the Universe. A detailed investigation 
            of the statistical significance of the low-multipole values of the 
            temperature  anisotropies can be found in \cite{efstathiou03}. 
            
            A CMB ``imager'' like Planck, i.e. an instrument capable to 
            image the primordial CMB anisotropies being signal dominated even at 
            its highest resolution of a few arcmin, will allow a deep investigation 
            of the physical nature of the concordance model assessed by WMAP
            and other astrophysical observations. 
            In particular, the detail in the CMB anisotropy pattern will spread 
            light on the possible primordial non-Gaussianity, allowing to
            discriminate between inflationary models, as well as on the 
            nature of the 96\% dark component in the Universe, through 
            the shear induced by forming structures at the onset of the 
            dark energy dominated era. 
            
            \begin{figure}[here!]
              \begin{center}
                \resizebox{16. cm}{!}{\includegraphics{f12_spectrum_s.eps}}
              \end{center}
              \caption{TT and TE cross power spectra for WMAP, COBE, 
                CBI and ACBAR data; the solid line corresponds to the WMAP best fit model 
                \cite{spergel03} (with permission, courtesy of the NASA/WMAP science team).}
              \label{WMAP_cl}
            \end{figure}
            
    \subsection{Planck}
    \label{subsec:planck}

        \subsubsection{Objectives, and expected performances}
        \label{subsubsec:planck_objectives}

            The Planck mission of the European Space Agency is the third generation space mission 
            dedicated to the measurement of CMB anisotropies. 
            Planck is designed to {\em fully}
            extract the cosmological information contained in CMB anisotropy
            by setting angular resolution, spectral coverage and sensitivity
            such that the power spectrum reconstruction will be limited by
            unavoidable cosmic variance and astrophysical foregrounds. Planck will also provide a precise measurement
            of the TE correlation and of the E-mode polarisation power spectrum and possibly offer a first B-mode detection.

            In order to achieve these ambitious goals, several, highly stringent performance requirements
            need to be satisfied by the instruments and the satellite (see Tab.~\ref{tab:planck_performances}):
            full sky coverage (to
            minimise sample variance at low multipole values), angular resolution $\lesssim 10\,'$ (to 
            measure the power spectrum up to the ``Silk damping'' region),
            noise per pixel $\lesssim 10\,\mu$K (to minimise the statistical uncertainties in the whole multipole
            range), wide frequency coverage (to ensure accurate foreground separation), 
            control of systematic effects at the $\mu$K level.

            \begin{center}
              \begin{threeparttable}
                \caption{
                  Main instrumental performances of Planck mission.
                }\label{tab:planck_performances}
                \begin{tabular*}{10.85cm}{|l|l|c|c|}
                  \cline{3-4}
                  \multicolumn{2}{c|}{ }
                                                                            &   {\it LFI}           &  {\it HFI}           \\
                  \hline \hline
                  
                  \multicolumn{2}{|l|}{Angular resolution}                  &     33$'$--13$'$      &   9.2$'$--5$'$       \\
                  \hline {Avg. $\Delta T /T$ per pixel}\tnote{a} &
                  Intensity                                                 & 6.5$\times 10^{-6}$   &  2.0$\times 10^{-6}$ \\
                  \cline{2-3} {(1 year mission)}\tnote{b}             &
                  Polaris.\tnote{c}
                                                                            & 9.2$\times 10^{-6}$   & 4.2$\times 10^{-6}$  \\
                  \hline \multicolumn{2}{|l|}{Baseline mission
                    lifetime}                                               & 14 months             & 14 months            \\
                  \hline \multicolumn{2}{|l|}{Spectral
                    coverage}                                               & 30--70 GHz            & 100--857 GHz         \\
                  \hline \multicolumn{2}{|l|}{Detector
                    technology}                                             & HEMT                  & Bolometers           \\
                  \hline \multicolumn{2}{|l|}{Detector
                    temperature}                                            & 20 K                  & 0.1 K                \\
                  \hline
                  \multicolumn{2}{|l|}{Cooling}                             & Active                & Active               \\
                  
                  \hline \hline
                \end{tabular*}
                \begin{tablenotes}
                  \small
                  \item[a] The sensitivity in thermodynamic temperature and
                    refers to the 70~GHz LFI and to the 100~GHz HFI
                    for on a 10$'$ pixel. A pixel is a square whose
                    side is the FWHM extent of the beam.
                  \item[b] These figures are calculated for the average integration
                    time per pixel. In general the integration time will be inhomogeneously
                    distributed on the sky and will be much higher in certain regions.
                  \item[c] Refers to E-mode polarisation (Stokes parameters U and Q).
                \end{tablenotes}
              \end{threeparttable}
            \end{center}

            The angular resolution is obtained with a 1.5~m aperture primary mirror,
            while the optimal optical response is offered by a combination of an off-axis dual reflector aplanatic telescope 
            and dual-profiled corrugated feed horns in the focal plane to maximise beam symmetry and
            minimise sidelobe pick-up (see Sect.~\ref{subsec:optical_pointing}).

            The most effective removal of astrophysical foregrounds will be guaranteed by a 
            wide frequency range, spanning from 30~GHz to $\sim$ 900~GHz.
            Such a wide frequency range calls for the implementation of different detector technologies, i.e. 
            HEMT radiometric receivers
            for the frequency range 30--70~GHz (the LFI instrument) 
            and bolometric detectors for the frequency range 100--900 GHz (the HFI instrument).

            The required sensitivity will be reached by a combination of wide-band components (20\% of the centre frequency
            for Planck-LFI, 30--40\% for HFI), long integration times (14 months of nominal mission time) 
            and multiple cryogenically-cooled detectors. In particular the front-end amplifiers of the LFI 
            instrument will be actively cooled at $\sim 20$~K, while the
            HFI bolometers will operate at $\sim 0.1$~K.

            The presence of a complex cryogenic chain and the need
            of a tight control of systematic effects has imposed the development of a satellite and instrument
            thermal design with high stability performances. The thermal design and orbit selection
            will guarantee the necessary stability to maintain the residual systematic effects at levels
            well below the instrument sensitivity.

        \subsubsection{Mission concept}
        \label{subsubsec:planck_mission_concept}


            \paragraph{Satellite and scanning strategy design.}
        
\begin{figure}[h!]
              \begin{center}
                \resizebox{15.5 cm}{!}{\includegraphics{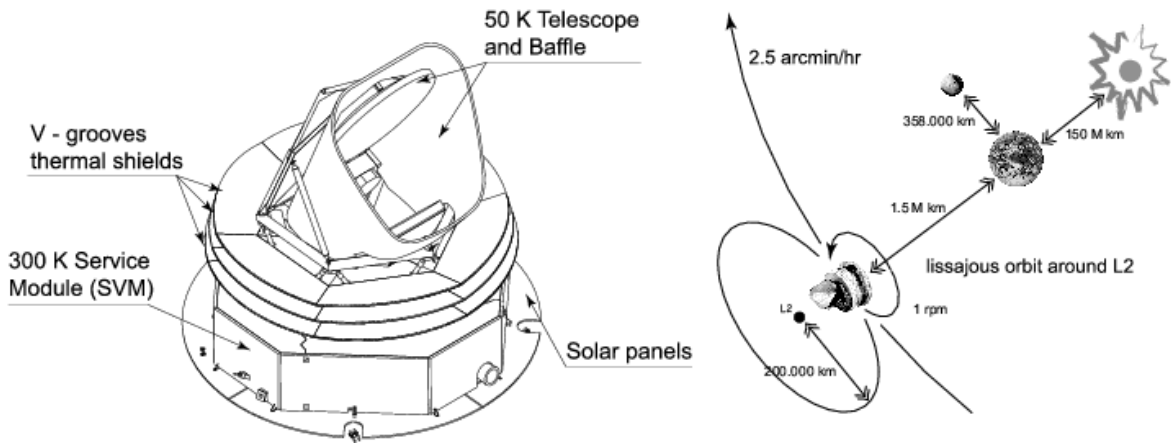}}
              \end{center}
              \caption{
                Left panel: a schematic of the Planck satellite showing the payload (with
                telescope and focal plane instruments), the warm service module and the
                three V-grooves that thermally decouple the cold from the warm satellite 
                stages. 
                The telescope includes the 
                primary reflector with its supporting structure, the secondary reflector with its 
                fixation struts, the telescope frame (holding the two mirror support structures, the focal plane 
                unit and the interfaces with the payload module struts), the payload module straylight baffle 
                (interfaced with the coldest V-groove shield at 50~K), the telescope inner baffle between the focal 
                plane array and the secondary mirror, the primary reflector extension baffle, the 
                instrumentation for the telescope hardware, and the hardware interfaces with the instruments. 
                Right panel: sketch of the nominal Planck scanning strategy.
              }
              \label{fig:schematic_planck_satellite_scanning_strategy}
            \end{figure}

            A schematic picture of the Planck satellite is shown in the left panel of 
            Fig.~\ref{fig:schematic_planck_satellite_scanning_strategy}. 
            The satellite is composed by two main modules: the payload module (PLM) with the telescope and the
            instruments in the focal plane and a Service Module (SVM) that houses the compressor
            systems of the on-board coolers and all the instrument and satellite warm electronics
            and subsystems.
            Three thermal shields of conical shape (called V-Grooves) 
            at $\sim$~150~K, $\sim$~100~K and $\sim$~50~K thermally
            decouple the passively cooled payload (at $\lesssim$~60~K) from the warm SVM (at $\sim$~300~K). The
            necessary power is provided by the solar panels located on the back of the SVM which are always
            oriented toward the Sun during the survey.
            
            After launch Planck will reach a nearly circular orbit ($\sim$~200.000~Km diameter) around the L2 Lagrangian
            point of the Sun-Earth system, at about 1.5 MKm from Earth,
            to maintain the focal plane instruments constantly in shadow with the sky in full view and the
            satellite with a very high degree of thermal stability.
            
            In the right panel of Fig.~\ref{fig:schematic_planck_satellite_scanning_strategy} 
            we show a sketch of the nominal Planck scanning strategy.
            Planck will spin at $\sim 1$~r.p.m., scanning the sky in nearly great
            circles with the telescope line-of-sight pointing at 85$^\circ$ with respect to the spin axis.
            Hourly maneuvers will re-point the telescope by about $\sim 2.5$~arcmin to maintain the spin
            axis constantly aligned with the Sun and the Earth. With this strategy each sky circle is scanned
            approximately sixty times before repointing (thus providing large redundancy over each scanned circle)
            and the whole sky is covered in about six
            months of continuous observations. The satellite also offers the possibility to 
            oscillate or precess the spin axis in order to increase the number of ``crossings''\footnote{
              With the term ``crossings'' we denote the pixels in the sky that are measured by a single detector
              during two different scan circles. These crossings are useful during data reduction to 
              recognise and possibly remove instrumental systematic effects that may be present in the
              measured data stream.
            } 
            and facilitate the removal of residual systematic effects from the data stream during data 
            processing. The definitive details of the scanning strategy 
            are not frozen yet. Because Planck will cover the sky twice, it is possible that the
            scanning laws of the two surveys will be different in order to optimise coverage and systematics
            rejection.
            
            \paragraph{Telescope optical design.}

\begin{figure}[h!]
              \begin{center}
                \resizebox{15.5 cm}{!}{\includegraphics{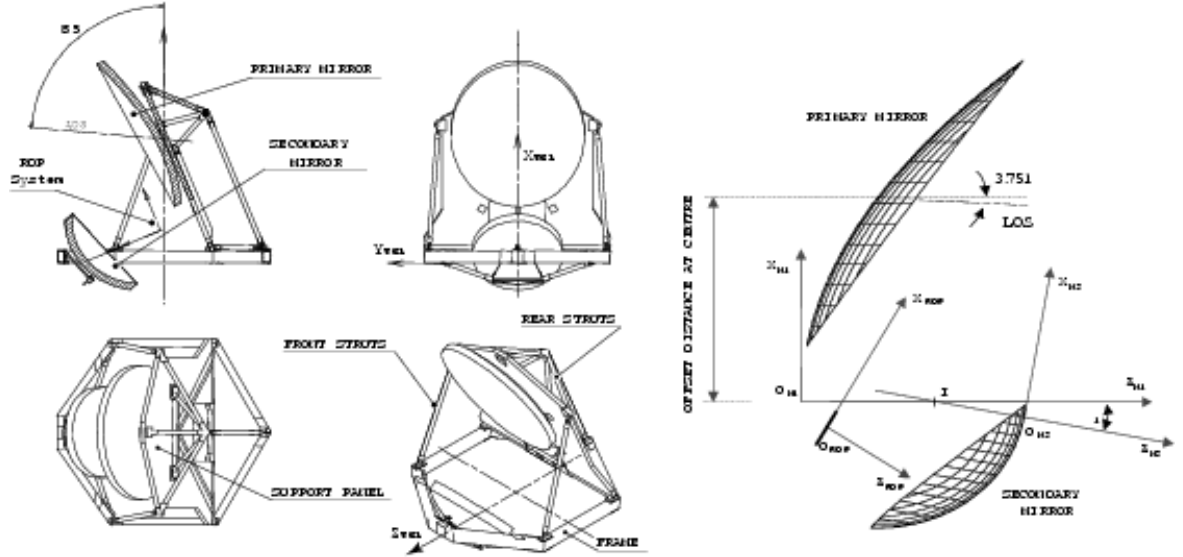}}
              \end{center}
              \caption{
               Telescope Design. Left: different views of the telescope mechanical layout. Right:
               telescope coordinate systems.
              }
              \label{fig:telescope}
            \end{figure}

            The Planck telescope represents a challenge for telescope technology and optical design, 
            because of the wide frequency coverage, the high performances required 
            by both HFI and LFI instruments sharing a very large 400$\times$400~mm 
            focal region, and the cryogenic environment 
            (40--65 K) in which the telescope will operate. 
            Its optical design is based on a two--mirror off--axis scheme (see Fig. \ref{fig:telescope}) 
            which offers the advantage to accommodate large focal plane instruments with an unblocked aperture, 
            maintaining the diffraction by the secondary mirror and struts at very low levels. 

            Both the primary and the secondary mirrors have an ellipsoidal shape as in the case of the Gregorian 
            aplanatic design \cite{dubruel00, villa02}. The primary mirror physical dimensions are 
            $\sim 1.9\times 1.5$~m, allowing a projected circular aperture of 1.5~m diameter. 
            The secondary reflector has been oversized up to approximately 1~m diameter to avoid any additional 
            under illumination of the primary. 

            The mirrors will be fabricated using Carbon Fiber (CFRP) technology. 
            The baseline consists of an all--CFRP rounded triangular tubes sandwich array arranged in a honeycomb-like 
            structure. This kind of structure has been chosen to satisfy the requirements of low mass ($< 120$~Kg 
            including struts and supports), high stiffness, high mechanical accuracy, and low thermal expansion 
            coefficient. The sandwich concept consists of a thick (4--10~cm) honeycomb-like core with the desired 
            shape, and to which two thin ($1-1.5$~$\mu$m) reflecting skins are bonded.
 
            \paragraph{Focal plane instruments.}
            
\begin{figure}[here!]
              \begin{center}
                \resizebox{14. cm}{!}{\includegraphics{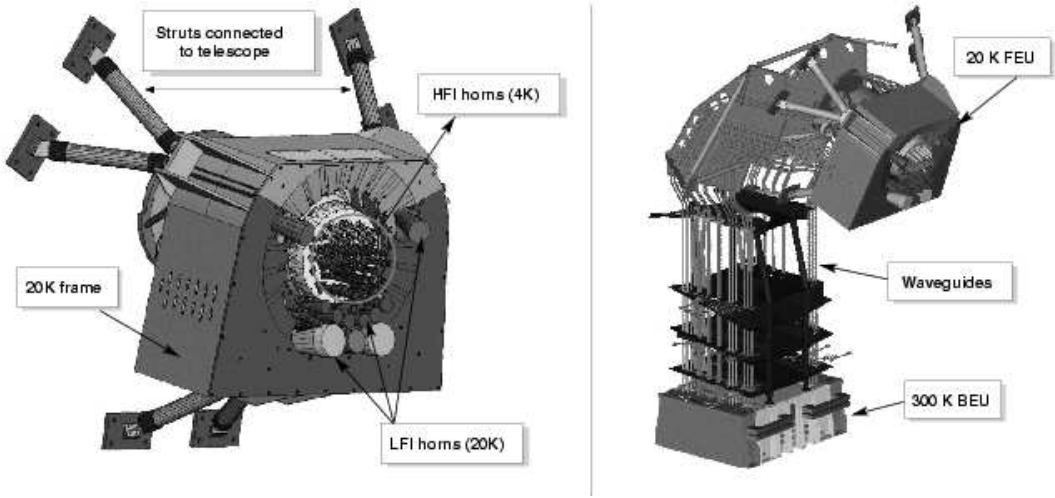}}
              \end{center}
              \caption{
                Left: view of the instruments in the Planck focal plane. The LFI horns surround the HFI instrument
                located in the centre of the focal surface. Right: view of the integrated LFI--HFI assembly. A set of 
                48 waveguides connect the front-end part of the LFI radiometers (FEU) to the warm back-end unit (BEU).
              }
              \label{fig:planck_focal_plane}
            \end{figure}

            Two complementary instruments are housed in the Planck focal plane (see 
            Fig.~\ref{fig:planck_focal_plane}): the Low Frequency Instrument
            (LFI) covering the 30--70~GHz range and the High Frequency Instrument (HFI) covering the
            100--857~GHz range. LFI is an array of coherent, differential radiometers
            based on cryogenic InP HEMT amplifiers operated at $\sim20$~K. To minimise power dissipation
            in the focal plane the radiometers are split into two sub-assemblies connected by a set of 
            custom designed waveguides. 

            The radiometers use a pseudo-correlation scheme to suppress
            1/$f$ noise induced by amplifier gain and noise temperature
            fluctuations \cite{seiffert02}. The differential measurement is taken comparing the sky signal against
            a stable internal cryogenic reference load cooled at $\sim$~4~K, taking
            advantage of the pre-cooling stage of the bolometric instrument.
            In addition, in LFI the effects of the residual input offset
            ($\lesssim 2$~K in nominal conditions) is compensated by introducing a
            {\it gain modulation factor} \cite{mennella03} which balances the output in the
            on-board signal processing and greatly improves the stability of the measured signal.
            
            Current LFI prototypes establish world-record performances in the
            30--100~GHz range for noise, bandwidth and low power consumption.
            The amplifiers at 30 and 44~GHz are incorporated into a microwave
            integrated circuit (MIC) \cite{roddis03}. At these frequencies the parasitics and
            uncertainties introduced by the bond wires in a MIC amplifier are
            controllable, particularly given the relatively low number of
            channels (total of 10), and the additional tuning flexibility
            facilitates optimisation for low noise. The LFI amplifiers have
            demonstrated noise temperatures $\sim 7.5$~K at 30~GHz with 20\%
            bandwidth. At 70~GHz there will be 12 channels \cite{sjoman03};
            amplifiers at these frequencies will use MMICs
            (monolithic microwave integrated circuits), which incorporate all
            circuit elements and the HEMT transistors on a single InP chip, so that the amplifiers
            can be mass-produced in a controlled process.
            Cryogenic
            MMIC amplifiers have been demonstrated at 75--115~GHz which
            exhibit $<35$~K over the required bandwidth. The LFI will thus
            fully exploit both MIC and MMIC technologies at their best. The
            LFI is ``naturally'' polarisation-sensitive in all of its
            channels, with a sensitivity to the Q and U Stokes parameters
            $\sim \sqrt{2}$ lower compared to temperature anisotropy. 


            The spectral region above 100~GHz is covered by the High
            Frequency Instrument (HFI), consisting of an array of 48
            spider-web bolometers operated at 0.1~K, and distributed in 6
            frequency bands from 100 to 850~GHz \cite{lamarre97}.
            Cooling the bolometers down to 0.1~K in space is a key driver in the
            design of the instrument and spacecraft. After the 50--60~K
            passive cooling and the 18--20~K stage (shared with LFI), the HFI cryo-chain uses a
            mechanical cooler to provide a 4~K stage (also
            used by LFI to cool the reference loads). The 4~K cooler is
            designed to minimise microphonic effects, and it operates at a
            frequency near 40~Hz adjustable in-flight to minimise resonances.
            Finally, a $^3$He - $^4$He closed-cycle dilution cooler provides
            0.1~K temperature to the bolometers. 

            The focal plane is
            constituted by an array of back-to-back horns at 4~K with
            spectral filters placed at $\sim 1.6$~K, and a third horn
            re-images radiation onto the bolometric detector \cite{church96}.
            For the HFI 100, 143 and 217~GHz channels, single-mode
            propagation is used. For these channels, as well as for LFI, the
            feed horns use double-profiled corrugated design to optimise telescope
            illumination and edge taper. At higher frequencies the angular
            resolution is achieved without diffraction-limited conditions,
            and multi-moded horns are being employed.

            Extraordinary sensitivities ($NEP \sim 1 \times
            10^{-17}$~W$\times$Hz$^{-1/2}$) are achievable with HFI, and have
            been demonstrated by prototype measurements. Thermal stability
            requirements are proportionally stringent, at sub-$\mu$K in the
            0.1~K stage environment. The use of ``spider-web'' bolometer
            technology \cite{bock96, lange96}, with silicon nitride
            micro-mesh, can provide nearly background-limited performance.
            Originally designed as
            a temperature sensitive instrument only, HFI has been modified to
            include polarisation sensitive bolometers (PSBs) \cite{delabrouille02}
            in a subset of the channels (100--350~GHz),
            allowing sensitive
            polarisation measurements at these frequencies.
            The HFI PSBs are
            built so that two absorbers made with parallel wires coupled to
            orthogonal polarisation modes are located in the same cavity and
            share the same feed horn, filter and optical path.

            \paragraph{Cryo-chain.}

            The performance of the ultra-high sensitivity detectors required for the Planck 
            mission is strongly coupled to their temperature. This implies that operating temperatures need 
            to be reached and maintained for the whole mission duration while temperature stability and possible 
            thermal noise due to the cryogenic systems must be controlled to the most accurate level. 

\begin{figure}[h!]
              \begin{center}
                \resizebox{15.5 cm}{!}{\includegraphics{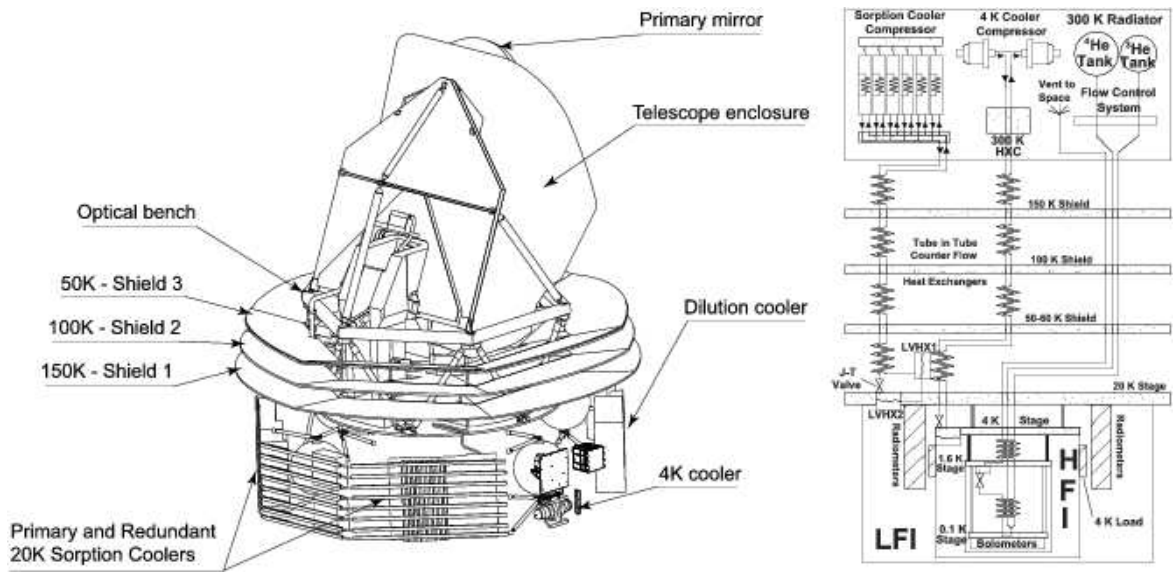}}
              \end{center}
              \caption{
                Left: The Planck spacecraft passive radiators scheme and the location on the spacecraft
                of the three active coolers. Right: schematic of the Planck Cryogenic Chain.
              }
              \label{fig:cryo-chain}
            \end{figure}

            The Planck thermal and cryogenic 
            design is one of the most complex ever conceived.
            The global architecture of the Planck cooling system (shown in right panel of 
            Fig.~\ref{fig:cryo-chain}) can be summarised as follows:

            \begin{itemize}
              \item Solar Array and SVM shield at 300~K to shield the payload from the sun. 
              \item Pre-cooling for all active coolers from 300~K to $\sim$~60~K by means of passive 
                radiators in three stages ($\sim 150$ K, $\sim 100$ K, $\sim 60$ K) \cite{bard84}.
              \item Cooling to 18--20~K for LFI and pre-cooling for the HFI 4~K cooler 
                with a $\mbox{H}_2$ Sorption Cooler
                \cite{wade00, prina02, pearson03a, bowman03}.
              \item Cooling to 4~K with a Helium Joule-Thomson cooler with mechanical compressors, 
                as pre-cooling stage for the dilution refrigerator \cite{bradshaw97}.
              \item Cooling down to 1.6 and 0.1~K with an open loop $^4\mbox{He}$-$^3\mbox{He}$ 
                dilution refrigerator \cite{benoit91, benoit97}.
            \end{itemize}

            In the thermal environment of the Earth-Sun Lagrangian L2 orbit, the Sun is the major 
            source of radiation: the combined action of the sun-shield/solar 
            array and the SVM shield 
            allows to keep the spacecraft in constant shadow.
            The cold payload is thermally decoupled from the warm SVM
            by low conductance struts and thermal radiators (the so-called V-Grooves, see left panel in 
            Fig.~\ref{fig:cryo-chain}). 

            The three V-Grooves are characterised by 
            low-emissivity specular surfaces with  
            an open angle of a few degrees between 
            each shield, allowing an extremely efficient heat rejection to 
            space \cite{bard84}. The coldest radiator, with a surface of only 3-4 m$^2$, 
            is capable of rejecting up to 2 W at 60 K.
            On the other side, V-Groove shields impose geometrical constraints and increase 
            integration complexity.

            Below 60~K a dedicated 
            active refrigeration system allows to reach and maintain the operating temperatures for 
            the two instruments, while minimising oscillations and thermal effects.
            The Planck cryogenic chain is driven by the $\mbox{H}_2$ Sorption Cooler \cite{wade00}, 
            a closed-cycle cryo-cooler designed to provide 1.2 Watt of heat lift at a temperature of 
            $\sim 18~\mbox{K}$ using isenthalpic expansion of hydrogen through a Joule-Thompson (JT) valve. 
            The Sorption Cooler performs a simple thermodynamic cycle based on hydrogen compression \cite{prina02}, 
            gas pre-cooling by three passive radiators, further cooling due to the heat recovery by the cold 
            low pressure gas stream, expansion through a J--T expansion valve and evaporation at the cold stage. 
            The engine of the Planck Sorption Cooler is the compressor system, based on the intrinsic property 
            of a metal hydride alloy (namely $\mbox{La}_{1.0}\mbox{Ni}_{4.78}\mbox{Sn}_{0.22}$) 
            to absorb large quantities of low-pressure hydrogen 
            and to desorb it at high pressure when heated in a limited volume \cite{pearson03a}. 
            The compressor system, with no moving parts,
            is made of six elements, each containing the sorbent material which is periodically cycled 
            between heating and cooling phases: their cycle phases are staggered in order to produce a 
            continuous stream of liquid refrigerant \cite{bowman03}. In such a system, there 
            is a basic clock time period over which each step of the process is conducted: since each 
            phase lasts 667 seconds, the cooler total cycle time is $\sim4000~\mbox{s}$.

            The gas leaves the compressor with a flow of 6.5 mg/sec, pressurised at 4.8 MPa: this high-pressure 
            hydrogen at $\sim 300$ K is then pre-cooled below the inversion temperature by the combination of cold 
            low-pressure return flow and the three V-Grooves. The temperature 
            of the coldest V-Groove determines the available heat lift of the cooler. After Joule-Thompson expansion, 
            the gas will partially liquefy, producing low pressure liquid refrigerant. The 
            saturated liquid-vapour hydrogen mixture is collected into two Liquid-Vapour Heat Exchangers, 
            LVHX-1 and LVHX-2, which interface to HFI and LFI, respectively. Heat 
            evaporates liquid hydrogen and the low-pressure gaseous hydrogen is re-circulated back to the sorbent for 
            compression. In order to 
            minimise the temperature oscillations at the interface plate with the LFI, a 
            Temperature Stabilisation Assembly (TSA), 
            based on an active PID control loop, is implemented on the LVHX2.

            The Planck lowest temperature stages (1.6 K and 0.1 K) are reached by the 0.1 K Open Cycle 
            Dilution/Joule-Thomson Refrigerator, which requires the gas to be pre-cooled at a temperature lower 
            than 10 K. This task is performed by a closed cycle 4 K Joule Thomson (JT) Cooler \cite{bradshaw97}. 
            This refrigerator uses $^4\mbox{He}$ as a fluid, pressurised by a pair of mechanical compressors mounted 
            back-to-back and controlled by low vibration drive electronics with force transducers and a servo 
            feedback loop to minimise the transmitted vibrations.
            Part of the heat lift produced by the Sorption Cooler is used to cool helium down to 18~K
            by high-efficiency heat exchangers. 

            The dilution refrigerator exploits a new dilution principle based on friction that does 
            not need gravity to operate \cite{benoit91}. Its cooling power depends on the low gas flow, which 
            allows enough gas storage to achieve long mission life \cite{benoit97}. For a total 
            ($^3$He and $^4$He) flow rate of $12~\mu\mbox{mole/s}$, a cooling power of 100 nW at 0.1 K has been 
            demonstrated. In the same process, the mixture is expanded in a Joule-Thomson valve, producing a cooling power 
            of several hundreds of mW at 1.6 K. This is enough to ensure a proper insulation of the 0.1 K stage from the 
            radiative and conductive thermal loads coming from the 4 K stage. The 1.6 K stage supports filters and intercepts 
            heat from the 4 K stage. The 0.1 K stage refrigerates the bolometers, thermometers, heaters, and filters. 
            Its temperature is controlled by a closed loop active system. The tubes from and to each stage are attached to 
            form heat exchangers for all circulating fluids in order to minimise thermal losses.

            \paragraph{Calibration.}


            Photometric calibration is necessary to 
            to convert the receiver output 
            (arbitrary digitised  telemetry units) 
            to the corresponding observed temperature of the sky (``absolute calibration'')
            and to monitor variations of such conversion with time 
            (``relative calibration'') to compensate for instrumental drifts.

            Calibration needs to be performed accurately and frequently, 
            to control the instruments behaviour.
            It is a crucial operation 
            since errors in this procedure may create artifacts and errors in the
            amplitude of detected structures in the final maps.
            Calibration uncertainty is one of the most significant systematic errors 
            affecting current CMB anisotropy experiments. The measurement accuracy sought by precision
            space experiments requires
            a calibration accuracy at the level of $\sim 1\%$ in order to derive the CMB power
            spectrum with comparable accuracy \cite{cappellini03}.
            
            Calibration is typically performed
            by observing celestial sources with known intensity, such as
            planets, strong radio sources or the CMB dipole.
           
            For the Planck instrument, relative calibration is performed every hour,
            thanks to the adopted scanning strategy. 
            Bright sources 
            (e.g. {\sc Hii} regions at lower frequencies, or planets) 
            are very useful in this respect, when available.
            This allows to establish calibration between individual scans, before
            and independently of establishing an absolute calibration for all data.
            Absolute calibration is performed on different time scales.
            On short time scales, it mainly relies on the CMB dipole induced 
            by the motion of the solar system, which is known at 0.4\% level thanks to 
            the COBE-FIRAS measurement \cite{fixsen96} and on the strong signal from the galactic plane
            \cite{cappellini03}.
            On long time scales ($\sim$ 3 months),
            it relies on the dipole modulation of the signal induced by the motion of the Earth 
            (and consequently of the satellite) around the Sun.


            \paragraph{Foregrounds removal.}
            As already mentioned in Sect.~\ref{sec:astrophysical_limitations}, any CMB experiment observes
            together with the CMB other foreground sources of astrophysical origin. These sources
            have in general a different spectral and spatial behaviour than the CMB and this fact will, 
            in principle, allow proper separation of the various contributions. However the strong accuracy 
            requirements of present and future CMB missions makes this a non trivial issue.
            In this respect a multi-frequency experiment with optimal spacing between frequencies is the
            first step for a proper separation of the sky components. Furthermore algorithms that allow
            effective foreground separation from measured maps are needed. 

            These algorithms can be divided into two main
            categories: the so-called ``blind'' algorithms and those that require some a-priori knowledge
            on the components to be separated. The Maximum Entropy Method (MEM) \cite{hobson98, stolyarov02}
            is an example of the latter category that 
            has been developed
            and already applied, both in real and in harmonic space, to Planck-like data-sets.
            The priors can be in the form of spectral indexes of the components or their
            spatial behaviour. 
            The Expectation Maximisation (EM) \cite{patanchon03} and the 
            Fast Independent Component
            Analysis ({\sc FastICA}) \cite{maino02}, instead, are blind schemes that have also
            been applied to astrophysical data-sets. {\sc FastICA}, in particular, 
            needs
            no assumptions either on the spectral or on the spatial dependences of the components but it requires
            only their statistical independence and the non Gaussianity of all (but one) components.
            This condition is easily fulfilled in CMB foreground separation where all the components are
            independent and CMB is expected to be Gaussian while
            foregrounds are not.
            {\sc FastICA} has been applied to both 
            simulated Planck-like
            data \cite{maino02} as well as to real data (e.g. on COBE-DMR \cite{maino03}). 
            An implementation in harmonic space is under way with the possibility of a further 
            improvement in the separation.
            
            \paragraph{ Data Processing Centres.} To correctly perform the calibration and
            foreground removal phases, fundamental for the proper scientific
            exploitation of the mission, a whole set of preparatory and merging work on
            the Planck data needs to be accomplished. These activities range from the
            receipt of the telemetry sent by the satellite to the actual delivery of the
            final scientific results of the mission to the European Space Agency,
            through a number of processing steps defining a ``pipeline''. Such
            activities are carried out by two separate but coordinated Data Processing
            Centres (DPCs), one for each of the Planck instruments.
            
            The DPCs are responsible for the operations of their own instrument, and for
            the production, delivery and archiving of the scientific data products,
            which can be considered as the deliverables of the mission. Such products
            include calibrated time series data for each receiver after removal of
            systematic features and attitude reconstruction; photometrically and
            astrometrically calibrated maps of the sky in the observed bands; sky maps
            of the main astrophysical components; catalogues of sources detected in the
            sky maps of the main astrophysical components. The DPCs are furthermore
            responsible for the production of realistic data simulating the behaviour of
            the instrument in flight, and for the support to instrument testing on the
            ground.
            
            During the pre-launch era, the DPC is built by integrating and making
            operational at a single site, within a data processing pipeline, a set of
            input algorithms implemented by scientists distributed throughout a dozen
            different countries. During flight, the DPC main efforts will
            be concentrated on controlling their instruments and achieving the best possible
            performances, through a detailed analysis of their
            behaviour and preliminary processing of scientific data. In the
            post-operations phase, the processing procedures will be finalised, and the
            whole pipeline will be run on all of the data, from telemetry handling down
            to the scientific analysis of the mission results, and delivery to the
            scientific community.

        \subsection{Current status}
        \label{subsec:current_status}

            With the release of the 1 year results from WMAP we enter the era of precision 
            cosmology. WMAP
            clearly detected acoustic oscillations in the CMB temperature power spectrum with 
            cosmic-variance accuracy
            up to $\ell \simeq 350$. Correlation between temperature and polarisation $E$-modes have 
            been also reported which
            is able to better constraint the re-ionisation epoch of the Universe. The main result is 
            that a flat $\Lambda$-dominated
            Universe seeded by a nearly scale-invariant adiabatic Gaussian fluctuations is a very good 
            fit to the WMAP data
            and a series of other astronomical observations.
            When combining WMAP data with those from smaller scales CMB experiments 
            (CBI \cite{mason03} and ACBAR \cite{kuo02})
            with large scale structure measurements from 2dFGRS \cite{percival01} and 
            Ly$\alpha$ power spectrum 
            \cite{croft02, gnedin02}, the data yield the values for the basic cosmological 
            parameters reported in Tab.~\ref{wmap_param}.
            
            \begin{table}[here!]
              \begin{center}
                \caption{Best-fit parameters from WMAP, 2dFGRS and Ly$\alpha$ data}
                \label{wmap_param}
                \begin{tabular}{lc}
                  \hline\hline
                  \ Parameter & Value \\
                  \hline
                  \ $A$                &  $0.83^{+0.09}_{-0.08}$ \\
                  \ $n_s$              &  $0.93\pm0.03$ \\
                  \ $dn_s/d{\rm ln}k$  &  $-0.031^{+0.016}_{-0.017}$ \\
                  \ $tau$              &  $0.17\pm0.06$ \\
                  \ $h$                &  $0.71^{+0.04}_{-0.03}$ \\
                  \ $\Omega_m h^2$     &  $0.135^{+0.008}_{-0.009}$ \\
                  \ $\Omega_b h^2$     &  $0.0224\pm 0.0009$ \\
                  \hline
                \end{tabular}
              \end{center}
            \end{table}
      
            The situation is expected to improve as long as WMAP will collect data for a 
            total of 4 years of operations. This
            will increase the signal-to-noise ratio by a factor of 2 with respect to 1 year 
            data yielding better estimation
            of cosmological parameters. Although the picture is clear in its main features it is not 
            complete yet. There are indeed
            some open issues raised by the WMAP data: as usual a new powerful experiment 
            will answer to many questions but it will also invariably open many
            new ones. There is, for example, the issue of low level of quadrupole and the 
            ``strange'' alignment of quadrupole and
            octupole \cite{deoliveira03} for which an explanation is not at hand. It is 
            possible that these low multipoles
            are still affected by not-subtracted foreground emissions for which WMAP is not 
            optimally designed.
            
            These questions will be possibly answered by a 
            real full-sky imaging experiment with the capability of detecting CMB anisotropy up to the
            maximum relevant multipole corresponding to the physical dimension of the Last Scattering 
            Surface, which is the goal
            of the ESA Planck mission.
            The high quality of Planck data is expected to reduce the error bars in cosmological 
            parameters at the level of
            few percent thanks to the combination of temperature and polarisation data, and
            for a wide set of cosmological parameters \cite{popa01} the final accuracy will be 
            largely independent from auxiliary informations coming from 
            other classes of astronomical observations.
            
            What do we expect after Planck? Recently it has become clear the the future of CMB 
            experiments is in the
            imaging of the polarisation anisotropy. We are in the same situation as we were before COBE-DMR 
            with respect temperature
            anisotropy detection. Polarisation is a powerful tool probing directly the Last 
            Scattering Surface, in particular
            the detection of $B$-modes is a key issue since it will strongly constrain the picture of 
            our Universe.

\section{FUTURE CHALLENGES}
\label{sec:future_challenges}

    In analogy with what happened after FIRAS measurement of the
    frequency spectrum \cite{mather96}, after Planck one might expect a
    decrease of activity in traditional anisotropy projects. However, in some
    areas Planck will act as a pioneering mission rather than as a
    conclusive one. Two main research directions can be anticipated: precision
    measurements of CMB polarisation and deep imaging at sub-arcminute
    scales. On these objectives, a number of sub-orbital programs are ongoing
    or are being planned for the coming several years.

    \subsection{Polarisation}

        Though extraordinary, the cosmological information obtainable with
        temperature anisotropy alone is far less than what in principle can be
        achieved with high-precision CMB polarimetry. Linear polarisation in the
        CMB arises from Thomson scattering of anisotropic radiation at last
        scattering \cite{rees68}, and it encodes information complementary to
        temperature anisotropies.

        Even the remarkable sensitivity of Planck to $Q$ and $U$ Stokes
        parameters will be far from fully extracting the information encoded in
        the cosmic polarised signal. In particular higher sensitivity will be
        needed to measure in detail the B-mode spectrum, related to gravitational
        effects. Deep polarisation maps may give indications on the inflation
        energy scale, thus probing ultra-high energy physics to levels beyond what
        can be obtained with any conceivable terrestrial particle accelerator.
        Even more subtle, but not less interesting, will be the exploration of
        effects of weak gravitational lensing through the distortion of the CMB
        polarisation on small scales \cite{zaldarriaga97}.

        Until recently, only upper limits to CMB polarisation existed (see
        \cite{hedman01} and references therein) at a level 10--15 $\mu$K.  A major
        recent breakthrough has been the detection of E-mode polarisation by the
        ground-based interferometry experiment DASI \cite{kovac02} at few
        $\mu$K level, and the detection by both WMAP and DASI of the T-E
        correlation \cite{kogut03, kovac02}. The full-sky data of
        WMAP allow to trace the correlation up to the largest scales, while DASI
        is limited by sample variance. The T-E correlation is a consequence of the
        fact that the fluid velocities which lead to polarisation are produced by
        the same density perturbations which are responsible for temperature
        anisotropy. These results provide confirmation of the standard model and,
        as discussed in Sect.~\ref{subsec:wmap_discussion_results},
        the T-E correlation on large angular scales
        observed by WMAP gives evidence of reionisation at redshifts $11<z<30$.
        
        At the same time, these first polarisation detections open up a new stage
        in CMB research. Both DASI and WMAP are still ongoing, and a number of new
        CMB polarisation experiments are being prepared from ground, balloon and
        eventually space.

        \subsubsection{Ongoing and planned Polarisation experiments}

            A few ground-based polarisation instruments are currently taking data (see
            e.g. \cite{carlstrom03}).  The CBI is operative at the Atacama plateau
            in Chile \cite{padin02} and it is based on Ka-band low-noise HEMT
            amplifiers. CBI achieves a sensitivity similar to that of DASI, but it
            improves the angular resolution by a factor $\sim 3$. The CAPMAP \cite{barkats03}
            was operated in 2003 with four W-band correlation receivers coupled
            to the Bell Labs 7-meter telescope leading to $\sim 5'$ resolution. In the
            future, 16 radiometers in W band and four in Q band are foreseen. The
            Beast instrument \cite{meinhold03}, currently taking data from the
            White Mountain Research Station, is planned to be converted into a W-band
            polarimeter. In addition, two new PSB bolometric arrays are being
            developed, both expected for 2005: BICEP (Back-ground Imaging of Cosmic
            Extragalactic Polarisation), to be operated from the South Pole \cite{keating03},
            and QUEST (Q and U Extra-galactic Survey Telescope) to be
            coupled to the DASI telescope. BICEP will be sensitive to scales $10
            <\ell<200$ and QUEST in the overlapping range $100 <\ell<2000$, both with
            sensitivity adequate to mapping the E-mode angular power spectrum, and
            possibly detection of B-modes.
            
            Ongoing balloon-borne polarisation programs include Boomerang \cite{montroy03}
            which made an Antarctic 12-day LDB flight in January 2003, and
            MAXIPOL which had a successful flight in May 2003 from Fort Sumner, New
            Mexico. The focal plane of Boomerang was reconfigured for its 2003 flight
            with PSBs developed at JPL, similar to those foreseen for Planck HFI. The
            MAXIPOL polarimeter was obtained as a refurbishment of MAXIMA by adding a
            rotating half-wave plate and fixed polarizing grids in front of the horns.
            Both Boomerang and MAXIPOL have the sensitivity to detect E-mode
            polarisation at degree scales. A space experiment to search for
            polarisation at $>7^\circ$ scales is SPOrt (Sky Polarisation Observatory)
            \cite{cortiglioni00} proposed for implementation on the International
            Space Station. The instrument is based on HEMT radiometers cooled at $\sim
            90$~K and operating at 4 frequencies in the 20--90~GHz range.  The
            balloon-borne counterpart of SPOrt, called BaR-SPOrt (Balloon-borne Radiometers for Sky POlarisation)
            \cite{zannoni02} has a
            multi-feed focal plane of polarimeters in W band at the focus of an
            on-axis telescope reaching sub-degree angular resolution.
            
            After its first detection, rapid progress can be expected in the
            observation of the E-mode power spectrum.  All these measurements are
            extremely demanding in terms of control of instrumental systematic errors
            and foreground contamination. The challenge will be to obtain results
            which are actually limited by the extraordinary sensitivity of the
            cryogenically cooled detectors. The present generation of instruments will
            measure CMB polarisation with previously unreached sensitivity, and will
            also explore new techniques and detectors, possibly uncovering unexpected
            challenges in terms of instrumental and astrophysical limitations.
            
        \subsubsection{Polarisation: the future}

            The polarisation amplitude, now known to be at few $\mu$K level, imposes
            new requirements on instrument sensitivity.  Advances in both bolometer
            and coherent receiver technology are promising, particularly in the
            direction of large optimised focal plane arrays. Compact and relatively
            low cost correlation receivers may be produced in the near future based on
            integrated circuit technologies. Arrays of hundreds of radiometers may be
            assembled either as independent radiometer channels or as future
            interferometers. Similarly, large arrays for bolometers are being
            experimented (see, e.g., \cite{dowell03}) based on monolithic arrays
            developed at Caltech/JPL. Arrays of $\sim 1000$ or more channels are
            possible with this technology.  An example is the planned experiment
            Polarbear, a 3 meter off-axis telescope dedicated to CMB polarisation
            measurements with as many as 1000--3000 elements of polarisation-sensitive
            antenna-coupled bolometers.
            
            Multi-frequency observations, possibly with matching beam sizes, would be
            needed to tackle the subtle problem of polarised foreground separation. It
            is difficult to anticipate which systematic effects could represent
            limiting factors at sub-$\mu$K levels; it is likely that thermal stability
            at the cryogenic temperatures needed by the ultra-high sensitivity
            detectors (either radiometers or bolometers) will be a major challenge;
            also, a multi-channel off-axis instrument, pushed at extreme
            sensitivities, would place critical requirements for ultra-low
            cross-polarisation.
            
            The sensitivity requirements for a full-sky, high-precision polarisation
            survey are extremely demanding.  The expected B-mode amplitude induced by
            gravitational waves is $\sim 0.1\%$ to 1\% of temperature $\Delta T/T$: a
            high signal-to-noise imaging requires a noise per pixel $\sim 0.05$~$\mu$K,
            i.e. about 300 times better than Planck. This is beyond the
            foreseeable future, but an intermediate step aiming at a full measurement
            of the power spectra, with sensitivity 0.5 $\mu$K per $10'$ resolution
            element, can be approached extrapolating existing technology.  The growing
            CMB community is now already in the process of discussing plans for a
            possible fourth-generation space mission dedicated to CMB polarisation. As
            in the case of Planck for temperature anisotropy, such mission would be
            expected to provide polarisation observations limited by cosmic variance
            and unavoidable foregrounds.

    \subsection{Fine-scale temperature signatures}

        On angular scales $\sim 1'$ and below the CMB is highly influenced by
        interaction with intervening ionised material. The CMB passed through the
        cosmic ``dark ages'', before star and quasar formation, from which direct
        observation is extremely difficult to obtain.  Using the CMB as a high
        redshift back-light, the study of its arcmin scale features may be one of
        the most powerful techniques to observe the very early processes of
        structure formation.  Furthermore, accurate S-Z measurements can be made
        to very high redshifts \cite{stebbins97}, all the way back to the cluster
        formation era \cite{aghanim97}.  Planck angular resolution is not
        sufficient to fully cover this promising area of CMB studies.  The S-Z
        effect from clusters is by far the dominant secondary source. Once this
        will be well identified and mapped, sky regions free from S-Z and other
        local foregrounds could be searched at $\sim 1'$ scales for fainter
        secondary signatures, such as those from gravitational collapse of large
        scale ($\sim 100$\,Mpc) structures, bulk motion of plasma (the
        ``Ostriker-Vishniac effect''), effects of the evolution of gravitational
        potentials on CMB photons (the ``integrated Sacks-Wolf effect'' and the
        ``Rees-Sciama effect''), lensing-induced signatures from clusters
        \cite{seljak00}, signature of local ionisation events \cite{platania02},
        and details of the ionisation history of the Universe.
        
        It is easy to expect that the WMAP and the future Planck surveys
        will trigger observations in selected sky areas searching for detailed,
        physically interesting features.  New instruments and observing strategies
        will be, and are already being, developed for the purpose.

        \subsubsection{Ongoing and planned fine-scale observations}

            The brightest clusters give a S-Z thermal component $\sim 1$~mK, while the
            kinetic effects is expected to be $\sim 10$ times lower.  Signatures such
            as filaments from in-falling clusters are expected at $\sim 10$~$\mu$K
            level.  Current bolometric or HEMT-based instruments, either filled
            aperture arrays or interferometers, are able to approach sub-$\mu$K
            sensitivity in very localised areas.
            
            Most ground based projects currently active or planned are based on
            interferometer instruments, capable of very high angular resolution.  As
            an example, the interferometer experiment called AMiBA (Array for
            Microwave Background Anisotropy), is expected to be soon operative
            \cite{kesteven02}.  The instrument is being built for fine-scale CMB
            temperature imaging, for measurements of the polarisation anisotropy, as
            well as for observations of the Sunyaev-Zel'dovich Effect. This will be a
            19 elements, W-band array by the Academia Sinica, Institute of Astronomy
            and Astrophysics, and National University of Taiwan.  Two sets of array
            dishes are planned, with 1.2 meter and 0.3 meter size. Initial
            observations are planned to start in 2004.
            
            Antarctic winter fine-scale observations may also have a long future. The
            Arcminute Cosmology Bolometer Array Receiver (ACBAR) is a 16 element array
            of bolometers cooled at 0.23~K designed for observation of small scale
            anisotropy and Sunyaev-Zel'dovich effect. The experiment, run by the Centre
            for Astrophysical Research in Antarctica (CARA), was deployed at the South
            Pole in December 2000. Preliminary results in the range $200 < \ell <
            2200$ based on data taken in the austral winter of 2001 were presented
            \cite{kuo02} including serendipitous discoveries of high redshift clusters
            via their S-Z effect.
            
        \subsubsection{Fine-scale: the future}

            At wavelengths $\lambda \approx 1~$mm, a $\lesssim 1'$ resolution
            translates in a telescope aperture typically $D \sim 10$~m.  New
            telescopes, to be coupled with ultra-high sensitivity arrays, are required
            to ensure sub-arcmin resolutions. Some of these are already in an advanced
            stage of planning. The South Pole Telescope (SPT) is an off-axis precision
            telescope with an aperture of 8 meters. The SPT is expected to be
            installed at the South Pole site in 2006, together with a large
            1000-element bolometric radiometer, and possibly with a large focal plane
            polarimeter. Another planned high-resolution
            millimetre telescope is the Atacama Cosmology Telescope (ACT), a 6-meter
            off-axis to be deployed in Chile.
            
            Although not primarily designed for CMB studies,
            the extraordinary resolution and sensitivity at $\nu \lesssim 20$~GHz
            of the Square Kilometre Array 
            (SKA)\footnote{http://www.skatelescope.org/ska\_concept.shtml}
            will allow an extremely accurate control of radiosource fluctuations 
            at very large multipoles useful for precise re-analysis of CMB dedicated 
            surveys and will represent at the same time a good 
            opportunity for studying CMB secondary anisotropies relevant at 
            multipoles $\approx 104-105$ (e.g. induced by thermal  
            SZ, Ostriker-Vishniac and Rees-Sciama effects), 
            and to detect and map thermal SZ effects at sub-arcmin scales 
            associated to gas heated at its virial temperature in proto-galaxies
            or to strong energy injections from quasars \cite{dezotti03}.
            
            Focal plane arrays for deep arcmin imaging are likely to be
            simultaneously sensitive to linear polarisation.  As mentioned,
            polarisation-sensitive arrays with up to 1000 elements ore more are being
            envisioned. In the future, high precision fine-scale CMB imaging would
            call for very wide frequency coverage, with ancillary monitoring at low
            frequency ($<5$~GHz) and high frequency ($>5$~THz) to safely remove
            unrelated foregrounds as well as backgrounds (such as primary CMB
            anisotropy, regarded as a source of confusion in this context).

             These observations can be expected to progress for several years with
            ground-based instrumentation.  A future sky survey of clusters velocities
            based on their S-Z features throughout the entire Hubble volume could, in
            principle, map the evolution of velocity fields over much of the history
            of the Universe. Eventually, a fine-scale, full-sky survey will only be
            possible with a space programme, currently out of reach.

\section{CONCLUSIONS}

    After the discovery of CMB temperature anisotropies by COBE, a great
    experimental effort has been accomplished with ground-based and
    balloon-borne experiments and recently by the WMAP space mission, leading
    to a good determination of the angular power spectrum up to sub-degree
    scales. The acoustic nature of the spectrum, expected by theory, is now
    beautifully matched by observation. This is a truly remarkable
    achievement. The recent detection of polarisation E-modes and the TE
    correlation represents a further striking confirmation of the soundness of
    standard cosmology and provides new incentive to the experimental growth
    of CMB research. However, these rewarding results are probably
    only a foretaste of the precision observation attainable by the
    forthcoming generations of experiments, culminating with the Planck
    mission. The Planck survey will generate an unprecedented set of
    multi-frequency temperature and polarisation data, leading to major
    improvement in the determination of cosmological parameters and
    to a new profound verification of our cosmological understanding.
    
    Beyond Planck, it is likely that CMB observations will concentrate
    on precision polarisation measurements (especially searching for
    gravitational wave signatures in the B-mode spectrum) and deep sub-arcmin
    imaging of secondary anisotropies. Large arrays of cryogenic detectors,
    now under study, should be capable of reaching the extreme instrument
    sensitivity required. However, this will not be enough. The experience of
    the past decades has shown that CMB measurements are typically limited by
    systematic effects, either of instrumental or astronomical nature, some of
    which have been discussed in this review. These limitations will need to
    be explored at a much deeper level in future high-precision enterprises.
    
    Precision measurements of the CMB are able to shed light on very high
    energy phenomena occurring in the primordial cosmic environment, as well
    as on the physical history of the Universe. Today, observations of the CMB
    promise to remain one of the most powerful cosmological probes for yet
    many years in the future.
\\

\noindent{\bf Acknowledgements}

    The completion of this review has been greatly
    helped, either directly or indirectly, by the work of many
    people, in particular by the Planck Science Team, by the
    Planck-LFI consortium and by the Boomerang and MAXIMA teams. 
    We would also like to thank the WMAP science team for figure
    permission and  L.A. Popa, D. S\'aez, L. Toffolatti,
    L. Danese and G. De Zotti for useful discussions.

\bibliographystyle{custom} 
\bibliography{references} 

\hyphenation{Post-Script Sprin-ger}
\begin{thebibliography}{100}

\bibitem{penzias65}
{Penzias}, A.~A. and {Wilson}, R.~W., 1965, \apj, 142, 419--421.

\bibitem{reber40}
{Reber}, G., 1940, \apj, 91, 621--624.

\bibitem{reber44}
{Reber}, G., 1944, \apj, 100, 279.

\bibitem{hubble29}
{Hubble}, E., 1929, Proceedings of the National Academy of Science, 15,
  168--173.

\bibitem{gamow46}
{Gamow}, G., 1946, Physical Review, 70, 572--573.

\bibitem{fixsen96}
{Fixsen}, D.~J., {Cheng}, E.~S., {Gales}, J.~M., et~al., 1996, \apj, 473, 576.

\bibitem{bersanelli94}
{Bersanelli}, M., {Bensadoun}, M., {de Amici}, G., et~al., 1994, \apj, 424,
  517--529.

\bibitem{deamici91}
{de Amici}, G., {Limon}, M., {Smoot}, G.~F., et~al., 1991, \apj, 381, 341--347.

\bibitem{bensadoun93}
{Bensadoun}, M., {Bersanelli}, M., {de Amici}, G., et~al., 1993, \apj, 409,
  1--13.

\bibitem{smoot87}
{Smoot}, G.~F., {Bensadoun}, M., {Bersanelli}, M., et~al., 1987, \apjl, 317,
  L45--L49.

\bibitem{sironi91}
{Sironi}, G., {Bonelli}, G., and {Limon}, M., 1991, \apj, 378, 550--556.

\bibitem{salvaterra02}
{Salvaterra}, R. and {Burigana}, C., 2002, \mnras, 336, 592--610.

\bibitem{smoot92}
{Smoot}, G.~F., {Bennett}, C.~L., {Kogut}, A., et~al., 1992, \apjl, 396,
  L1--L5.

\bibitem{bersanelli02}
{Bersanelli}, M., {Maino}, D., and {Mennella}, A., 2002, Riv. Nuovo Cimento,
  25(9).

\bibitem{sachs67}
R.K., S. and A.M., W., 1967, Astrophysical Journal, 147, 73.

\bibitem{silk68}
{Silk}, J., 1968, \apj, 151, 459.

\bibitem{bond00}
{Bond}, J.~R., {Jaffe}, A.~H., and {Knox}, L., 2000, \apj, 533, 19--37.

\bibitem{verde03}
{Verde}, L., {Peiris}, H.~V., {Spergel}, D.~N., et~al., 2003, \apjs, 148,
  195--211.

\bibitem{madau03}
{Madau}, P., 2003, IAU Symposium, 687.

\bibitem{Becker01}
{Becker}, R.~H., {Fan}, X., {White}, R.~L., et~al., 2001, \aj, 122, 2850--2857.

\bibitem{sugiyama93}
{Sugiyama}, N., {Silk}, J., and {Vittorio}, N., 1993, \apjl, 419, L1.

\bibitem{debernardis97}
{de Bernardis}, P., {Balbi}, A., {de Gasperis}, G., {Melchiorri}, A., and
  {Vittorio}, N., 1997, \apj, 480, 1.

\bibitem{hu00}
{Hu}, W., 2000, \apj, 529, 12--25.

\bibitem{kogut03}
{Kogut}, A., {Spergel}, D.~N., {Barnes}, C., et~al., 2003, \apjs, 148,
  161--173.

\bibitem{spergel03}
{Spergel}, D.~N., {Verde}, L., {Peiris}, H.~V., et~al., 2003, \apjs, 148,
  175--194.

\bibitem{valageas01}
{Valageas}, P., {Balbi}, A., and {Silk}, J., 2001, \aap, 367, 1--17.

\bibitem{sunyaev72}
{Sunyaev}, R.~A. and {Zeldovich}, Y.~B., 1972, Comments on Astrophysics, 4,
  173.

\bibitem{komatsu02}
{Komatsu}, E. and {Seljak}, U., 2002, \mnras, 336, 1256--1270.

\bibitem{blanchard87}
{Blanchard}, A. and {Schneider}, J., 1987, \aap, 184, 1--2.

\bibitem{seljak96}
{Seljak}, U., 1996, \apj, 463, 1.

\bibitem{linde97}
{Kallosh}, R. and {Linde}, A., 1997, \prd, 56, 3509--3514.

\bibitem{contaldi99}
{Contaldi}, C.~R., {Bean}, R., and {Magueijo}, J., 1999, Physics Letters B,
  468, 189--194.

\bibitem{martin00}
{Martin}, J., {Riazuelo}, A., and {Sakellariadou}, M., 2000, \prd, 61, 83518.

\bibitem{bartolo02}
{Bartolo}, N., {Matarrese}, S., and {Riotto}, A., 2002, \prd, 65, 103505.

\bibitem{gangui02}
{Gangui}, A., {Martin}, J., and {Sakellariadou}, M., 2002, \prd, 66, 83502.

\bibitem{gupta02}
{Gupta}, S., {Berera}, A., {Heavens}, A.~F., and {Matarrese}, S., 2002, \prd,
  66, 43510.

\bibitem{liguori03}
{Liguori}, M., {Matarrese}, S., and {Moscardini}, L., 2003, Submitted to \apj,
  astro-ph/0306248.

\bibitem{bartolo03}
{Bartolo}, N., {Matarrese}, S., and {Riotto}, A., 2003, ArXiv Astrophysics
  e-prints, astro-ph/0309692.

\bibitem{gangui01}
{Gangui}, A., {Pogosian}, L., and {Winitzki}, S., 2001, \prd, 64, 43001.

\bibitem{levin02}
{Levin}, J., 2002, \physrep, 365, 251--333.

\bibitem{eriksen02}
{Eriksen}, H.~K., {Banday}, A.~J., and {G{\' o}rski}, K.~M., 2002, \aap, 395,
  409--415.

\bibitem{gott90}
{Gott}, J.~R.~I., {Park}, C., {Juszkiewicz}, R., et~al., 1990, \apj, 352,
  1--14.

\bibitem{schmalzing98}
{Schmalzing}, J. and {Gorski}, K.~M., 1998, \mnras, 297, 355--365.

\bibitem{novikov00}
{Novikov}, D., {Schmalzing}, J., and {Mukhanov}, V.~F., 2000, \aap, 364,
  17--25.

\bibitem{heavens01}
{Heavens}, A.~F. and {Gupta}, S., 2001, \mnras, 324, 960--968.

\bibitem{dore03}
{Dor{\' e}}, O., {Colombi}, S., and {Bouchet}, F.~R., 2003, \mnras, 344,
  905--916.

\bibitem{barreiro01}
{Barreiro}, R.~B., {Mart{\'{\i}}nez-Gonz{\' a}lez}, E., and {Sanz}, J.~L.,
  2001, \mnras, 322, 411--418.

\bibitem{phillips01}
{Phillips}, N.~G. and {Kogut}, A., 2001, \apj, 548, 540--549.

\bibitem{winitzki00}
{Winitzki}, S. and {Wu}, J. H.~P., ArXiv Astrophysics e-prints,
  astro-ph/0007213.

\bibitem{hu01}
{Hu}, W., 2001, \prd, 64, 83005.

\bibitem{kunz01}
{Kunz}, M., {Banday}, A.~J., {Castro}, P.~G., {Ferreira}, P.~G., and {G{\'
  o}rski}, K.~M., 2001, \apjl, 563, L99--L102.

\bibitem{komatsu01}
{Komatsu}, E. and {Spergel}, D.~N., 2001, \prd, 63, 63002.

\bibitem{komatsu02a}
{Komatsu}, E., {Wandelt}, B.~D., {Spergel}, D.~N., {Banday}, A.~J., and {G{\'
  o}rski}, K.~M., 2002, \apj, 566, 19--29.

\bibitem{detroia03}
{De Troia}, G., {Ade}, P.~A.~R., {Bock}, J.~J., et~al., 2003, \mnras, 343,
  284--292.

\bibitem{chiang02}
{Chiang}, L.-Y., {Christensen}, P.~R., {J{\" o}rgensen}, H.~E., et~al., 2002,
  A\&A, 392, 369--376.

\bibitem{marinucci02}
{Marinucci}, D. and {Piccioni}, M., 2002, Annals of Statistics, in press.

\bibitem{hansen02}
{Hansen}, F.~K., {Marinucci}, D., {Natoli}, P., and {Vittorio}, N., 2002, \prd,
  66, 63006.

\bibitem{hansen03}
{Hansen}, F.~K., {Marinucci}, D., and {Vittorio}, N., 2003, \prd, 67, 123004.

\bibitem{barreiro01a}
{Barreiro}, R.~B. and {Hobson}, M.~P., 2001, \mnras, 327, 813--828.

\bibitem{barreiro00}
{Barreiro}, R.~B., {Hobson}, M.~P., {Lasenby}, A.~N., et~al., 2000, \mnras,
  318, 475--481.

\bibitem{martinezgonzalez02}
Gonzalez, E.~M., Gallegos, J.~E., Argueso, F., Cayon, L., and Sanz, J.~L.,
  2002, MNRAS, 336, 22.

\bibitem{mukherjee00}
{Mukherjee}, P., {Hobson}, M.~P., and {Lasenby}, A.~N., 2000, \mnras, 318,
  1157--1163.

\bibitem{ferreira98}
{Ferreira}, P.~G., {Magueijo}, J., and {Gorski}, K.~M., 1998, \apjl, 503, L1.

\bibitem{banday00}
{Banday}, A.~J., {Zaroubi}, S., and {G{\' o}rski}, K.~M., 2000, \apj, 533,
  575--587.

\bibitem{komatsu03}
{Komatsu}, E., {Kogut}, A., {Nolta}, M.~R., et~al., 2003, \apjs, 148, 119--134.

\bibitem{chiang03}
{Chiang}, L., {Naselsky}, P.~D., {Verkhodanov}, O.~V., and {Way}, M.~J., 2003,
  \apjl, 590, L65--L68.

\bibitem{naselsky03a}
{Naselsky}, P.~D., {Verkhodanov}, O.~V., {Chiang}, L., and {Novikov}, I.~D.,
  2003, Submitted to \apj, astro-ph/0300235.

\bibitem{coles03}
{Coles}, P., {Dineen}, P., {Earl}, J., and {Wright}, D., 2003, Submitted to
  \mnras, astro-ph/0300252.

\bibitem{park03}
{Park}, C., 2003, Submitted to \mnras, astro-ph/0307469.

\bibitem{vielva03}
{Vielva}, P., {Martinez-Gonzalez}, E., {Barreiro}, R.~B., {Sanz}, J.~L., and
  {Cayon}, L., 2003, ArXiv Astrophysics e-prints, astro-ph/0300273.

\bibitem{smoot99}
{Smoot}, G.~F., 1999, ASP Conf. Ser. 181: Microwave Foregrounds, 61.

\bibitem{haslam82}
{Haslam}, C.~G.~T., {Stoffel}, H., {Salter}, C.~J., and {Wilson}, W.~E., 1982,
  \aaps, 47, 1.

\bibitem{reich86}
{Reich}, P. and {Reich}, W., 1986, \aaps, 63, 205--288.

\bibitem{jonas98}
{Jonas}, J.~L., {Baart}, E.~E., and {Nicolson}, G.~D., 1998, \mnras, 297,
  977--989.

\bibitem{platania03}
{Platania}, P., {Burigana}, C., {Maino}, D., et~al., 2003, Accepted for
  publication in \aap, astro-ph/0303031.

\bibitem{bennett03}
{Bennett}, C.~L., {Halpern}, M., {Hinshaw}, G., et~al., 2003, \apjs, 148,
  1--27.

\bibitem{tegmark96}
{Tegmark}, M. and {Efstathiou}, G., 1996, \mnras, 281, 1297--1314.

\bibitem{lasenby97}
{Lasenby}, A.~N., 1997, Microwave Background Anistropies, 453.

\bibitem{baccigalupi01}
{Baccigalupi}, C., {Burigana}, C., {Perrotta}, F., et~al., 2001, \aap, 372,
  8--21.

\bibitem{baccigalupi03}
{Baccigalupi}, C., {Perrotta}, F., {De Zotti}, G., et~al., 2003, Accepted for
  publication in \mnras.

\bibitem{paladini03a}
{Paladini}, R., {Burigana}, C., {Davies}, R.~D., et~al., 2003, \aap, 397,
  213--226.

\bibitem{paladini03b}
{Paladini}, R., {DeZotti}, G., and {Davies}, R.~D., 2003, Submitted to \mnras,
  astro-ph/0309350.

\bibitem{finkbeiner03}
{Finkbeiner}, D.~P., 2003, \apjs, 146, 407--415.

\bibitem{dickinson03}
{Dickinson}, C., {Davies}, R.~D., and {Davis}, R.~J., 2003, \mnras, 341,
  369--384.

\bibitem{kogut96}
{Kogut}, A., {Banday}, A.~J., {Bennett}, C.~L., et~al., 1996, \apj, 460, 1.

\bibitem{schlegel98}
{Schlegel}, D.~J., {Finkbeiner}, D.~P., and {Davis}, M., 1998, \apj, 500, 525.

\bibitem{banday03}
{Banday}, A.~J., {Dickinson}, C., {Davies}, R.~D., {Davis}, R.~J., and
  {Gorski}, K.~M., 2003, Accepted for publication in \mnras, astro-ph/0302181.

\bibitem{maino03}
{Maino}, D., {Banday}, A.~J., {Baccigalupi}, C., {Perrotta}, F., and {Gorski},
  K.~M., 2003, \mnras, 344, 544.

\bibitem{deoliveira97}
{de Oliveira-Costa}, A., {Kogut}, A., {Devlin}, M.~J., et~al., 1997, \apjl,
  482, L17.

\bibitem{leitch97}
{Leitch}, E.~M., {Readhead}, A.~C.~S., {Pearson}, T.~J., and {Myers}, S.~T.,
  1997, \apjl, 486, L23.

\bibitem{draine98}
{Draine}, B.~T. and {Lazarian}, A., 1998, \apjl, 494, L19.

\bibitem{franceschini89}
{Franceschini}, A., {Toffolatti}, L., {Danese}, L., and {De Zotti}, G., 1989,
  \apj, 344, 35--45.

\bibitem{toffolatti98}
{Toffolatti}, L., {Argueso Gomez}, F., {De Zotti}, G., et~al., 1998, \mnras,
  297, 117--127.

\bibitem{dezotti99}
{De Zotti}, G., {Toffolatti}, L., {Arg{\" u}eso}, F., et~al., 1999, AIP Conf.
  Proc. 476: 3K cosmology, 204.

\bibitem{mennella03}
{Mennella}, A., {Bersanelli}, M., {Seiffert}, M., et~al., 2003, \aap, 410,
  1089--1100.

\bibitem{bersanelli98}
{Bersanelli}, M., {Mattaini}, E., {Santambrogio}, E., et~al., 1998,
  Experimental Astronomy, 8, 231--238.

\bibitem{barnes02}
{Barnes}, C., {Limon}, M., {Page}, L., et~al., 2002, ApJ Suppl. Series, 143,
  567--576.

\bibitem{murphy02}
{Murphy}, J.~A., {Colgan}, R., {Gleeson}, E., et~al., 2002, AIP Conf. Proc.
  616: Experimental Cosmology at Millimetre Wavelengths, 282--289.

\bibitem{villa03}
{Villa}, F., {Sandri}, M., {Mandolesi}, N., et~al., 2003, Experimental
  Astronomy, in press.

\bibitem{welford78}
{Welford}, W.~T. and {Winston}, R., 1978, {The optics of nonimaging
  concentrators - Light and solar energy}, New York: Academic Press, 1978.

\bibitem{burigana98}
{Burigana}, C., {Maino}, D., {Mandolesi}, N., et~al., 1998, \aaps, 130,
  551--560.

\bibitem{fosalba02}
{Fosalba}, P., {Dor{\' e}}, O., and {Bouchet}, F.~R., 2002, Physical Review D,
  65, 63003.

\bibitem{mandolesi00}
{Mandolesi}, N., {Bersanelli}, M., {Burigana}, C., et~al., 2000, A\& A Suppl.,
  145, 323--340.

\bibitem{page03a}
{Page}, L., {Barnes}, C., {Hinshaw}, G., et~al., 2003, Apj Suppl. Series, 148,
  39--50.

\bibitem{arnau00}
{Arnau}, J.~V. and {S\'aez}, D., 2000, New Astronomy, 5, 121--135.

\bibitem{burigana03a}
{Burigana}, C. and {S{\' a}ez}, D., 2003, \aap, 409, 423--437.

\bibitem{dubruel00}
{Dubruel}, D., {Fargan}, G., {Cornut}, M., et~al., 2000, ESA Conf. Proc. SP444:
  AP2000 Millennium Conference on Antennas \& Propagation.

\bibitem{villa02}
{Villa}, F., {Bersanelli}, M., {Burigana}, C., et~al., 2002, AIP Conf. Proc.
  616: Experimental Cosmology at Millimetre Wavelengths, 224--228.

\bibitem{page03b}
{Page}, L., {Jackson}, C., {Barnes}, C., et~al., 2003, ApJ, 585, 566--586.

\bibitem{sandri03a}
{Sandri}, F., {Bersanelli}, M., {Burigana}, C., et~al., 2003, ESA Conf. Proc.
  WPP-212: 3rd ESA Workshop on Millimeter Wave Technology and Applications,
  199.

\bibitem{sandri03b}
{Sandri}, M., {Villa}, F., {Nesti}, R., et~al., 2003, Submitted to A\& A,
  astro-ph/0305152, also in astro-ph/0305152.

\bibitem{burigana03}
{Burigana}, C., {Sandri}, M., {Villa}, F., et~al., 2003, Submitted to A\& A,
  astro-ph/0303645, also in astro-ph/0303645.

\bibitem{ziel76}
{van der Ziel}, A., 1976, {Noise in Measurements}, New York: John Wiley and
  Sons, 1976.

\bibitem{einstein06}
Einstein, A., 1906, Ann. Phys, 19, 289.

\bibitem{johnson28}
{Johnson}, J.~B., 1928, Physical Review, 32, 97--109.

\bibitem{nyquist28}
{Nyquist}, H., 1928, Physical Review, 32, 110--113.

\bibitem{schottky18}
{Schottky}, W., 1918, Ann. Phys, 57, 541.

\bibitem{wong03}
{Wong}, H., 2003, Microelectron. Reliab., 43, 585--599.

\bibitem{smoot90}
{Smoot}, G., {Bennett}, C., {Weber}, R., et~al., 1990, \apj, 360, 685--695.

\bibitem{jarosik03}
{Jarosik}, N., {Bennett}, C.~L., {Halpern}, M., et~al., 2003, \apjs, 145,
  413--436.

\bibitem{seiffert02}
{Seiffert}, M., {Mennella}, A., {Burigana}, C., et~al., 2002, \aap, 391,
  1185--1197.

\bibitem{maris00}
{Maris}, M., {Maino}, D., {Burigana}, C., and {Pasian}, F., 2000, \aaps, 147,
  51--74.

\bibitem{maris03}
{Maris}, M., {Maino}, D., {Burigana}, C., et~al., 2003, Accepted for
  publication in \aap, astro-ph/0304089.

\bibitem{maino03a}
{Maino}, D., {Burigana}, C., and {Pasian}, F., 2003, New Astronomy, 8,
  711--718.

\bibitem{lineweaver94}
{Lineweaver}, C.~H., {Smoot}, G.~F., {Bennett}, C.~L., et~al., 1994, \apj, 436,
  452--455.

\bibitem{wright96}
{Wright}, E.~L., {Hinshaw}, G., and {Bennett}, C.~L., 1996, \apjl, 458,
  L53--L55.

\bibitem{hinshaw03}
{Hinshaw}, G., {Barnes}, C., {Bennett}, C.~L., et~al., 2003, \apjs, 148,
  63--95.

\bibitem{tegmark97}
{Tegmark}, M., 1997, \apjl, 480, L87.

\bibitem{natoli01}
{Natoli}, P., {de Gasperis}, G., {Gheller}, C., and {Vittorio}, N., 2001, \aap,
  372, 346--356.

\bibitem{dore01}
{Dor{\' e}}, O., {Teyssier}, R., {Bouchet}, F.~R., {Vibert}, D., and {Prunet},
  S., 2001, \aap, 374, 358--370.

\bibitem{maino02a}
{Maino}, D., {Burigana}, C., {G{\' o}rski}, K.~M., {Mandolesi}, N., and
  {Bersanelli}, M., 2002, \aap, 387, 356--365.

\bibitem{burigana01a}
C.~{Burigana}, C., {Natoli}, P., {Vittorio}, N., {Mandolesi}, N., and
  {Bersanelli}, M., 2001, Experimental Astronomy, 12, 87--106.

\bibitem{terenzi02}
{Terenzi}, L., {Bersanelli}, M., {Burigana}, C., et~al., 2002, AIP Conf. Proc.
  616: Experimental Cosmology at Millimetre Wavelengths, 245--247.

\bibitem{cappellini03}
{Cappellini}, P., {Maino}, D., {Albetti}, G., et~al., 2003, \aap, 409,
  375--385.

\bibitem{maris03a}
{Maris}, M., {Burigana}, C., {Cremonese}, G., {Marzari}, F., and {Fogliani},
  S., 2003, to be published in Proc. of the 5th Italian Conference of Planetary
  Sciences, Gallipoli, Lecce, Italy, September 15-19.

\bibitem{maris03b}
{Maris}, M., {Burigana}, C., {Cremonese}, G., et~al., 2003, to be published in
  Proc. XLVII National Conference of the Italian Astronomical Society, Trieste,
  Italy, April 14-17.

\bibitem{kogut96a}
{Kogut}, A., {Banday}, A.~J., {Bennett}, C.~L., et~al., 1996, \apj, 470, 653.

\bibitem{gorski96}
{Gorski}, K.~M., {Banday}, A.~J., {Bennett}, C.~L., et~al., 1996, \apjl, 464,
  L11.

\bibitem{hinshaw96}
{Hinshaw}, G., {Banday}, A.~J., {Bennett}, C.~L., et~al., 1996, \apjl, 464,
  L25.

\bibitem{romeo01}
{Romeo}, G., {Ali}, S., {Femen{\'{\i}}a}, B., et~al., 2001, \apjl, 548, L1--L4.

\bibitem{davies96}
{Davies}, R.~D., {Watson}, R.~A., and {Gutierrez}, C.~M., 1996, \mnras, 278,
  925--939.

\bibitem{dicker99}
{Dicker}, S.~R., {Melhuish}, S.~J., {Davies}, R.~D., et~al., 1999, \mnras, 309,
  750--760.

\bibitem{gundersen95}
{Gundersen}, J.~O., {Lim}, M., {Staren}, J., et~al., 1995, \apjl, 443,
  L57--L60.

\bibitem{platt97}
{Platt}, S.~R., {Kovac}, J., {Dragovan}, M., {Peterson}, J.~B., and {Ruhl},
  J.~E., 1997, \apjl, 475, L1.

\bibitem{coble99}
{Coble}, K., {Dragovan}, M., {Kovac}, J., et~al., 1999, \apjl, 519, L5--L8.

\bibitem{peterson00}
{Peterson}, J.~B., {Griffin}, G.~S., {Newcomb}, M.~G., et~al., 2000, \apjl,
  532, L83--L86.

\bibitem{wollack97}
{Wollack}, E.~J., {Devlin}, M.~J., {Jarosik}, N., et~al., 1997, \apj, 476, 440.

\bibitem{devlin98}
{Devlin}, M.~J., {de Oliveira-Costa}, A., {Herbig}, T., et~al., 1998, \apjl,
  509, L69--L72.

\bibitem{leitch00}
{Leitch}, E.~M., {Readhead}, A.~C.~S., {Pearson}, T.~J., et~al., 2000, \apj,
  532, 37--56.

\bibitem{baker99}
{Baker}, J.~C., {Grainge}, K., {Hobson}, M.~P., et~al., 1999, \mnras, 308,
  1173--1178.

\bibitem{partridge97}
{Partridge}, R.~B., {Richards}, E.~A., {Fomalont}, E.~B., {Kellermann}, K.~I.,
  and {Windhorst}, R.~A., 1997, \apj, 483, 38.

\bibitem{subrahmanyan98}
{Subrahmanyan}, R., {Kesteven}, M.~J., {Ekers}, R.~D., {Sinclair}, M., and
  {Silk}, J., 1998, \mnras, 298, 1189--1197.

\bibitem{holzapfel97}
{Holzapfel}, W.~L., {Wilbanks}, T.~M., {Ade}, P.~A.~R., et~al., 1997, \apj,
  479, 17.

\bibitem{halverson02}
{Halverson}, N.~W., {Leitch}, E.~M., {Pryke}, C., et~al., 2002, \apj, 568,
  38--45.

\bibitem{kovac02}
{Kovac}, J.~M., {Leitch}, E.~M., {Pryke}, C., et~al., 2002, \nat, 420,
  772--787.

\bibitem{pearson03}
{Pearson}, T.~J., {Mason}, B.~S., {Readhead}, A.~C.~S., et~al., 2003, \apj,
  591, 556--574.

\bibitem{kuo02}
{Kuo}, C.~L., {Ade}, P.~A.~R., {Bock}, J.~J., et~al., 2002, Submitted to \apj,
  astro-ph/0202289.

\bibitem{meyer91}
{Meyer}, S.~S., {Cheng}, E.~S., and {Page}, L.~A., 1991, \apjl, 371, L7--L9.

\bibitem{debernardis94}
{de Bernardis}, P., {de Gasperis}, G., {Masi}, S., and {Vittorio}, N., 1994,
  \apjl, 433, L1--L4.

\bibitem{tucker97}
{Tucker}, G.~S., {Gush}, H.~P., {Halpern}, M., and {Towlson}, W., 1997,
  Microwave Background Anistropies, 167.

\bibitem{tanaka96}
{Tanaka}, S.~T., {Clapp}, A.~C., {Devlin}, M.~J., et~al., 1996, \apjl, 468,
  L81.

\bibitem{debernardis00}
{de Bernardis}, P., {Ade}, P.~A.~R., {Bock}, J.~J., et~al., 2000, \nat, 404,
  955--959.

\bibitem{netterfield02}
{Netterfield}, C.~B., {Ade}, P.~A.~R., {Bock}, J.~J., et~al., 2002, \apj, 571,
  604--614.

\bibitem{hanany00}
{Hanany}, S., {Ade}, P., {Balbi}, A., et~al., 2000, \apjl, 545, L5--L9.

\bibitem{balbi00}
{Balbi}, A., {Ade}, P., {Bock}, J., et~al., 2000, \apjl, 545, L1--L4.

\bibitem{benoit03}
{Beno{\^ i}t}, A., {Ade}, P., {Amblard}, A., et~al., 2003, \aap, 399, L19--L23.

\bibitem{bennett03a}
{Bennett}, C.~L., {Bay}, M., {Halpern}, M., et~al., 2003, \apj, 583, 1--23.

\bibitem{gorski98}
{Gorski}, K., {Hivon}, E., and {Wandelt}, B., 1998, MPA/ESO Cosmology
  Conference "Evolution of Large-Scale Structure", eds. A.J. Banday, R.S. Sheth
  and L. Da Costa, PrintPartners Ipskamp, NL, 37--42.

\bibitem{bennett03b}
{Bennett}, C.~L., {Hill}, R.~S., {Hinshaw}, G., et~al., 2003, \apjs, 148,
  97--117.

\bibitem{efstathiou03}
{Efstathiou}, G., 2003, \mnras, 343, L95--L98.

\bibitem{roddis03}
{Roddis}, N., {Kettle}, D., {Winder}, F., et~al., 2003, 3$^{\rm rd}$ ESA
  Workshop on millimetre wave technology and applications, 81.

\bibitem{sjoman03}
{Sj$\ddot{\rm o}$man }, P., {Hughes}, N., {Jukkala}, P., et~al., 2003, 3$^{\rm
  rd}$ ESA Workshop on millimetre wave technology and applications, 75.

\bibitem{lamarre97}
{Lamarre}, J.-M., 1997, Microwave Background Anistropies, 31.

\bibitem{church96}
{Church}, S., {Philhour}, B., {Lange}, A., et~al., 1996, Proceedings of the
  30th ESLAB Symposium ``Submillimetre and Far-Infrared Space
  Instrumentation'', 77.

\bibitem{bock96}
{Bock}, J., {DelCastillo}, H., {Turner}, A., et~al., 1996, Proceedings of the
  30th ESLAB Symposium ``Submillimetre and Far-Infrared Space
  Instrumentation'', 119.

\bibitem{lange96}
{Lange}, A., {Church}, S., {Mauskopf}, P., et~al., 1996, Proceedings of the
  30th ESLAB Symposium ``Submillimetre and Far-Infrared Space
  Instrumentation'', 105.

\bibitem{delabrouille02}
{Kaplan}, J. and {Delabrouille}, J., 2002, AIP Conf. Proc. 609: Astrophysical
  Polarized Backgrounds, 209.

\bibitem{bard84}
Bard, S., 1984, J. Spacecraft \& Rockets, 21, 150--155.

\bibitem{wade00}
Wade, L., Bhandari, P., Bowman, Jr., R., et~al., 2000, vol. 45A, 499--506.

\bibitem{prina02}
Prina, M., Bhandari, P., Bowman, R., et~al., 2002, Advances in Cryogenic
  Engineering, AIP Conference Proceeding Series, New York, vol. 613,
  1201--1208.

\bibitem{pearson03a}
Pearson, D., Prina, M., Borders, J., et~al., 2003, {Test Performance of a
  Closed Cycle Continuous Hydrogen Sorption Cryocooler}, Kluwer Academic/Plenum
  Publishers, vol.~12.

\bibitem{bowman03}
Bowman, Jr., R., Prina, M., Barber, D., et~al., 2003, {Evaluation of Hydride
  Compressor Elements for the Planck Sorption Cooler}, Kluwer Academic/Plenum
  Publishers, vol.~12.

\bibitem{bradshaw97}
Bradshaw, T. and Orlowska, A., 1997, Proc. 6th European Symposium on Space
  Environmental Control Systems, ESA, vol. SP400, 465--470.

\bibitem{benoit91}
Benoit, A. and Pugeol, S., 1991, Physica B, 169, 457--458.

\bibitem{benoit97}
{Beno{\^ i}t et al.}, A., 1997, ESA SP-400: Sixth European Symposium on Space
  Environmental Control Systems, 497--502.

\bibitem{hobson98}
{Hobson}, M.~P., {Jones}, A.~W., {Lasenby}, A.~N., and {Bouchet}, F.~R., 1998,
  \mnras, 300, 1--29.

\bibitem{stolyarov02}
{Stolyarov}, V., {Hobson}, M.~P., {Ashdown}, M.~A.~J., and {Lasenby}, A.~N.,
  2002, \mnras, 336, 97--111.

\bibitem{patanchon03}
{Patanchon}, G., {Snoussi}, H., {Cardoso}, J.~F., and {Delabrouille}, J., 2003,
  ArXiv Astrophysics e-prints, astro-ph/0302078.

\bibitem{maino02}
{Maino}, D., {Farusi}, A., {Baccigalupi}, C., et~al., 2002, \mnras, 334,
  53--68.

\bibitem{mason03}
{Mason}, B.~S., {Pearson}, T.~J., {Readhead}, A.~C.~S., et~al., 2003, \apj,
  591, 540--555.

\bibitem{percival01}
{Percival}, W.~J., {Baugh}, C.~M., {Bland-Hawthorn}, J., et~al., 2001, \mnras,
  327, 1297--1306.

\bibitem{croft02}
{Croft}, R.~A.~C., {Weinberg}, D.~H., {Bolte}, M., et~al., 2002, \apj, 581,
  20--52.

\bibitem{gnedin02}
{Gnedin}, N.~Y. and {Hamilton}, A.~J.~S., 2002, \mnras, 334, 107--116.

\bibitem{deoliveira03}
{de Oliveira-Costa}, A., {Tegmark}, M., {Zaldarriaga}, M., and {Hamilton}, A.,
  2003, ArXiv Astrophysics e-prints, astro-ph/0307282.

\bibitem{popa01}
{Popa}, L.~A., {Burigana}, C., and {Mandolesi}, N., 2001, \apj, 558, 10--22.

\bibitem{mather96}
{Mather}, J.~C., 1996, IAU Symp. 168: Examining the Big Bang and Diffuse
  Background Radiations, 419.

\bibitem{rees68}
{Rees}, M.~J., 1968, \apjl, 153, L1.

\bibitem{zaldarriaga97}
{Zaldarriaga}, M., 1997, \prd, 55, 1822--1829.

\bibitem{hedman01}
{Hedman}, M.~M., {Barkats}, D., {Gundersen}, J.~O., {Staggs}, S.~T., and
  {Winstein}, B., 2001, \apjl, 548, L111--L114.

\bibitem{carlstrom03}
{Carlstrom}, J.~E., {Kovac}, J., {Leitch}, E.~M., and {Pryke}, C., 2003, To be
  published in the proceedings of ``The Cosmic Microwave Background and its
  Polarization'', New Astronomy Reviews, (eds. S. Hanany and K.A. Olive),
  astro-ph/0308478.

\bibitem{padin02}
{Padin}, S., {Shepherd}, M.~C., {Cartwright}, J.~K., et~al., 2002, \pasp, 114,
  83--97.

\bibitem{barkats03}
{Barkats}, D., 2003, To be published in the proceedings of ``The Cosmic
  Microwave Background and its Polarization'', New Astronomy Reviews (eds. S.
  Hanany and K. A.Olive), astro-ph/0306002.

\bibitem{meinhold03}
{Meinhold}, P.~R., {Bersanelli}, M., {Childers}, J., et~al., 2003, Submitted to
  \apj, astro-ph/0302034.

\bibitem{keating03}
{Keating}, B.~G., {Ade}, P.~A.~R., {Bock}, J.~J., et~al., 2003, Polarimetry in
  Astronomy. Edited by Silvano Fineschi . Proceedings of the SPIE, Volume 4843,
  pp. 284-295 (2003)., 284--295.

\bibitem{montroy03}
{Montroy}, T., {Ade}, P.~A.~R., {Balbi}, A., et~al., 2003, To be published in
  the proceedings of ``The Cosmic Microwave Background and its Polarization'',
  New Astronomy Reviews, (eds. S. Hanany and K.A. Olive), astro-ph/0305593.

\bibitem{cortiglioni00}
{Cortiglioni}, S., {Carretti}, E., {Orsini}, M., et~al., 2000, AIP Conf. Proc.
  504: Space Technology and Applications International Forum, 91.

\bibitem{zannoni02}
{Zannoni}, M., {Baralis}, M., {Bernardi}, G., et~al., 2002, Astrophysical
  Polarized Backgrounds, held 9-12 October, 2001 in Bologna Italy. Edited by
  Stefano Cecchini, Stefano Cortiglioni, Robert Sault, and Carla Sbarra. AIP
  Conference Proceedings, Vol. 609. Melville, NY: American Institute of
  Physics, 115.

\bibitem{dowell03}
{Dowell}, C.~D., {Allen}, C.~A., {Babu}, R.~S., et~al., 2003, Millimeter and
  Submillimeter Detectors for Astronomy. Edited by Phillips, Thomas G.;
  Zmuidzinas, Jonas. Proceedings of the SPIE, Volume 4855, pp. 73-87 (2003).,
  73--87.

\bibitem{stebbins97}
{Stebbins}, A., 1997, astro-ph/9709065.

\bibitem{aghanim97}
{Aghanim}, N., {de Luca}, A., {Bouchet}, F.~R., {Gispert}, R., and {Puget},
  J.~L., 1997, \aap, 325, 9--18.

\bibitem{seljak00}
{Seljak}, U. and {Zaldarriaga}, M., 2000, \apj, 538, 57--64.

\bibitem{platania02}
{Platania}, P., {Burigana}, C., {De Zotti}, G., {Lazzaro}, E., and
  {Bersanelli}, M., 2002, \mnras, 337, 242--246.

\bibitem{kesteven02}
{Kesteven}, M., 2002, Astrophysical Polarized Backgrounds, held 9-12 October,
  2001 in Bologna Italy. Edited by Stefano Cecchini, Stefano Cortiglioni,
  Robert Sault, and Carla Sbarra. AIP Conference Proceedings, Vol. 609.
  Melville, NY: American Institute of Physics, 2002., p.156, 156.

\bibitem{dezotti03}
{De Zotti}, G., {Burigana}, C., {Cavaliere}, A., et~al., 2003, to be published
  in Proc. of International Symposium Plasmas in the Laboratory and in the
  Universe: new insights and new challenges, Como, Italy, September 16-19.

\end{thebibliography}

\end{document}